\documentclass[aps,prx,onecolumn,notitlepage,superscriptaddress]{revtex4-2}

\usepackage{bm,dsfont} \usepackage{graphicx} \usepackage{amsmath,amsthm,stmaryrd,amssymb} 
\usepackage{hyperref}
\usepackage{enumitem,comment}
\usepackage{tikz}
\usetikzlibrary{quantikz2}
\usepackage[flushleft]{threeparttable} \usepackage{booktabs}

\DeclareMathOperator*{\Tr}{Tr}
\DeclareMathOperator*{\ave}{\mathds E}
\DeclareMathOperator*{\argmin}{argmin}

\newcommand{\bs}[1]{\boldsymbol{#1}}

\newtheorem{thm}{Theorem}[section]
\newtheorem{lemma}[thm]{Lemma}
\newtheorem{coro}[thm]{Corollary}

\usepackage{titlesec}

\titleformat*{\section}{\Large\bfseries\sffamily}
\titleformat*{\subsection}{\large\bfseries\sffamily}
\titleformat*{\subsubsection}{\normalsize\bfseries\sffamily}

\makeatletter
\renewcommand{\p@subsection}{}
\renewcommand{\p@subsubsection}{}
\makeatother

\begin{document}

\title{Statistical Complexity of Quantum Learning}

\author{Leonardo Banchi} \email{leonardo.banchi@unifi.it}
\affiliation{Department of Physics and Astronomy, University of Florence, 
via G. Sansone 1, I-50019 Sesto Fiorentino (FI), Italy}
\affiliation{ INFN Sezione di Firenze, via G. Sansone 1, I-50019, Sesto Fiorentino (FI), Italy }

\author{Jason Luke Pereira}
\affiliation{ INFN Sezione di Firenze, via G. Sansone 1, I-50019, Sesto Fiorentino (FI), Italy }

\author{Sharu Theresa Jose} 
\affiliation{
Department of Computer Science, 
University of Birmingham
Birmingham, B15 2TT, UK}

\author{Osvaldo Simeone}
\affiliation{
	King's Learning and Information Processing lab (KCLIP) \& \\
	Centre for Intelligent Information Processing Systems (CIIPS)\& \\
Department of Engineering, King's College London, \\  Strand, London, WC2R 2LS, UK
}



\begin{abstract}
	Recent years have seen significant activity on the problem of using data for the purpose of learning properties of quantum systems or of processing classical or quantum data via quantum computing. As in classical learning, quantum learning problems involve settings in which the mechanism generating the data is unknown, and  the main goal of a learning algorithm is to ensure satisfactory accuracy levels when only given access to data and, possibly, side information such as expert knowledge. This article has two main goals. First, it 
reviews and generalizes different results on the complexity of quantum
		learning, using information-theoretic techniques and focusing on data
	complexity and model complexity.  
Second, it introduces the notion of copy complexity, which quantifies  the number of copies of a quantum state required to achieve a target accuracy level.  Copy  complexity arises from the destructive nature of quantum measurements, which irreversibly alter the state to be processed, limiting the information that can be extracted about quantum data.  As a result, in a quantum system, unlike in classical machine learning, it is generally not possible to evaluate the training loss simultaneously on multiple hypotheses  using the same quantum data, making empirical risk minimization generally inapplicable. Furthermore, copy complexity also affects the performance at test time, as the information that can be obtained from a quantum system increases with the number of available copies.  The paper presents novel results on the copy complexity for both training and testing.  To make the paper self-contained and approachable by different research communities, we provide extensive background material on classical results from 
statistical learning theory, as well as on the distinguishability of quantum states. Throughout, we highlight the differences between quantum and classical learning by addressing both supervised and unsupervised learning, and we provide extensive pointers to the literature.

\end{abstract}

\maketitle

\tableofcontents

\section{Introduction and  Summary }\label{sec:summary}

In this section, we provide an introduction to the topic of statistical complexity of quantum learning and we highlight some key results that will be elaborated on in the rest of the paper.

\subsection{Scope}
 
Quantum information theory addresses the implications of quantum mechanics on the representation, storage, and transmission of information in microscopic physical systems. Mathematically, it centers on  the characterization of probabilistic and statistical aspects of observations made on quantum systems through the lens of information and uncertainty quantification. In this context, recent years have seen significant activity on the problem of using observations -- i.e., data -- for the purpose of learning properties of quantum systems or of processing classical or quantum data via quantum means \cite{biamonte2017quantum,gebhart2023learning,schuld2021machine,simeone2022introduction,ciliberto2018quantum,dunjko2018machine}.

In this article, we study the complexity of quantum learning using 
information-theoretic techniques. As in classical learning, quantum learning problems involve settings in which the mechanism generating the data is unknown, and  the main goal of a learning algorithm is to ensure satisfactory accuracy levels when only given access to data and, possibly, side information such as expert knowledge. 

The complexity of quantum learning is a multi-faceted concept, and in this article we focus on the following three aspects:
\begin{itemize}
    \item \emph{Data complexity}: As in classical machine learning, limitations on the amount of available data -- which may be classical or quantum -- play a key role in determining the achievable accuracy levels for data-driven methods. Data complexity refers to the requirements in terms of data set size on the performance of a learning algorithm. Throughout the paper, we will use the letter $N$ to indicate the number of training data examples.
   \item \emph{Training copy complexity}: The second form of complexity, first introduced in this work, arises due to 
    the destructive nature of quantum measurements, which 
irreversibly alter the state to be processed. As a result,  
the amount of information that can be extracted from quantum data depends on the number of copies available for each quantum state in the data set. This stands in contrast to classical data, which, barring computational complexity constraints, can be accessed, copied, and processed an arbitrary number of times. The disturbance caused by quantum measurements entails that, in general, in quantum machine learning it is not possible to evaluate the training loss simultaneously on multiple hypotheses  using the same quantum data. In fact,  each evaluation irreversibly modifies the quantum training data. Therefore, the training loss cannot be computed with arbitrary precision, and the \emph{training copy complexity} reflects the requirements on the number of copies of a training quantum state that are needed to ensure given accuracy levels. We will use the letter $S$ to indicate the number of copies of each training data example.
\item  \emph{Testing copy complexity}: Owing to the destructive nature of quantum measurements, the amount of information that can be extracted from a quantum state at test time also depends  on the number of copies of the state available to the quantum model as input. We will use the letter $V$ to denote the number of copies of a test state, and refer to the requirements on $V$ to achieve target accuracy levels as the \emph{testing copy complexity}.
    \item \emph{Model complexity}: Quantum algorithms are implemented via quantum circuits whose complexity can be measured in terms of number of qubits and number of gates. For a quantum learning algorithm to be successful, it is necessary to strike a balance between the expressivity enabled by a more complex model and trainability, which may be impaired by the adoption of larger models with more tunable parameters. 
\end{itemize}

Classical learning theory has traditionally focused on developing results that are agnostic to the distribution of the data, typically depending only on the complexity of the model \cite{shalev2014understanding}. In contrast, more recent information-theoretic analyses center on the interplay between learning algorithm, data distribution, and model complexity. In the context of quantum learning, data-dependent analyses are particularly useful in offering greater physical insights. In fact, since such analyses explicitly depend on the characteristics of the  
quantum data, they can reveal important physical aspects such as the features of quantum states that determine the data, copy, and/ or model complexity of learning.

Accordingly, in this article, we will quantitatively study data, copy, and model complexity by adopting an information-theoretic viewpoint. To make the paper self-contained, we will provide extensive background material on classical results from 
statistical learning theory, as well as on the distinguishability of quantum states. 
Then, we will extend some analyses from other papers, including
\cite{banchi2021generalization,arunachalam2017guest,huang2020power,huang2021information,caro2021encoding,caro2022generalization,du2022demystify,abbas2021power,haug2021capacity,gyurik2023structural,huang2022quantum}, 
presenting them in a unified framework, generalizing some key results, and presenting new applications.

\subsection{Examples}\label{sec: examples supervised unsupervised}     
In this article, we consider both supervised and unsupervised quantum learning problems. To illustrate the wide range of applications for the problems under study, this subsection lists some examples in both categories.

\textbf{Supervised learning:} In supervised learning problems, one is given data in the form of an input -- classical or quantum -- state and of a desired output, which is typically classical. Some examples of applications are as follows.

\begin{itemize}
	\item {\it Quantum classification of classical data}: Consider classical data $x$, e.g., images or text, that need to be classified. A quantum computer can be used for this purpose   by first embedding the classical data $x$ into a quantum state 
		$\rho(x)$, and by then using a quantum measurement to classify the resulting states. The embedding $\rho(x)$ is produced by a quantum circuit that may be optimized based on data along with the measurement, and the output of the circuit is a classification label $y$.

\item {\it Quantum classification of quantum sensed data}: Consider now the problem of detecting some physical property $y$ on the basis of quantum data collected by a quantum sensor. 
		For example, quantum light may be used to illuminate a sample $x$ under study, some photons are dispersed in the environment,  while others enter the detector. Here, quantum data is described by the quantum state $\rho(x)$ of the system collected by the detector, and $y$ represent some feature of the illuminated physical sample $x$ 
		that we want to extract. 
		Entangled light was shown to enhance the classification of sensed images 
		\cite{banchi2020quantum,zhuang2019physical}.

	\item {\it Classification of quantum phases of matter}. The phase of a quantum many-body system can be determined on the basis of the ground state, or some other equilibrium state, of the
		system, with classical input $x$ modeling the tunable external parameters, such as magnetic fields, coupling strengths among 
		particles and  the temperature \cite{carleo2019machine}.

	\item {\it Detecting entanglement}. Given a quantum state, one may be interested in assessing their entanglement structure, 
		e.g., whether they are separable or entangled  for a certain binary partition. This can be done by applying a quantum algorithm on a state $\rho(x)$, where the inputs $x$ model all the parameters in the (unitary) transformation used 
		to create the resulting state. 

	\item {\it Classification of noisy states.} Consider a source that emits two possible states, which are then 
		perturbed by an environment. The goal is to determine which of the two possible states was emitted by the source based on access to the   perturbed state $\rho$. 
\end{itemize}

\textbf{Unsupervised learning:} Unsupervised learning is a vast field that includes tasks as different as clustering and generative modelling. Classically, unsupervised learning refers generically to the task of inferring properties of the unknown data generation mechanism. In particular, in generative modeling, the ultimate goal is generating new data by sampling from the data distribution. Focusing on generative tasks, examples of unsupervised learning problems are as follows.

\begin{itemize}
   
    \item  {\it Generating classical data}: Quantum  models can be used to approximately sample from an unknown classical distribution. In this case, the statistical 
			description of the model is expressed as a quantum state, which has to be learnt from  data. 
			After training, new classical data are generated by applying measurements onto that state \cite{benedetti2019generative, situ2020quantum,romero2021variational}.

    \item {\it Loading a classical distribution into a quantum state}: Given an unknown data distribution $P(x)$, it may be of interest to generate  a quantum 
			superposition  $\ket{\psi} = \sum_x \sqrt{P(x)}\ket x$ that ``loads'' such distribution on a quantum system~\cite{scott2005learning}. This can enable the processing of classical distributions using quantum hardware. 

		\item {\it Quantum state approximation}: Given copies of a quantum state $\rho$, one may be interested in optimizing a quantum circuit that can produce systems in a state $\hat{\rho}$ that is approximately equal to  $\rho$, with applications 
			in \textit{quantum state compilation} \cite{ezzell2023quantum},
			quantum noise sensing \cite{braccia2022quantum},
			approximation of unknown pure \cite{benedetti2019adversarial} and  mixed states
			\cite{braccia2021enhance,ezzell2023quantum}.
\end{itemize}

\subsection{Learning Settings}\label{sec:settings}

 \begin{table}[t!]
	\centering
	\begin{tabular}{|p{1.5cm}|p{3.5cm}|p{6.4cm}|p{3.5cm}|}
		\hline
		& Unknowns &	Available information &  Processing inputs \\ 
		\hline
		Classical & Data distribution $P(x,y)$ & $N$ training samples $\{(x_n,y_n)\}_{n=1}^{N}$, test input $x$ & arbitrary copies of training data 
	$\{(x_n,y_n)\}_{n=1}^{N}$ and test input $x$ \\
		\hline
		Quantum &  Classical data distribution $P(x,y)$, 
 training states $\{\rho(x_n)\}_{n=1}^{N}$, test state $\rho(x)$  & classical inputs $\{x_n\}_{n=1}^{N}$, $S$ copies of training states $\{\rho(x_n)^{\otimes S}\}_{n=1}^{N}$, labels $\{y_n\}_{n=1}^N$, $V$ copies of test 
		state $\rho(x)^{\otimes V}$  & training data $\{(x_n,\rho(x_n)^{\otimes S},y_n)\}_{n=1}^{N}$ and test input $\rho(x)^{\otimes V}$ \\
		\hline
	\end{tabular}
	\caption{Unknowns, available information, and processing inputs for classical and quantum learning problems with a single test input, considering a representative case of quantum learning. 
		In classical learning, the data distribution $P(x,y)$ is unknown, and the available information amounts to  the training set and to the test input, which are assumed to be  independent samples from the unknown distribution $P(x,y)$. Arbitrary copies 
		of these data can be created and  manipulated during training and testing. 
		In the most general form of quantum learning, the quantum states $\rho(x_n)$ used for training and the quantum states $\rho(x)$ accessible for testing are 
		unknown. A quantum algorithm cannot make copies of unknown quantum states, which must be 
	provided as input. 
		However, unlike the inaccessible data distribution $P(x,y)$, quantum states can be manipulated via quantum 
		algorithms, e.g., via measurements. Classical inputs may or may not be available to the learner.  }
	
	\label{tab:known}
\end{table}

In this article, we study both supervised and unsupervised learning problems. A learner is given classical and/ or quantum data, and the goal is to extract information from data that can be used for tasks such as inference and data generation.

As illustrated in the first row of Table~\ref{tab:known},  in classical machine learning, a data point $(x,y)$ encompasses a classical input $x$, e.g., a vector of numbers, and a desired classical output $y$. Each pair  is assumed to be generated from an unknown data distribution $P(x,y)$. The learner has access to information in the form of $N$ training pairs $(x_n,y_n)$ with $n=1,...,N$ and to a test input $x$. The learner can copy this information at will to compute any number of times on the training data and test input. Note that, in the case of unsupervised learning, the classical label $y$ is not defined.

For quantum learning problems, a data point $(x,\rho(x),y)$ generally encompasses: a \emph{classical input} $x$;
a \emph{quantum input state} $\rho(x)$; 
and a \emph{classical output} $y$. 
Note that not all three elements must be present for all problems. For instance, in some settings, such as the classification of quantum sensed data or that of noisy states mentioned in the previous subsection, a data point may not include input $x$. Furthermore, as mentioned, for  unsupervised learning, the classical label $y$ is not present.

The classical input $x$ and the classical output $y$ -- jointly distributed according to an unknown joint distribution $P(x,y)$ -- can be processed and copied at will, while for the training quantum input state $\rho(x_n)$ we must distinguish between the following situations: 
\begin{itemize}
\item \emph{Known training quantum states}: In the first situation, the quantum states $\rho(x_n)$ from the training set are known, in the sense that classical descriptions of all density matrices $\rho(x_n)$ exist, either on the basis of a theoretical model or because of a prior full state tomography. Training may not involve any quantum hardware, as arbitrary copies of the training states can be made within a classical computer, just as in the classical setting. Accordingly, the performance of the learner is limited by the number $N$ of available examples, as for classical machine learning.

\item \emph{Unknown training quantum states, known state-preparation mechanism}: In the second type of problem, the training quantum states $\rho(x_n)$ are unknown, even when the classical inputs $x_n$ are available, and the learner can only gather \emph{partial} information about the states via measurements. However, a physical mechanism for creating training and test states is known, so a learner can repeat the same state-preparation procedure to create a number, $S$, of copies of each training state $\rho(x_n)$, and $V$ copies of the test state. Given the trade-off between disturbance and the amount of information extracted from a state by a measurement \cite{maccone2007entropic}, the number of copies available for the training and test states determine the degree to which the learner can access information about the training and test data, respectively.  Therefore, in this case, the performance of the learner is limited not only by the number $N$ of available examples, but also by the number of copies of each training state, $S$ and of the test state, $V$.

\item \emph{Unknown training quantum states, unknown state-preparation mechanism}: In this last type of problem, not only are the states $\rho(x_n)$ unknown, but 
the mechanism for creating them is also not available. For instance, in the problem of classifying noisy states described in the previous section, each state is generally different in the sense that  the specific perturbation applied to the state is typically not known. This extreme case may be reduced to the previous one by setting $S=1$, i.e., each training state is treated as an individual copy. In this case, it may still be meaningful to allow for a number $V$ of copies of the test state, which may be created by the unknown physical mechanism at test time.
\end{itemize}

Learning settings with unknown quantum states and known state-preparation mechanism are more general than those with known quantum states. This is in the sense that, as the number of copies available for each quantum state $\rho(x_n)$ or $\rho(x)$ grows, the learner can obtain a description of the corresponding states, which may hence considered as known. Accordingly, studying scenarios with known states is often useful as a stepping stone to address more challenging problems with unknown states. 

Referring to Table~\ref{tab:known}, a learner has generally access to 
\begin{itemize} 
\item a \emph{finite-copy}, or \emph{$S$-copy training set} $\mathcal{S}^{S}$ of $N$ \emph{training data} points $(x_n,\rho(x_n)^{\otimes S},y_n)$, for $n=1,\dots,N$, comprising classical input $x_n$, $S$ copies of the \emph{unknown} quantum state $\rho(x_n)$, and a classical output $y_n$;
\item and one or more \emph{test  inputs} $(x,\rho(x)^{\otimes V})$, for which one has $V$ copies of the quantum state $\rho(x)$,  to which the learner wishes to assign a classical label $y$. 
\end{itemize}
When the training states $\rho(x_n)$ are known, the learner is said to have access to the \emph{abstract training set}, denoted as $\mathcal{S}$, which contains a classical description of the states $\rho(x_n)$. The abstract training set  can be thought of, conceptually, as containing an infinite number of copies of the states.

\subsection{Inductive vs. Transductive Learning}

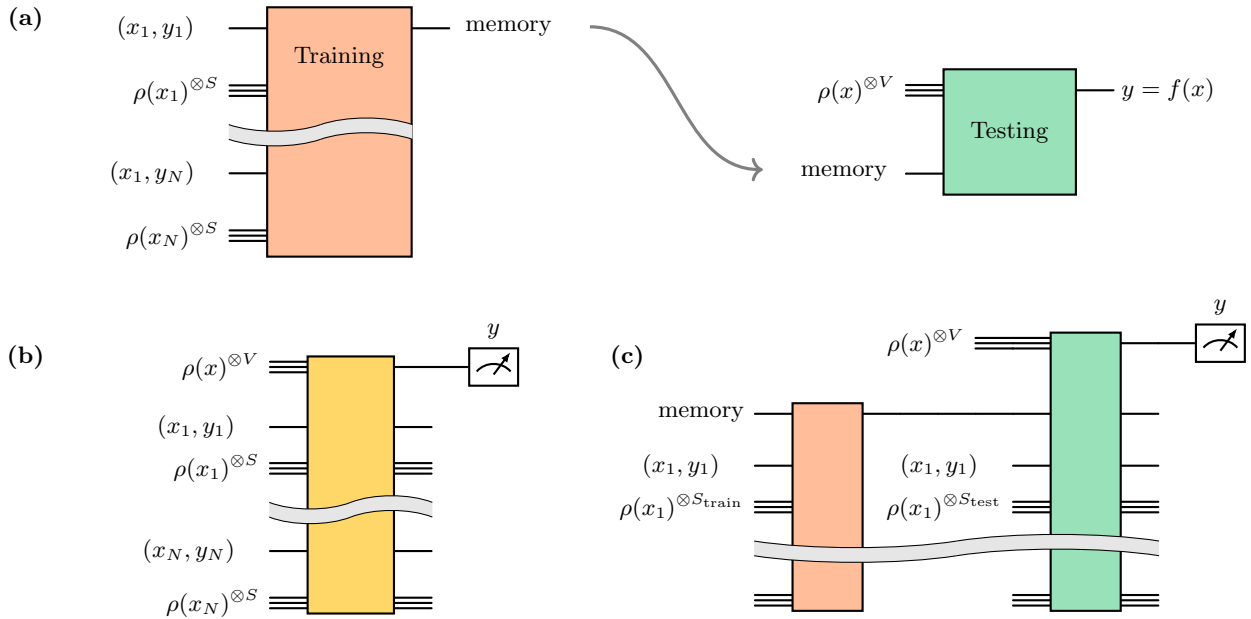
\begin{figure}[t!]
\begin{center}
 \begin{tikzpicture}[scale=1, transform shape]
 		\node at (-2,0) {
 				\begin{quantikz}[row sep=5.5mm,classical gap=0.07cm,wire types={q,b,q,q,b}]
					\lstick{$(x_1,y_1)$~~~} & \gate[5,label style={yshift=1cm},style={fill=orange!70!red!40}]{\rm ~~Training~~}& {\rm~~ memory}  \\
 					\lstick{ ${\rho(x_1)}^{\otimes S}$} &  \\
					\wave[black!10]&& \\
 					\lstick{$(x_1,y_N)$~~~} &\\ 
 					\lstick{ ${\rho(x_N)}^{\otimes S}$} &  
 				\end{quantikz}
 			}; 
 		\node at (7,0) {
 				\begin{quantikz}[row sep=5.5mm,classical gap=0.07cm,wire types={b,q}]
					\lstick{${\rho(x)}^{\otimes V}$} & \gate[2,style={fill=blue!30!green!40}]{~~\rm Testing~~} &\setwiretype{q}  ~{y=f(x)} \\
 					\lstick{memory~}&
 				\end{quantikz}
 			};
		\draw[very thick, gray, ->] (1.5,1.4) to [out=0,in=180] (3.8,-0.5);
		\node at (-6,1.5) {\bf (a)}; 
		\node at (-6,-3) {\bf (b)}; 
		\node at (2,-3) {\bf (c)}; 
 		\node at (-2,-4.5) {
 				\begin{quantikz}[row sep=5.5mm,classical gap=0.07cm,wire types={b,q,b,q,q,b}]
 					\lstick{${\rho(x)}^{\otimes V}$} & \gate[6,label style={yshift=0.3cm},style={fill=yellow!80!red!70}]{~~~~~~~~}& \setwiretype{q}& \meter[1]{y} \\
 					\lstick{$(x_1,y_1)$~~~} & & \\
 					\lstick{ ${\rho(x_1)}^{\otimes S}$} 
 										& & \\
 								\wave[black!10]&& \\
 					\lstick{$(x_N,y_N)$~~~} &&\\ 
 					\lstick{ ${\rho(x_N)}^{\otimes S}$}
 										& & 
 				\end{quantikz}
 			}; 
 		\node at (6,-4.3) {
 			\begin{quantikz}[row sep=5.5mm,classical gap=0.07cm,wire types={n,q,q,b,q,b},align equals at=1.5]
 				&&&&\lstick{${\rho(x)}^{\otimes V}$} &\setwiretype{b}& \gate[6,label style={yshift=0.3cm},style={fill=blue!30!green!40}]{~~~~~~}& \setwiretype{q}&  \meter[1]{y}\\
 				\lstick{memory} & \gate[5,label style={yshift=0.3cm},style={fill=orange!70!red!40}]{~~~~~~}&&&&&&\\
 				\lstick{$(x_1,y_1)$~~~} &&\setwiretype{n} & &&
 				\lstick{$(x_1,y_1)$~~~} & \setwiretype{q}& \\
 				\lstick{ ${\rho(x_1)}^{\otimes S_{\rm train}}$} 
 									&&\setwiretype{n} & &&
 				\lstick{ ${\rho(x_1)}^{\otimes S_{\rm test}}$} 
 									&\setwiretype{b} &\\
 							\wave[black!10]&&&&&&& \\
 							\lstick{ ~} &&\setwiretype{n} &&&& \setwiretype{b} &
 			\end{quantikz}
 		}; 

 \end{tikzpicture}
\end{center}
\caption{
		(a) In inductive learning, during the training phase (left), the training
		data $\mathcal{S}^S$ is used only once  to produce the classical
		description of an inference operation $f$, which is stored in a classical
		memory. During test (right), copies of the training data are no longer
		needed, and the inference operation $f$ can be used on an arbitrary number
		of test inputs $\rho(x)^{\otimes V}$
		to produce the prediction $y$. (b)
		In transductive learning, an $S$-copy training set $\mathcal{S}^S$ is
		jointly processed with the test input
		$\rho(x)^{\otimes V}$ to produce
		the prediction $y$. Therefore, new copies of the training set are required
		for each test input. 
		(c) Transductive learning with an induction step. Training proceeds as in (a), but 
		with less copies, $S_{\rm test}<S$. At test stage, the remaining copies $S_{\rm test}=S-S_{\rm train}$ 
		are processed together with the input $\rho(x)^{\otimes V}$ and the classical memory, created during 
		training, to produce the prediction $y$. 
}
\label{fig:strategies}
\end{figure}

We can distinguish two main ways to infer the output $y$ given the input test data and the training data, with the second being more general than the first.
\begin{itemize}
	\item \emph{Inductive learning}: As illustrated in Figure~\ref{fig:strategies}(a), inductive learning follows a two-step procedure, in which training data is only used in the first phase, and is no longer needed to make decisions on a new input. \begin{itemize} \item In the \emph{training phase}, inductive learning methods use training data to  obtain a general \emph{inference operation} $f\in\mathcal F$, among a chosen function class $\mathcal F$, which is classically described and stored in a classical memory. The inference operation may, e.g., amount to the specification of a quantum circuit or of a classical description of quantum states. Selection of the inference operation $f$ is typically done by comparing the performance of operation $f$ on the same available training data. For unknown states, quantum measurements require the ``consumption'' of different copies of the training states, and hence comparisons are inherently stochastic unless the number of copies $S$ is arbitrarily large. 
		\item During the \emph{testing phase}, the inference operation $f$ is applied to the test example to determine the output $y$.\end{itemize} 
			Accordingly, induction refers to the extraction of a general, classically describable, rule from examples. Importantly, the rule $f$ can be applied to any number of test examples, and training examples are no longer needed after the training phase. 
\item \emph{Transductive learning}: As illustrated in
	Figure~\ref{fig:strategies}(b), transductive learning strategies jointly
	process an $S$-copy version $\mathcal{S}^S$ of the training data set $\mathcal{S}$
	and the test example. Accordingly, producing the output $y$ for different test
	examples $(x,\rho(x)^{\otimes V})$ generally requires distinct copies of the
	training data.  This is unlike inductive methods, for which training data is
	no longer used for testing.  We will specifically target transductive
	strategies for the general case in which the quantum states $\rho(x)$ are
	\emph{unknown}. In this case, one needs to use the available copies
	of each training example as efficiently as possible, and the generality of
	transductive strategies may offer advantages in this regard. Transductive
	strategies may be ``universal'', in the sense that they amount to a single
	quantum operation that applies to any training data set and test input to
	produce a prediction. As a note of terminology,  one may still define an inference function for
	transductive learning, although such a function is not inferred from data, but
	rather enacted as a result of the single quantum operation on the training set and
	test example. More generally, as
	shown in  Figure~\ref{fig:strategies}(c), there
	are transductive strategies that also have inductive elements, storing some
	learned information in a classical memory. For such strategies, the inductive
	learning step consumes additional copies of the training set, allowing the
	quantum operation to be tailored to the given problem.
\end{itemize}

We end with two remarks on inductive and transductive learning. First, we observe that in classical machine learning, transductive schemes are similarly defined (see, e.g., \cite{vovk2005algorithmic}). Using classical transductive concepts, transduction may be useful to improve statistical efficiency in the case in which the quantum states $\rho(x)$ are known. For example, it could be used, in conjunction with conformal prediction strategies \cite{vovk2005algorithmic}, to enhance calibration (see also \cite{park2023quantum}).
Second, in principle, it is possible to replace the classical memory with a quantum memory in 
	the strategies of Figs.~\ref{fig:strategies}(a,c). However,
	in spite of the separation between training and testing, since a quantum memory 
	is destroyed after each measurement, there is no reusable information and 
	this approach can be cast in the purely transductive scheme of Figure~\ref{fig:strategies}(b).

\subsection{Architectures}\label{sec:architectures}

Table \ref{tab:strategies} provides  examples of inductive and transductive
quantum learning architectures, which are briefly reviewed next. We emphasize that transductive techniques only apply to the case of unknown states, in which copies
of the training states are needed to process each new input. 

 \begin{table}[t]
	\centering
	\begin{tabular}{|p{8cm}|c|}
		\hline
		Architectures 						& Learning type \\ 
		\hline
		Quantum neural networks 			& inductive \\ 
  Quantum reservoir computing 	& inductive \\
  Classical shadows 						& inductive \\
  Helstrom classifier						& transductive \\
		Quantum kernel methods 				& transductive \\
		\hline
	\end{tabular}
	\caption{Examples of architectures that will be reviewed in this article.
		Note that all strategies should be considered as inductive when a classical description of the training states 
		is available to the learner.
	}
	\label{tab:strategies}
\end{table}

\subsubsection{Inductive Learning} 

The following are popular instances of inductive learning.
\begin{itemize}
	\item \emph{Quantum neural networks} use a trainable parametric quantum circuit followed by 
		a fixed measurement to predict the outcome $y$. The label $y$ can be extracted by means of a single-shot measurement, via a classical 
		post-processing of multiple shots, or via the expectation value of a fixed observable \cite{schuld2021machine}.
		In this case, the model can be described classically via the circuit structure and the circuit parameters,
		while the circuit evaluation is done in a quantum hardware. Evaluation may be also implemented on classical hardware 
		for shallow circuits and few qubits.
		Training follows the inductive learning procedure of finding the optimal circuit parameters -- along with possibly other hyperparameters 
	such as the circuit structure. This is done via classical optimization. A typical approach is to leverage gradient descent schemes, whereby the gradient is evaluated by taking measurements of the output of the circuit, and variations thereof, thus requiring copies of the training data for each gradient evaluation. Furthermore, at testing time, the circuit with learnt parameters is employed to process, separately, any number of new inputs,  
		without requiring access to the training data.

	\item \emph{Quantum reservoir computing and extreme learning machines} represent an experimentally friendly 
		hybrid approach for quantum machine learning, which exploits the dynamics of a fixed complex quantum 
		system to process the inputs, followed by trainable post-processing
		\cite{fujii2017harnessing,mujal2021opportunities}, whose parameters can be stored in a classical memory. Training and testing follow the same general inductive learning approach as for quantum neural networks.
	\item \emph{Classical shadows} represent a powerful technique that has been applied to different 
		learning problems with quantum inputs. Notably, in the original formulation\cite{huang2020predicting}, 
		the task was the regression-like problem of predicting expectation values of a few observables,
		depending on the outcomes of random measurements. Following an inductive
		learning method, training consists of a two-step process, with the  
		first step consisting of the collection of  partial classical information about the unknown
		training states $\rho(x)$, followed by a classical post-processing of the measurement outcomes. 
		It should be mentioned that alternatives involving a quantum memory have also been proposed 
		\cite{huang2021information,huang2022quantum}.  
\end{itemize}

\subsubsection{Transductive Learning} Some instance of transductive learning architectures are as follows.

\begin{itemize}
\item 
		The \emph{Helstrom classifier} is the  optimal binary quantum state detector  under the most 
		general and unrestricted set of measurements \cite{lloyd2020quantum,banchi2021generalization}. It  has an explicit expression in terms of the training data, and hence, for the case of known training states, it can be inductively applied 
		to any test data. However, approximations to the classifier for unknown states can benefit from a transductive learning architecture that jointly processes  all training and testing data. One such architecture was proposed, which hinges on the state exponentiation and phase estimation algorithms, 
		with an algorithmic complexity that scales logarithmically with the dimension of the Hilbert space.\cite{lloyd2020quantum,kimmel2017hamiltonian}. 
		In this architecture, there are no learnt parameters to be stored in either quantum or classical memories, and each 
		classification requires processing different copies of the test state $\rho(x)$ as well as  different 
		copies of the training states.
\item \emph{Quantum kernel methods} produce a prediction $y$ for a new test
		state $\rho(x)$ by carrying out a comparison, using a quantum routine, with
		all examples in the training set\cite{schuld2021supervised,huang2020power}.
		Accordingly, following a transductive learning approach, training data can
		only be leveraged to process a single test example. The comparisons between
		test and training states produce classical outputs that are combined with
		the training labels in order to obtain the final decision $y$. To
		elaborate, define as $k(x,x')=\Tr[\rho(x)\rho(x')]$ 
		the kernel function that measures the similarity of states $\rho(x)$ and
		$\rho(x')$. Note that, when states $\rho(x)$ and $\rho(x')$ are unknown,
		estimating $k(x,x')$ requires access to  copies of the states. With this
		definition,  the prediction produced by quantum kernel methods is $f(x) =
		\sum_n \alpha_n k(x_n,x)$, where $x_n$ represents the $n$-th training input
		and the weight parameters $\alpha_n$ are obtained during training from the
		kernel matrix $k(x_n,x_m)$ between pairs of training data $(x_n,x_m)$. 
		Accordingly, training requires copies of the training data, but also 
		providing a prediction for a new input $(x,\rho(x))$ requires having access
		to copies of the test state $\rho(x)$ and of the training states
		$\rho(x_n)$.
		Overall, quantum kernel methods are transductive, but with an inductive component for estimating the classical coefficients $\alpha_n$.

\end{itemize}

Although it may naively appear better to focus on inductive strategies, since
they can can be applied an arbitrary number of times, it has been shown\cite{huang2022quantum} 
that there are transductive strategies  for
predicting properties of physical systems, performing quantum principal
component analysis, and learning about the physical dynamics of $n$-qubit
states that require exponentially fewer (in $n$) samples than inductive
strategies that measure each state separately. Note that the latter category is
more restrictive than allowing the most general inductive strategy.
Nonetheless, this result shows that transductive strategies can be advantageous
if the dimension of the inputs is large and we do not have many copies of the
test inputs .

\subsection{Optimality Gap and Generalization Error}\label{sec:optimality} 

In this section, we discuss how to define and evaluate the performance of
quantum learning algorithms as a function of the number of training samples,
$N$,  of the number of copies $S$ available for each training state, 
namely the training copy complexity, and of the
number of copies $V$ of the test state, namely the test copy complexity. 
As we detail next, we consider two different regimes.

\begin{itemize}
    \item \emph{Known training states}: In the first regime, complete knowledge of the training states is assumed, which can be viewed as the limit $S\to\infty$ of an infinite number of copies of the respective states.	This setting can be mapped to a classical problem, with inputs given by the classical description $\rho(x)$ rather than $x$, and one can use tools from classical statistical learning theory to define and bound the errors in terms of the training data set size $N$. 
 
\item \emph{Unknown training states}: In the second regime, we assume that the quantum states are unknown and the average number of copies $S$ is finite. 
When the state-preparation mechanism is known, one may assume that gathering new data, i.e., increasing $N$, is more costly
than increasing the number of copies of each available 
data, i.e., increasing $S$. Conversely, in order to address the case in which  the state-preparation mechanism is not known, one can study the special case in which a single sample, $S=1$, is available for the  training samples.

\end{itemize}

\subsubsection{Known Training States}\label{sec:known training states}

In supervised learning, as part of the problem definition, one specifies a \emph{loss function} to gauge the quality of a prediction. The overall performance of a learning algorithm, as well as its design, typically hinge on different types of averages of the loss function. Machine learning operates in the absence of information about the data-generation mechanism, apart from the available training data and, possibly, domain knowledge.  If the data generation mechanism, denoted as $\mathcal{P}$, were known, one could define an \emph{average loss} $L_{\mathcal P}(f)$, where the average is taken with respect to any randomness in the data generation process -- as dictated by $\mathcal{P}$ -- and in the processing steps. The average loss $L_{\mathcal P}(f)$ is also known as the \emph{test loss}, for reasons that will be clarified below. 

As illustrated in Figure  2, in the ideal case in which one can evaluate the average loss  $L_{\mathcal P}(f)$, it is in principle possible to 
optimize the inference function by minimising $L_{\mathcal P}(f)$. The \emph{minimum average loss} $L_{\mathcal P}(f_*)$ attained for the optimal inference function $f_*$ represents the  performance level 
accrued by an optimal processing of the test input given knowledge of the  mechanism $\mathcal{P}$. 

\begin{figure}[t]
\begin{center}
\begin{tikzpicture}[scale=1, transform shape]

	\tikzset{
		declare function = { 
			xa(\x) = 1+\x;
			xb(\x) = 3+\x;
			xba(\x) = -xa(-xb(\x));
			ya(\x) = 2 + \x*\x*4/9;
			yb(\x) = 1 + \x*\x;
		}
	}
	\coordinate (a) at ({xa(0)},{ya(0)}); 
	\coordinate (b) at ({xb(0)},{yb(0)}); 
	\pgfmathsetmacro{\ybup}{ya(xba(0))}
	\coordinate (b1) at ({xb(0)},\ybup);
	\pgfmathsetmacro{\yc}{yb(-0.7)}
	\coordinate (c) at ({xb(-0.7)},\yc); 
	\pgfmathsetmacro{\ycup}{ya(xba(-0.7))}
	\coordinate (c1) at ({xb(-0.7)},\ycup);
	\pgfmathsetmacro{\yd}{ya(-1)}
	\coordinate (d) at ({xa(-1)},\yd); 

	\draw[thick,->] (-5.5,0) -- (13,0);
	\node at (11.5,0.3) {$f\in\mathcal F$};
	\draw[thick,->] (-5.5,0) -- node [above,rotate=90] {Loss} ++(0,6);

	\draw[dotted] (a) -- (5.5,{ya(0)});
	\draw[dotted] (b) -- (10,{yb(0)});
	\draw[<->] (5,{ya(0)}) -- node [right] {Optimality gap $\mathcal E(f_{\mathcal S})$ ~} (5,\ybup);
	\draw[dotted] (c) -- (4.5,\yc);
	\draw[<->] (4.5,{yb(0)}) -- node [right] {~Knowledge gap $E^{S}_{\mathcal S}$} (4.5,\yc);
	\draw[dotted] (-1.5,\ybup) -- ++(11.5,0);
	\draw[dotted] (-1.5,\ycup) -- (c1);
	\draw[<->] (9.5,{yb(0)}) -- node [right,text width=3.2cm,align=center] {Generalization error \\ $\mathcal G(f_{\mathcal S})$} (9.5,\ybup);
	\draw[dashed] (b) -- (b1);
	\draw[dashed] (c) -- (c1);
	\draw[<->] (-1.3,\ycup) -- node [left] {Excess testing error $E^{S}_{\rm test}$}  (-1.3,\ybup);

	\draw[<->] (4,0) -- node [right,yshift=-5pt] {Training error} (4,\yc);
	\draw[<->] (0.5,0) -- node [left,yshift=-30pt] {Average testing error} (0.5,\ycup);

	\draw[very thick, blue] ({xa(-3)},{ya(-3)}) parabola bend (a) ({xa(3)},{ya(3)}) 
		node [right] { Average loss $L_{\mathcal P}(f)$ };

	\draw[very thick, red] ({xb(-2)},{yb(-2)}) parabola bend (b) ({xb(2)},{yb(2)}) 
		node [right] { Dataset loss $L(f,\mathcal S)$ };

	\fill[blue]  (a)  circle (0.08) node [below] {$f_*$};
	\fill[blue]  (b1)  circle (0.08);
	\fill[red]  (b)  circle (0.08) node [below] {$f_{\mathcal S}$};
	\fill[red]  (c)  circle (0.08) node [xshift=-8pt,yshift=-5pt] {$f^{S}_{\mathcal S}$};
	\fill[blue]  (c1)  circle (0.08);

\end{tikzpicture}
\end{center}
\caption{The error zoo expressed in terms of the average loss $L_{\mathcal{P}}(f)$  for a given inference function $f\in\mathcal F$ and of the dataset loss  $L(f,\mathcal S)$ based on the abstract training data set $\mathcal S$ with $N$ entries. Recall that availability of the abstract training data set entails knowledge of the quantum states in the training set. 
The true minimum of functions $L_{\mathcal{P}}(f)$ and $L(f,\mathcal S)$ are respectively denoted as $f_*$ and 
$f_{\mathcal S}$. When an $S$-copy version of the training data set is available, the learner obtains 
the approximate minimum $f_{\mathcal S}^{S}$, which achieves the dataset loss $L(f_{\mathcal S}^{S},\mathcal S)$ 
 and the average, or testing, loss $L_{\mathcal{P}}(f_{\mathcal S}^{S})$. 
}
\label{fig:gaps}
\end{figure}
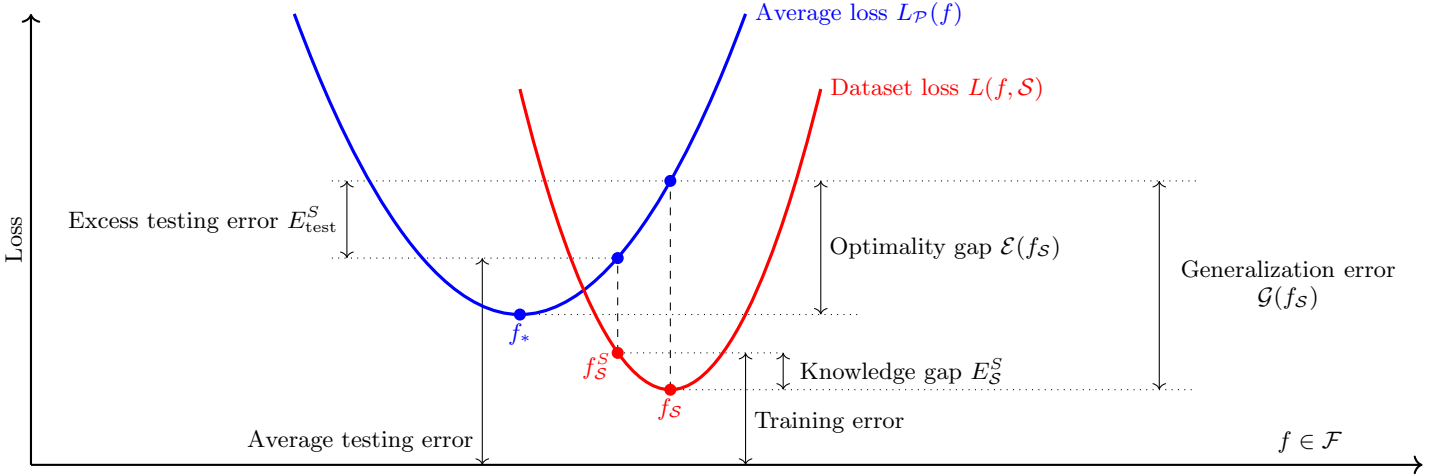

Given any, generally suboptimal, inference function $f$, the difference between the minimum average loss and the function's average loss is defined as the \emph{optimality gap} \begin{align}
    \mathcal{E}(f)= L_{\mathcal{P}}(f)-L_{\mathcal{P}}(f_*). \label{eq:excess_risk}
	\end{align} The optimality gap, illustrated in Figure~\ref{fig:gaps},  gauges the suboptimality of function $f$ with respect to the optimal inference function $f_*$. 

In this subsection, we focus on the case in which the data-generation mechanism is not known, i.e., the distribution $P(x,y)$ is not known, but the training states are known. In this case, one can estimate the average loss $L_{\mathcal{P}}(f)$ via an average over the \emph{abstract} training set $\mathcal{S}$. We refer to this estimate  as the  \emph{dataset loss} $L(f,\mathcal S)$, since it relies on the availability of what may be considered as a classical dataset -- the abstract data set $\mathcal{S}$ -- that contains samples from distribution $P(x,y)$ along with a classical description of the training set. In this setting, the dataset loss is an empirical \textit{training loss} since it is obtained as an empirical average over classical data points from the training set.

In the training phase, the learner minimizes the dataset loss $L(f,\mathcal S)$ to obtain an optimized inference function $f_\mathcal{S}$, which is generally different from the optimal inference function $f_*$. The inference function $f_\mathcal{S}$ achieves the \emph{average loss} $L_{\mathcal{P}}(f_\mathcal{S})$. This is also called the test loss, because the inference function $f_{\mathcal S}$ constructed during training is tested on the (abstract) data-generation mechanism $\mathcal P$. Note that the learner's average loss can be expressed in terms of the optimality gap as \begin{align}
    L_{\mathcal{P}}(f_{\mathcal{S}}) = L_{\mathcal{P}}(f_*) +\mathcal{E}(f_{\mathcal{S}}).  \label{eq:decomposition_sec0}
\end{align}

We are interested in studying how the optimality gap  $\mathcal{E}(f_\mathcal{S})$ for the learned inference function changes as a function of the amount of training data $N$ and of the the model complexity. In this regard, it is observed that, as the size of the data set, $N$, grows arbitrarily large, the optimality gap tends to zero for a well-designed learning strategy. In fact, in this regime, the learner can acquire full information about the distribution $P(x,y)$ and the training loss tends to the average loss, implying that the optimized function $f_\mathcal{S}$ approximates increasingly well the ideal  function $f_*$.

An important performance metric for an inference function $f$ is the {\it generalization error}\begin{equation}\label{eq:generrorfirst} \mathcal{G}(f)=L_{\mathcal{P}}(f)- L(f,\mathcal{S}),  \end{equation}which quantifies how well the training loss $L(f,\mathcal{S})$ approximates the average loss $L_{\mathcal{P}}(f)$. A smaller generalization error indicates that the learner can ``trust'' the training loss as a design criterion used as a proxy for the unknown average loss $L_{\mathcal{P}}(f)$. The average loss $L_{\mathcal{P}}(f_{\mathcal{S}})$ of the inference function $f_{\mathcal{S}}$ can be decomposed in terms of the generalization error as
\begin{align}
    L_{\mathcal{P}}(f_{\mathcal{S}}) =  L(f_{\mathcal{S}},\mathcal{S}) + \mathcal{G}(f_{\mathcal{S}}).\label{eq:decomposition_sec1}
\end{align}   From Eqs.~\eqref{eq:decomposition_sec0} and \eqref{eq:decomposition_sec1}, 
as illustrated in Figure~\ref{fig:gaps}, an inference function $f_{\mathcal{S}}$ that either minimizes the optimality gap, or that simultaneously minimizes the generalization error and the training error, ensures a small average loss.

\subsubsection{Unknown Training States}\label{sec:unknown states many copies}
When the training states are unknown, information about the data-generation mechanism $\mathcal{P}$ at the learner is limited to an $S$-copy version $\mathcal{S}^S$ of the abstract data set. Constraints on $S$ determine the copy complexity of a given learning problem. In this subsection, we elaborate on the notions of optimality gap and generalization error in this setting.

In general,  the goal of the learner is to implement an inference function, denoted as $f_\mathcal{S}^S$ with as small an average loss as possible. 
It is useful to start by noting that,
in the asymptotic limit $S\to\infty$ of an infinite number of copies, we return to the setting of Section~\ref{sec:known training states}, as the learner acquires the abstract data set and so perfect knowledge of the 
dataset loss $L(f,\mathcal{S})$. Hence, the optimized  function $f_{\mathcal S}^{S}$ tends to the function $f_{\mathcal S}$ designed in the case of known training states, and the performance is limited only by the number of data points, $N$.


In a general non-asymptotic regime, in the absence of knowledge of the training states, the learner may not carry out intermediate measurements, and so it may not have access to any empirical estimate of the average loss.
We refer to whatever loss function the learner  minimizes to obtain the optimized inference function $f_{\mathcal S}^{S}$ as the \emph{training loss}, regardless of whether it uses any empirical data. We now elaborate on how inductive and transductive learning schemes may operate in this regime.

\begin{itemize}
\item \emph{Inductive learning}: Following an inductive learning approach, the learner may
carry out measurements on the training set to construct a classical set of observations and define an empirical loss over that set.
This loss may be considered ``doubly-empirical'',
in the sense that it is conceptually obtained by approximating the average loss $L_{\mathcal{P}}(f)$ via two empirical averages. The first average is taken with respect to the $N$ samples of the abstract training set $\mathcal{S}$ (which is abstract as the training states are unknown), yielding the abstract dataset loss $L(f,\mathcal{S})$, 
while the second average is taken with respect to 
the outcomes of the measurements applied to the $S$ copies available for each training state, yielding an empirical loss that approximates $L(f,\mathcal{S})$.

\item \emph{Transductive learning}: An alternative approach consists in using a transductive strategy whereby a fixed  -- or partially programmable, for strategies with inductive elements -- circuit is applied jointly to both the new test input and the $S$-copy training set. In this case, as we will show in Section~\ref{sec:supervised learning}, whilst the learner never directly gains any empirical, i.e., measurement-based, knowledge of the dataset loss, the inference function $f^S_{\mathcal{S}}$ enacted by the learner to test a new input can approximate the inference function $f_S$ that minimizes the dataset loss.

\end{itemize}

Owing to the availability of a finite number of copies, $S$, of the training quantum states, for any finite value of $N$, there is generally a
difference -- which may be positive or negative --
between the test loss $L_{\mathcal P}(f_{\mathcal S}^{S})$ attained by the inference function $f^{S}_{\mathcal{S}}$ optimized by the learner and the test loss $L_{\mathcal P}(f_{\mathcal S})$ that the learner would have obtained with known training states as in the setting studied in the previous subsection. As illustrated in Figure~\ref{fig:gaps}, 
we refer to this difference as the {\it excess testing error}, which is defined as
\begin{equation}
E^{S}_{\rm test}=L_{\mathcal{P}}(f^{S}_{\mathcal{S}})-L_{\mathcal{P}}(f_{\mathcal{S}}).\label{eq:excess testing}
\end{equation}

In a similar way, the gap in performance between the functions $f_{\mathcal S}$ and $f^{S}_{\mathcal{S}}$ can be evaluated in terms of the dataset loss, yielding the {\it knowledge gap}
\begin{equation} E^{S}_{\mathcal{S}}=L(f^{S}_{\mathcal{S}},\mathcal{S})-L(f_{\mathcal{S}},\mathcal{S}).\label{eq: knowledge gap}
\end{equation}
The knowledge gap is always positive and depends on the learning strategy.

With these definitions, we can decompose the test loss for the inference function $f^S_{\mathcal{S}}$ as 
\begin{align}
    L_{\mathcal P}(f_{\mathcal S}^{S})  &= L_{\mathcal P}(f_*)+ \mathcal E(f^{S}_{\mathcal S})\label{eq:error optimality gap0} \\
    & = L(f_{\mathcal{S}},\mathcal{S})+\mathcal{G}(f_{\mathcal{S}})+E^{S}_{\rm test},\label{eq:error optimality gap_1}\\
    &= L(f_{\mathcal{S}},\mathcal{S})+\mathcal{G}(f_{\mathcal S}^{S})+E^{S}_{\mathcal{S}},\label{eq:error optimality gap_2}
\end{align}
These decompositions highlight the additional error terms $E^{S}_{\rm test}$ and $E^{S}_{\mathcal{S}}$ caused by the availability of an $S$-copy training set in lieu of the abstract training set. The analysis of these terms constitutes the main new contribution of this work.

The test loss $L_{\mathcal P}(f_{\mathcal S}^{S})$ can be in principle analyzed by using any of the decompositions highlighted above. Different techniques may indeed approach the problem in distinct ways depending on the particular optimization protocol being implemented. For instance, one may attempt to analyze the generalization error $\mathcal{G}(f_{\mathcal{S}})$ of the inference function  $f_{\mathcal{S}}$, alongside the excess testing error $E^S_\mathrm{test}$, while leveraging the third decomposition; or one may study the  generalization error $\mathcal{G}(f^S_{\mathcal{S}})$ of the inference function  $f^S_{\mathcal{S}}$, along with the knowledge gap $E^S_\mathcal{S}$, by using the last decomposition.

\subsection{Overview of Results with Unconstrained Operations }

In the next two subsections, we review results concerning the analysis of the optimality gap and generalization errors that will be further analyzed in the rest of the paper. Throughout these sections, we focus on classification problems 
with two possible equiprobable classes of quantum states, indexed by the label $y=\pm1$ with probabilities $P(y)=1/2$. We adopt  the standard linear 01 loss, 
which takes value $1$ when the predicted class $\hat{y}$ of a quantum state $\rho(x)$ differs from the true class $y$, and takes value $0$ otherwise. 
As such, the 01 loss has a natural interpretation 
as the probability of misclassification. We also assume that the classical input $x$ is not available to the learner.

As discussed, the classifier has access to 
$V$ copies of a given test state $\rho(x)^{\otimes V}$. 
Since, as long as the probability of misclassification with a single copy is 
no larger than $1/2-O(V^{-1})$, a majority rule over the $V$ copies 
ensures that the average 01 loss decreases exponentially with $V$ (see Sections~\ref{sec:majority rule} and~\ref{sec:rademacher majority}), here we neglect the test copy complexity and focus on the $V=1$ case. Furthermore, in this subsection, we consider general classification strategies, without imposing  
any constraints on the allowed quantum and classical operations, while in the next subsection we discuss the role played by restrictions on the class of feasible models.

We begin this subsection by analyzing the ideal case in which the learner knows the overall data-generation mechanism $\mathcal{P}$, including the classical distribution $P(x,y)$ and the training states $\rho(x_n)$. In this regime, no training data are required. Then, we address the learning problems presented in the previous subsection with known or unknown training states. 

\textbf{1) Known data-generation mechanism $\mathcal P$, known states:} 
In this first setting, detailed in Section~\ref{sec:qsd}, we assume that the learner knows the statistical description $P(x,y)$ of the classical input $x$ and output $y$, as well as the quantum states $\rho(x)$ in the training set.
Thanks to the linearity of the 01 loss, this setting reduces to the 
discrimination of the average states 
\begin{equation}\label{eq:denstrue}
	\bar{\rho}_{y}=\int P(x|y)\rho(x) dx
\end{equation} 
that describe the state of the system for classes  $y=+$ and $y=-$. Note that the average in (\ref{eq:denstrue}) captures the 
lack of knowledge at the learner concerning the classical input $x$. When the average states are not distinguishable, namely 
$\bar{\rho}_{+}=\bar{\rho}_{-}$, for instance because the two state classes 
uniformly populate the state space, it may still be possible to use 
copies to make the average states distinguishable -- 
see Section~\ref{sec:entanglement} for an example concerning the classification
of entanglement. 

Intuitively, the more distinct  states $\bar{\rho}_{+}$ and
$\bar{\rho}_{-}$ are, the easier it is to classify them. Under the $01$ loss, the minimum average,
or testing, loss $L_{\mathcal{P}}(f_*)$ for $V=1$ is given by 
\begin{align}
	L_{\mathcal{P}}(f_*) & = \frac{1}{2} -\frac{1}{4} \Vert \bar{\rho}_{+} - \bar{\rho}_{-} \Vert_1 \nonumber\\ & =
1-2^{-H^{\rm min}(Y|Q)} \nonumber\\
 	& \leq 1-2^{I(Y{:}Q)-H(Y)},
	\label{eq:tracedistanceI}
\end{align} 
which is conventionally expressed, as in the first equality,  in terms of the trace distance $\frac{1}{2}\Vert \bar{\rho}_{+} - \bar{\rho}_{-} \Vert_1$ between the mixed states $\bar{\rho}_{+}$
and $\bar{\rho}_{-}$ (see, e.g., \cite{Nielsen_Chuang} and Section~\ref{sec:Helstrom}). The second equality shows that the minimum average loss  can also be written using an information theoretic
quantity, namely the so-called \emph{conditional quantum min-entropy} $H^{\rm min}(Y|Q)$ of the classical label $Y$ given the quantum state $Q$ representing the test input $\bar{\rho}_y$. The conditional quantum min-entropy is a variant of the von Neumann conditional entropy -- within the family of R\'enyi entropies -- that relies on an alternative relative entropy metric \cite{konig2009operational}. Intuitively, it captures the residual uncertainty on the classical label $y$ given access to one copy of the corresponding input quantum test state $\bar{\rho}_y$.  

The inequality in (\ref{eq:tracedistanceI}) demonstrates that the minimum average loss can be also bounded in terms of the  von Neumann  mutual
information $I(Y{:}Q)$, which is a measure of correlation between test input quantum state and classical label $y$. Accordingly, the minimum average loss decreases as more information can be extracted about label $y$ from state $\bar{\rho}_y$. The optimal measurement for binary classification, which attains the minimum average loss 
$L_{\mathcal P}$, is known as the Holevo-Helstrom (HH) measurement, and it involves the eigendecomposition of the difference  $\bar{\rho}_{+} - \bar{\rho}_{-}$ (see  \cite{konig2009operational} for generalizations beyond the binary case).

\textbf{2) Unknown data-generation mechanism $\mathcal P$, known states:} 
In this second, more challenging, setting, studied in Section~\ref{sec:supervised learning}, the learner does not fully know the data-generation mechanism $\mathcal{P}$. Specifically, it is not aware of the true joint classical distribution $P(x,y)$, but it knows the quantum states $\rho(x)$ in the training set. Accordingly, the learner has access to the abstract training set $\mathcal{S}=\{(x_n,y_n,\rho(x_n))\}_{n=1}^N$ of $N$ data samples, where the states $\rho(x_n)$ are known.

Given the available information, 
the learner can choose a measurement that minimizes the training loss $L(f,\mathcal{S})$. The resulting optimized measurement setting $f_\mathcal{S}$ for binary classification turns out to be the HH measurement for the empirical quantum states  
\begin{equation}
\label{eq:densest}
\bar\rho^{\mathcal S}_y = \frac{1}{N_y} \sum_{n=1}^N \delta_{y,y_n}\rho(x_n)
\end{equation} 
for the two classes $y=\pm$, where $N_\pm$ is the number of samples with $y_n=\pm$ in the training set ($\delta_{y,y_n}$ equals 1 if $y=y_n$ and zero otherwise). The densities (\ref{eq:densest}) provide an empirical estimate of the true per-class densities (\ref{eq:denstrue}). Note that these matrices can be computed at the learner given the learner's knowledge of the quantum states $\rho(x_n)$  in the training set.

From the decomposition \eqref{eq:decomposition_sec1} for the test loss, i.e., $L_{\mathcal{P}}(f_{\mathcal{S}}) =  L(f_{\mathcal{S}},\mathcal{S}) + \mathcal{G}(f_{\mathcal{S}})$,  where 
$L(f_{\mathcal{S}},\mathcal{S})$ depends on the 
trace distance between $\bar\rho^{\mathcal S}_\pm$, in a similar way to Eq.~\eqref{eq:tracedistanceI}, and
$\mathcal{G}(f_{\mathcal{S}})$ is the corresponding generalization error. 
Intuitively, the latter 
must decrease with the data set size $N$, as the empirical density matrices  (\ref{eq:densest}) approximate increasingly well the true density matrices  (\ref{eq:denstrue}). Furthermore, for a fixed value $N$, the quality of this approximation must depend on how much the density matrices $\rho(x)$ change with input $x$. In fact, a larger variability generally requires more observations, i.e., a larger $N$, in order to ensure an accurate estimate of the per-class density matrices (\ref{eq:denstrue}).

To formalize this intuition, it turns out that a useful definition of correlation between input $x$ and state $\rho(x)$ is given by the R{\'e}nyi quantum mutual information (see Section~\ref{sec:informationtheoretic})
\begin{equation}
	I^{\mathcal S}_{1/2}(X{:}Q)= 2\log_2\left({\rm Tr}\sqrt{\frac{1}{N}\sum_{n=1}^N \rho(x_n)^2} \right). 
	\label{eq:mutual info}
\end{equation} Note that this mutual information is evaluated with respect to the empirical distribution of the data points in the abstract training set $\mathcal{S}$. 
With this definition, the generalization error can be shown to decrease with the training data set size $N$  as 
\begin{equation}
	\mathcal{G}(f_{\mathcal{S}})\lessapprox\mathcal{O}\left(\sqrt{\frac{2^{I^{\mathcal S}_{1/2}(X{:}Q)}}{N}}\right). \label{eq:generror_upperbound}
\end{equation} 
When $V$ copies of the test states are employed, the above bound increases by a factor $\mathcal O(\sqrt V)$ -- 
see Section~\ref{sec:rademacher majority}. 
The above result is proved   by analyzing the so-called Rademacher complexity of classes involving unconstrained generalized measurements or unconstrained observables. The approximate inequality \eqref{eq:generror_upperbound} shows that generalization with few training examples $N$ is possible when either the input space is small or the quantum states $\rho(x)$ do not depend too much on the input $x$.

\textbf{3) Unknown data-generation mechanism $\mathcal P$, unknown states:} 
We now consider a learner lacking information about both the classical joint
distribution $P(x,y)$ and the quantum states $\rho(x)$ in the training
set. As discussed, in this case, the learner's performance depends 
also on the number of
copies, $S$, of the quantum states in the training data set $\mathcal{S}^S$. To evaluate the effect of the number $S$ of copies, we can adopt the decomposition \eqref{eq:error optimality gap_2} of the average loss, i.e., $L_{\mathcal P}(f_{\mathcal S}^{S})  = L(f_{\mathcal{S}},\mathcal{S})+\mathcal{G}(f_{\mathcal S}^{S})+E^{S}_{\mathcal{S}}$, in which the knowledge gap $E^{S}_{\mathcal{S}}$ specifically captures the impact of the availability of a finite number of copies.

We know from the previous discussion that the HH measurement defined by 
the empirical average states \eqref{eq:densest} is optimal, so a learner 
should in principle try to implement this solution even without knowledge of such states. 
There are both inductive and transductive strategies to approximate 
the HH measurement.

The simplest  inductive  strategy is based on quantum state tomography. 
The learner first uses the $S$-copy training
set to estimate the quantum densities (\ref{eq:densest}), and then
defines a HH measurement based on the reconstructed states. 
As we detail in Section~\ref{sec:tomography classification}, this results in a knowledge gap due to the errors in the tomographic reconstruction
of such states that grows as $E^S_{\mathcal{S}}=\mathcal{O}(d/\sqrt{NS})$,
where $d$ is the Hilbert space dimension. 
If the states in the training set 
are pure, this can be reduced to $E^S_{\mathcal{S}}=\mathcal{O}(\sqrt{d/S})$.
Accordingly, tomographic approaches are not practical for large dimensional
systems, since the dimension of the Hilbert space $d$ grows exponentially with the number of qubits. 

A more favorable scaling in terms of the Hilbert space dimension $d$ can be obtained with a transductive strategy
that uses a combination of phase estimation and the state exponentiation
algorithms\cite{lloyd2020quantum}. 
This strategy, described in Section~\ref{sec:learning with helstrom}, requires a coherent manipulation
of the $S$-copy training data set $\mathcal{S}^S$ and a new test input $\rho(x)$ to produce the binary classification $y$. 
Ignoring logarithmic corrections, the knowledge gap scales as 
$E^S_{\mathcal{S}}=\mathcal{O}((NS)^{-1/3})$, without any dependence 
on the dimension $d$ -- see Section~\ref{section: scaling with N and S} for 
a more precise analysis. Therefore, the transductive strategy  is the preferred choice when the Hilbert space dimension $d$ is large, such as in many-qubit systems. That said, the scaling of the knowledge gap with $N S$ is worse than for the tomography-based inductive approach.

From the above analysis, we understand that if we let $N\to\infty$ and keep $S$ fixed, both the generalization error and the knowledge gap 
tend to zero. This is because, in such an asymptotic limit, the learner observes an infinite number of copies of any state $\rho(x)$ that has a non-zero probability under the joint distribution $P(x,y)$, while also acquiring complete information about $P(x,y)$. 

Overall, using the  decomposition \eqref{eq:error optimality gap_2} of the average loss and recalling the inequality (\ref{eq:generror_upperbound}) for the generalization error,   depending on the specific setting given by number of data points, $N$, number of copies, $S$, and correlation between input and input quantum state, $I^{\mathcal S}_{1/2}(X{:}Q)$, the generalization error or the knowledge gap terms may dominate. 

\textbf{Computing Generalization Bounds.} As a note regarding the computation of the generalization error  bound (\ref{eq:generror_upperbound}), 
we observe that, when all the states in $\mathcal S$ are pure,
we can alternatively express the mutual information $I^{\mathcal S}_{1/2}(X{:}Q)$ as 
\begin{equation}
	I^{\mathcal S}_{1/2}(X{:}Q) = H_{1/2}(\bar \rho_{\mathcal S}) = H_{1/2}(\bs \eta_{\mathcal S}),
	\label{B kernel main}
\end{equation}
where $H_\alpha(\rho)=(1-\alpha)^{-1}\log_2\Tr[\rho^\alpha]$ is the quantum R\'enyi entropy, 
$\bar\rho_{\mathcal S}=\frac1N\sum_{n=1}^N \rho(x_n)$ is the average state from the training set, 
$H_\alpha(\bs p) = (1-\alpha)^{-1}\log_2\sum_i p_i^\alpha$ is the classical  R\'enyi entropy,  
and $(\eta_{\mathcal S})_j$ are the eigenvalues of the kernel matrix\cite{banchi2021generalization}
\begin{align}
	\bs\eta_{\mathcal S} &= {\rm spectrum}(K_{\mathcal S}), 
					& 
					(K_{\mathcal S})_{nm} = \frac1N\bra{\psi(x_n)}\psi(x_m)\rangle,
\end{align}
which satisfies $K_{\mathcal S}\geq 0$ and $\Tr K_{\mathcal S}=1$. Accordingly, vector 
$\bs\eta_{\mathcal S}$ defines a probability distribution,
and $H_{1/2}(\bs \eta_{\mathcal S})$ is its R\'enyi entropy. 
The second equality in Eq.~\eqref{B kernel main} follows by defining the state
$\ket{\psi_{AB}} = N^{-1/2} \sum_n \ket{n}_A\ket{\psi(x_n)}_B$, and noting that the 
entanglement of the two subsystems $A$ and $B$ is equal, namely $H_{1/2}(A)= H_{1/2}(B)$. 

Since the overlap $\bra{\psi(x_n)}\psi(x_m)\rangle$ between two quantum states can be evaluated efficiently with a shallow 
quantum circuit, the calculation of the vector $\bs \eta_{\mathcal S}$ is possible even for a 
large Hilbert spaces. This allows one to estimate the bound (\ref{eq:generror_upperbound}) based on data.

\subsection{Overview of Results with Constrained Operations }\label{sec:constraints}

In this subsection, we analyze settings in which the space of inference
functions available at the learner is constrained. As we will demonstrate, and
as is also the case for classical machine learning, controlling the capacity of
the space of inference functions is in practice essential in order to guarantee
generalization. This can be done by leveraging domain knowledge about the
problem under study, such as symmetries or locality properties of the involved quantum states.

To understand the connection between model capacity and generalization, let us revisit the bound in \eqref{eq:generror_upperbound}. As we will detail in the next section, this bound is derived  based on uniform-deviation arguments 
from statistical learning theory, which quantify how different the average loss and the dataset loss can be over the entire space of possible inference functions. More precisely, uniform-deviation bounds  gauge the worst-case generalization error   $\mathcal G(f)$ over all possible inference functions $f\in\mathcal F$ in the function class $\mathcal{F}$. If the function class is very large, the uniform deviation can be 
large even when the generalization error $\mathcal G(f_{\mathcal S})$ of the inferred function $f_{\mathcal{S}}$ itself is not. Consequently, uniform-deviation bounds 
may predict overfitting, and hence poor generalization, even when the model generalize well  in practice. 

As a relevant example, 
for classical neural networks, uniform-deviation bounds on the  generalization 
error scale with the number of trainable parameters in network. Accordingly, the bounds predict large generalization errors for deep neural networks with billions of parameters, thereby failing to explain its exceptional generalization performance observed  in practice.  
Recent variants of uniform-deviation bounds\cite{bartlett2021deep} obtain tighter bounds on 
generalization error by imposing suitable constraints on the set of inference functions. Indeed, all successful applications of classical machine learning models rely on some understanding of the structure of the data to support the selection of a class of models that matches well the data-generation mechanism.

A similar approach can be employed in quantum learning methods. 
In fact, all the strategies  
from Table~\ref{tab:strategies}, with the exception of the Helstrom classifier,  introduce
some constraints in the set of measurements or in the set of observables. 
For instance, in quantum neural networks, the circuit ansatz  
effectively constrains the inference function, quantum kernel methods constrain
the underlying observable being evaluated with a regularization term,
while quantum reservoir computing and classical shadows use fixed randomized measurements followed 
by classical postprocessing.

As in the classical case, constraints favour generalization by limiting the expressivity of
the inference function, namely by selecting a reduced function class, 
which might result in a larger training error. However, in the quantum case,
constraints also limit the learner's ability to discriminate quantum states, and might also 
increase the knowledge gap. 
In this section, we assume that generalization error is the dominant source of 
error, and 
investigate the impact of such restrictions on the generalization 
performance of a quantum model. 

In order to use the state-dependent bounds from the previous sections, 
we map the constraints on sets of measurements or observables to a data processing of the quantum inputs,
i.e., a completely positive trace preserving quantum map $\mathcal N : Q \mapsto Q_{\mathcal N}$, mapping the original state space $Q$ into a lower-dimensional 
space $Q_{\mathcal N}$. 
By the data processing inequality, this operation reduces both the mutual information between quantum state and output label, which dictates the minimum average loss (\ref{eq:tracedistanceI}), and the mutual information between quantum state and classical input, which in turn determines the bound (\ref{eq:generror_upperbound}) on the generalization error. In particular, we have the inequalities
$I(Y{:}Q_{\mathcal N}) \leq I(Y{:}Q)$ and 
$I_{1/2}(X{:}Q_{\mathcal N}) \leq I_{1/2}(X{:}Q)$. Therefore, constraining the output space generally causes the minimum average loss to increase, since some important information about the output label may be lost, while the generalization error is expected to decrease.

Choosing a model class that is tailored to the problem at hand should hence ensure that only information that is not relevant to the classification task is discarded by the channel that describes it. This situation may be accounted for by imposing the conditions
\begin{align}
	I(Y{:}Q_{\mathcal N}) &= I(Y{:}Q), &
	I_{1/2}(X{:}Q_{\mathcal N}) &\ll I_{1/2}(X{:}Q).
	\label{eq:data processing}
\end{align}
In this way, the capacity of the model to minimize the average loss is not impaired, while reducing  the generalization error and hence the data complexity. 
To this end, prior domain information can be leveraged to define a map $\mathcal N$ that retains only infomation that is critical for the task of interest. We next provide three examples.

\textbf{Leveraging symmetries:} As a first example, suppose that 
the states are known to be invariant with respect to some symmetry group. 
The problem can be mapped, without any approximation, into a lower dimensional space 
where learning is simpler. For instance, it is possible to define a parametric quantum circuit with the same 
symmetries as the problem \cite{ragone2022representation}, which can be used to classify 
the states with fewer parameters and less training data. 
Moreover, if the actual data or measurement strategy breaks the symmetry by introducing 
some noise, a ``symmetric classifier'' will directly ignore these differences without 
first having to learn that these differences are indeed irrelevant for classification. 
Therefore, as can be expected, exploiting prior information allows learning with less data. 
We will discuss an explicit example in Section~\ref{sec:entanglement}, where 
we study the problem of classifying entanglement.

\textbf{Leveraging locality:} As another example, suppose that the states can be classified using $k$-local observables, namely 
with observables acting at most on $k$ qubits. We may exploit this prior knowledge to define the maps 
\begin{align}
	\bar{\mathcal N}(\rho(x)) &= \frac1{N_k}\sum_{s_k} \Tr_{\bar s_k}[\rho(x)],
														&
	\mathcal N^{\oplus}(\rho(x)) &= \frac1{N_k}\bigoplus_{s_k} \Tr_{\bar s_k}[\rho(x)],
	\label{eq:local maps}
\end{align}
where $s_k$ are all the possible subsets of $k$ qubits, $N_k$ is the total number of such subsets,
and $\Tr_{\bar s_k}$ is the partial trace 
over all qubits, except those in $s_k$. Then,  since the mutual information $I_{1/2}(X{:}Q_{\mathcal N})$ is at most equal to the
logarithm of the Hilbert space dimension, we get 
\begin{align}
	I_{1/2}(X{:}Q_{\bar{\mathcal N}}) &\leq k, &
	I_{1/2}(X{:}Q_{\mathcal N^\oplus}) &\leq k\log_2(N_k).
\end{align}
In other words, if the states can be classified using local observables,  at most 
$N\approx \mathcal O(2^k)$ samples are required to ensure low generalization error
when $I(Y{:}Q_{\bar{\mathcal N}}) \simeq I(Y{:}Q)$ 
-- respectively $\mathcal O(N_k^k)$ samples 
when $I(Y{:}Q_{\mathcal N^\oplus}) \simeq I(Y{:}Q)$.
A variant of the projection $\mathcal N^\oplus$ was employed in Ref.~\cite{huang2020power} to define a 
hybrid quantum kernel method with favorable generalization properties.

\textbf{Parametric Quantum Circuits: } Complexity can also be limited by adopting  a quantum neural network (QNN) with a suitably small number of qubits and quantum gates (see Section 1.4). Details for this scenario can be found in Section 3 and Section 4. Recent works \cite{caro2022generalization,caro2021encoding, du2022efficient} have  characterized the generalization error of QNNs. Notably, the work \cite{caro2022generalization} shows that the generalization error for QNNs can be bounded as \begin{equation}
    \mathcal{G}(f_{\mathcal{S}})\lessapprox\mathcal{O}\left(\sqrt{\frac{N_g \log N_g}{N}}\right), \label{eq:generror_upperboundQNN}
\end{equation} where $N_g$ is the number of trainable gates in the QNN.

\subsection{Applications} 
In this subsection, we discuss three applications of the concepts reviewed in this section.

\subsubsection{Parametric Quantum Circuits} \label{sec:fourier}

As a first example, we focus on classifying classical data $x$ using a quantum algorithm. 
	This is done by first embedding $x$ into a quantum state 
via a parametric quantum circuit, followed by measurements and post-processing. 
It was shown \cite{schuld2021effect} that,
for many popular choices from the literature, the state produced by the circuit can be expressed
as a Fourier-like series 
\begin{equation}
	 \ket{\psi(x)} = \frac{1}{\sqrt{|\Omega|}}\sum_{\omega\in\Omega} e^{i\omega x} \ket{\phi_\omega},
\end{equation}
where the \emph{different} frequencies $\omega$ belong to a set $\Omega$ with 
cardinality $|\Omega|$. In Appendix~\ref{app:fourier} we derive 
an expression similar to \eqref{B kernel main}, but where the 
probability distribution $\bs\eta_{\mathcal S}$ is replaced by the 
eigenvalues of the $|\Omega|\times|\Omega|$ ``Fourier matrix'' $F_{\mathcal S}$ 
\begin{align}
	\bs\varphi_{\mathcal S} 
	&= {\rm spectrum}(F_{\mathcal S}),  &
	(F_{\mathcal S})_{\omega,\omega'} &= \frac{1}{N|\Omega|}\sum_n  e^{i x_n (\omega-\omega')}.
\end{align}
Since the entropy cannot be larger than the logarithm of the dimension, 
we get 
\begin{align}
2^{I^{\mathcal S}_{1/2}(X{:}Q)} 
	&= 2^{H^{\mathcal S}_{1/2}(\bs \varphi_{\mathcal S})} \leq |\Omega|,
								&
	I^{\mathcal S}_{1/2}(X{:}Q) &\leq \log_2(|\Omega|). 
	\label{B frequencies}
\end{align}
Our bound shows that  a large number of Fourier frequencies in the model entails that more data are required to 
ensure generalization. 
A similar result was already known for regression problems \cite{caro2021encoding,canatar2022bandwidth}. 
Note though that in \cite{caro2021encoding}, $\Omega$ is the set of frequency 
differences, rather than the set of frequencies. 
Our result \eqref{B frequencies} differs from the ones in the literature, as it can be applied to 
classification, rather than regression problems. Different bounds on $|\Omega|$ from 
\cite{caro2021encoding} for  different circuit structures can be employed, thanks to \eqref{B frequencies},
to classification problems as well.
For instance, for strategies based on the repetition of arbitrary Pauli 
encodings, it was found that for $D$-dimensional data $x$ and $N_{ge}$ encoding gates, the cardinality of frequencies scales as
$|\Omega|=\mathcal O((N_{ge}/D)^D)$.

\subsubsection{Learning to Classify Phases of Matter}\label{sec:phases of matter}

Quantum phase transitions describe the abrupt change, in the thermodynamic limit, of the 
ground state of a quantum many-body system when some external parameters $x$ are continuously 
varied across a critical region\cite{sachdev1999quantum}. 
Common examples of second-order quantum phase transitions can be detected by measuring a 
local observable, called order parameter. 

In spin systems, this order parameter is often the magnetization 
\begin{equation}
	M^\alpha = \left\langle \sum_n \sigma_n^\alpha \right\rangle_\rho = 
	\Tr\left(\sigma^\alpha \bar{\mathcal N}[{\rho}]\right),
	\label{eq:magnetization}
\end{equation}
where $\alpha=x,y,z$ and $\sigma_n^\alpha$ are the Pauli operators on the $n$th spin,
and in the second inequality we used the map from \eqref{eq:local maps} to transform 
the local observables on an $n$ qubit state into a single-qubit observable on the 
projected average state $\mathcal N(\rho)$, with $k=1$. 

Suppose, on the other hand, that the order parameter is unknown, but that physical wisdom 
tells us that it is expected to be a combination of $k$-local observables. 
By generalizing Eq.~\eqref{eq:magnetization}, we can turn the problem into a quantum learning one 
with states $\mathcal N(\rho(x))$,  and by the analysis of the previous section we expect that this 
model can be learnt efficiently, using a number of data which scales at most as $2^k$. This was also observed in numerical experiments with a simple Ising model\cite{banchi2021generalization}, 
where, without any projections, the only errors were close to the phase transition point, where 
quantum fluctuations may confuse the learner without an explicit projection. 
Other examples were considered in the literature, e.g., \cite{dong2019machine,huembeli2018identifying,rem2019identifying}.

For quantum phase transitions with global order parameters, or for topological phase transitions, generalization 
is less trivial, and whether the information theoretic bounds can provide an explanation is an open question. 
It was found  that, in some cases \cite{huang2022provably}, a learner needs 
 ``non-linear'' functions of the density matrix, e.g., observables acting on $c$ copies $\Tr[A\rho^{\otimes c}]$,
 but also that $k$-local reduced density matrices may suffice. We can model this observation by generalizing 
 Eq.~\eqref{eq:local maps} as $\mathcal N^{\oplus}(\rho(x)^{\otimes c})$, where the 
 input $\rho(x)^{\otimes c}$ is the space with sufficient information to detect the phases with a
 non-linear function of the density matrix, and the map $\mathcal N^{\oplus}$ performs a projection onto 
 a reduced space that contains a combination 
 of linear and non-linear functions of the reduced density matrices. If such a projection contains enough information 
 about the phases, then generalization is expected with at most $\mathcal O((\rm const)^k)$ data.

\subsubsection{Learning to Classify Entanglement} \label{sec:entanglement}

As another instructive example, we focus on the toy problem of classifying two 
classes of quantum states on a composite Hilbert space $AB$, namely 
separable states and maximally entangled states. We generate them as 
\begin{align}
	\ket{\psi_{\rm sep}(x)} &= U_{A}(x)\ket 0_A\otimes U_{B}(x)\ket 0_B, &
\ket{\psi_{\rm ent}(x)} &= [U_{A}(x)\otimes U_{B}(x)]\ket{\Phi}_{AB},
\label{eq:entangled states}
\end{align}
where $\ket{\Phi}_{AB} = \sum_{n=1}^d \ket{n}_A\ket{n}_B/\sqrt d$ is a maximally entangled state, $\ket 0$ is a 
reference state in a $d$-dimensional Hilbert space, and $x$ models the parameters in the 
(possibly different) local unitaries $U_A$ and $U_B$. We assume that $U_A$ and $U_B$ form 
a 2-design, so their average is indistinguishable from a Haar (uniform) average up to the second moments 
\cite{scott2005learning}. 

By construction, linear observables cannot discriminate the two states. Indeed, their average is the same, i.e., 
\begin{equation}
	\int dx 	\ket{\psi_{\rm sep}(x)}\!\! \bra{\psi_{\rm sep}(x)} 
	= \int dx 	\ket{\psi_{\rm ent}(x)}\!\! \bra{\psi_{\rm ent}(x)} = \frac{\openone\otimes\openone}{d^2},
\end{equation}
namely they both uniformly ``populate'' the Hilbert space. To make their average distinguishable, we need 
to focus on non-linear functions, and the simplest one involves two copies. Therefore, we define the available states as
\begin{equation}
\rho(x) = 
\begin{cases}
	\ket{\psi_{\rm sep}(x)}\!\! \bra{\psi_{\rm sep}(x)}^{\otimes 2},  & y=+1, \\
	\ket{\psi_{\rm ent}(x)}\!\! \bra{\psi_{\rm ent}(x)}^{\otimes 2},  & y=-1.
\end{cases}
\end{equation}

With such two-copy states, we can  construct a simple measurement that can unambiguously distinguish between the two classes. In fact, recalling that, per Eq.~(\ref{eq:entangled states}), each copy consists of two subsystems ($A$ and $B$), we can carry out a SWAP measurement (see Figure~\ref{fig:swap}) on the $A$ subsystems of both copies of our test state. If the state we are testing is $\ket{\psi_{\rm sep}(x)}^{\otimes 2}$, the measurement result will always be $1$, whilst if it is $\ket{\psi_{\rm ent}(x)}^{\otimes 2}$, the result will only be $1$ with probability $1/d$.

Given that this is a problem that admits a simple  analytical solution -- namely the SWAP measurement -- for any Hilbert space dimension, it is an interesting question to address how difficult it is for a machine learning model to learn such classification rule  without knowing the structure of the problem. From Eq.~\eqref{B kernel main}, we know that generalization error bounds depend on the entropy of the average state, which can be computed explicitly, using Haar integrals, to get $2^{I_{1/2}(X{:}Q)}=\mathcal O(d^2)$. Therefore, by (\ref{eq:generror_upperbound}), for multi-qubit systems where $d=2^n$ grows exponentially with the number of qubits $n$, the number of samples to learn to classify entanglement  grows exponentially with $n$.

In stark contrast, if we have prior information about the structure of the problem, applying our physical understanding of the situation, we may be able to drastically simplify the learning problem.
In particular, since entanglement is invariant under local operations and classical communications\cite{horodecki2009quantum}, 
we can directly average out all local details via the map $\mathcal N_A\circ \mathcal N_B(\rho)$ 
where $\mathcal N_A = \mathcal N_B = \int dU U^{\otimes 2}\rho U^{\otimes 2,\dagger}$ are local 
twirling channels that act collectively on both A/both B subsystems of the two-copy states. Owing to the Schur-Weyl duality, such channels can be decomposed as projections 
onto the symmetric and anti-symmetric subspaces. In other words, without losing relevant information,
we can transform the problem via the map 
$\mathcal N(\rho) = \ket0\!\bra0 \Tr[\mathcal P_S\rho] + \ket 1\bra 1 \Tr[\mathcal P_A\rho]$ where 
$\mathcal P_{S/A}$ are the projectors onto the symmetric/antisymmetric subspaces. This map 
satisfies \eqref{eq:data processing}, namely it removes a large number of irrelevant degrees of freedom,
mapping a $d^4$-dimensional state to a single-qubit state, 
without altering the capacity to predict the class $Y$.
 Since, after the mapping, we have a single-qubit state regardless of the initial dimension, the mutual information is a constant, 
$I_{1/2}(X{:}Q_{\mathcal N})=\mathcal O(1)$, and we recover the
expected result that few training data are required to learn this process.

\subsection{Organization of the Paper}

In the rest of the paper, we will thoroughly analyse the generalization performance of quantum machine learning in both settings of supervised and unsupervised learning, and for both known and unknown quantum states. The general goal is to understand how the generalization performance is affected by the properties of available data, the number of copies of each datum, the performance loss criterion, and the class of inference operations $f$ under optimization. Results are typically given in the form of scaling laws, describing the decrease of the different errors shown in Figure~\ref{fig:gaps}  as a function of parameters such as size of the training set, or the number of copies. 

To start, Section \ref{sec:classical} presents the necessary background on classical statistical learning theory, the branch of statistics that addresses the dependence of generalization on the properties of data, loss measure, and inference operation class. 
Section \ref{sec:qsd} describes the baseline quantum information processing
task of quantum state discrimination with known quantum states. This problem is introduced as a
benchmark, as it corresponds to an idealized situation for quantum supervised
learning in which the statistical description $\mathcal{P}$ of the test input
is available. 
Section \ref{sec:learning discriminate}  studies quantum state discrimination
of unknown quantum states, but known data distribution.
In this
section, we also discuss transductive learning solutions, which are compared to
more conventional inductive learning methods.
Section \ref{sec:supervised learning} covers quantum supervised learning problems, where both 
the data distribution and the quantum states are unknown.
Here we merge the techniques from Sections~\ref{sec:classical} and~\ref{sec:learning discriminate}
to study the generalization error and optimality gap, due to the unknown data distribution, 
and excess errors due to the unknown quantum states. 
Section \ref{sec:unsupervised learning} moves on to the generalization analysis of quantum unsupervised learning for generative modelling, distinguishing applications where quantum generative models either approximate a classical description $P(x)$ or a quantum state $\rho$.
Conclusions and outlooks are drawn in Sec \ref{sec:outlook}.

\section{Classical Statistical Learning Theory}\label{sec:classical}
In this section, we elaborate on generalization theory for classical supervised and unsupervised learning within the general framework presented in the previous section. Most of the material is adapted from  \cite{shalev2014understanding,bartlett2021deep,kawaguchi_bengio_kaelbling_2022, arora2017generalization, zhang2017gendiscr}. This material will form the background for the extensions to quantum systems to be studied in the following sections.

\subsection{Supervised Learning}\label{sec:classical supervised}
In supervised learning, a data instance is given as a pair $(x,y)$ of input feature  $x$ and output label $y$, which are related according to an unknown input-output mapping. In a deterministic setting, the input-output relationship is assumed to be described by an unknown function $y=f(x)$; while, in a stochastic setting, it is described by an unknown conditional probability distribution $f(y|x)$ of label $y$ given input $x$. From now on, we use the same symbol $f$ to
denote either deterministic or stochastic mappings without making an explicit distinction. The goal of supervised learning is to infer the unknown mapping from input to output based on the observation of a finite  training set of data instances. This requires \emph{generalizing} the observed input-output pairs outside the training set, a process known as \emph{induction} (see, e.g., \cite{simeone2022machine}).

Depending on whether the output label $y$ is discrete or continuous, supervised machine learning problems amount to as classification or regression tasks. In classification, the set of outputs is discrete, e.g., the next word in a text, while in
regression the output is continuous, e.g., the next market value of a certain stock.

From a statistical perspective, we assume that each data instance $(x,y)$ is drawn from an \emph{unknown probability distribution} $P(x,y)$, with training and test  data instances being independent samples 
from $P(x,y)$. The first step of the learning process is to fix a function  class, known as \emph{model class},  $\mathcal F$ of candidate input-output mappings, or models. This class can, for example, consist of neural networks with trainable parameters.  Learning then amounts to  finding the best function $f\in \mathcal F$  that can most faithfully predict the output label of any feature input $x$, with the quality of the prediction evaluated with respect to the underlying, unknown, distribution $P(x,y)$. 

The predictive performance of a function $f \in \mathcal{F}$ on a data instance $(x,y)$ can be measured via a \emph{loss function} $\ell(f,x,y)$ that introduces a penalty dependent on the degree to which the predicted output $f(x)$ is different 
from the real output $y$. Common losses for regression with deterministic function classes include 
the square loss $\ell(f,x,y) = (y-f(x))^2$ and its extension to multivariate outputs via $\ell_2$-norms. For classification, typical loss functions include the probability of misclassification 
$\ell(f,x,y) = \sum_{\bar y} (1-\delta_{\bar y,y})f(\bar y|x) = 1-f(y|x)$ when the function class is stochastic; and the hinge loss $\ell(f,x,y) = \max\{0,1-yf(x)\}$  with $y=\pm1$ for binary classification problems with a deterministic function class. 

The goal of the supervised learner is to find the candidate function that minimizes the {\it average loss}  
\begin{equation}
	L(f) = \ave_{(x,y)\sim P}[\ell(f,x,y)],
	\label{risk}
\end{equation} which is the average loss with respect to the abstract data distribution $P(x,y)$. We denote as
\begin{equation}
	f_* = \argmin_{f\in \mathcal F}L(f)
	\label{f star}
\end{equation}  the optimal function that minimizes $L(f)$. Recall that this optimal solution is a conditional probability distribution for stochastic model classes. Note that, as compared to the general notation introduced in the previous section we drop the dependence of the average loss (\ref{risk}) on the statistical description $\mathcal{P}$, here the distribution $P(x,y)$, in order to simplify the presentation.

Finding the optimal solution \eqref{f star} is impossible in practice, since the average loss \eqref{risk} cannot be evaluated due to the unknown data distribution $P(x,y)$. In fact, the only information available to the learner about $P(x,y)$ is via a data set $\mathcal{S}= \{(x_n,y_n)\}_{n=1}^N$ of $N$ examples sampled i.i.d.~according to $P(x,y)$.
Hence, the learner replaces the average loss in \eqref{f star} with an empirical {\it training loss} 
\begin{equation}
	L(f,\mathcal S) = \frac 1N \sum_{n=1}^N \ell(f,x_n,y_n),
	\label{emp risk}
\end{equation} which is the empirical average of the losses incurred on examples in the training set $\mathcal{S}$. This results in the following approximation to the optimal function $f_*$:
\begin{equation}
	f_{\mathcal{S}} = \argmin_{f\in \mathcal F}L(f,\mathcal S)
	\label{emp f star}.
\end{equation}

If the training loss $L(f_{\mathcal{S}}, \mathcal S)$ corresponding to the learnt function $f_{\mathcal{S}}$ is sufficiently small, it indicates that the assumed function class $\mathcal{F}$ is sufficiently complex to capture the input-output relationship. However, ensuring small training loss $L(f_{\mathcal{S}}, \mathcal S)$ on observed training data does not ensure that the learnt function performs well on previously unseen data drawn from distribution $P(x,y)$. The learnt function $f_{\mathcal{S}}$ is said to {\it generalize} if it also performs well on new, previously unseen data $\mathcal{S}_{\rm test}$ drawn from $P(x,y)$, i.e., more precisely, if the loss 
$L(f_{\mathcal{S}}, \mathcal S_{\rm test})$, is also small.

\subsection{Generalization Error in Supervised Learning}\label{sec:Rad}

As discussed in the previous subsection, the ability to generalize is a crucial desired performance criterion for any machine learning algorithm. 
If the learnt function performs well on the training set, but poorly on test set, we say that the function \emph{overfits} the training data and fails to generalize. 
In general, the generalization ability of a learnt function $f_{\mathcal{S}}$ is determined by three main factors:
\begin{enumerate}[leftmargin=45pt,align=left,labelwidth=45pt,labelsep=0pt]
	\item[F.1] The function class $\mathcal F$ should be sufficiently complex 
		to contain a ``good'' approximation of the optimal, unknown input-output mapping. 
	\item[F.2] The training set $\mathcal S$ should be sufficiently comprehensive, so that the 
		empirical training loss $L(f,\mathcal S)$ provides a ``good'' approximation to 
		the (unknown) average loss \eqref{risk}. 
	\item[F.3] The learning algorithm that optimizes the empirical training loss $L(f,\mathcal S)$ should be sufficiently powerful to yield a solution close to the training loss-minimizing model $f_{\mathcal{S}}$. 
\end{enumerate}

Given a model class $\mathcal{F}$, one is generally interested in finding a model $f$ that ensures a small optimality gap  $\mathcal{E}(f)$ as defined in \eqref{eq:excess_risk}.
For the learnt model $f_{\mathcal S}$, which minimizes the training loss, 
the optimality gap \eqref{eq:excess_risk} can be decomposed as
\begin{align}
 \mathcal{E}(f_{\mathcal S}) =
 \underbrace{L(f_{\mathcal S})-L(f_{\mathcal S},
	 \mathcal{S})}_{:=\mathcal{G}(f_{\mathcal S},\mathcal{S})} + 
	 \underbrace{L(f_{\mathcal S}, \mathcal{S})-
	L_{\mathcal{P}}(f^{*})}_{:=\mathcal{A}(f_{\mathcal S},\mathcal{S})}, 
	\label{eq:decomposition}
\end{align} 
where the first term,  $\mathcal{G}(f_{\mathcal S},\mathcal{S})$,  is the \emph{generalization error} 
of the empirical inference operation,  and the second term $\mathcal{A}(f_{\mathcal S},\mathcal{S})$ is the excess empirical error.
In fact, for any model $f\in \mathcal{F}$, the generalization error $\mathcal{G}(f,\mathcal{S})$ can be upper bounded as  
\begin{align}
\mathcal{G}(f,\mathcal{S}) =  L(f) - L(f,\mathcal S)\leq \mathcal D(\mathcal F,\mathcal S) := \sup_{f\in \mathcal F}|L(f) - L(f,\mathcal S)|,
	\label{uniform deviation}
\end{align} by maximizing over all models in the class $\mathcal{F}$. In (\ref{uniform deviation}),  we have defined the  {\it uniform deviation} $\mathcal D(\mathcal F,\mathcal S)$ of the function class $\mathcal{F}$ with respect to data set $\mathcal{S}$. On the other hand, an upper bound on the excess empirical error $\mathcal{A}(f_{\mathcal{S}},\mathcal{S})$ can be derived as
\begin{align}
    \mathcal{A}(f_{\mathcal{S}},\mathcal{S})= L(f_{\mathcal{S}}, \mathcal{S})-L(f_*,\mathcal{S})+ L(f_*,\mathcal{S})-L(f_*) \leq \mathcal{D}(\mathcal{F},\mathcal{S}) \label{gen bound}
\end{align}
where we have used the uniform deviation bound, as well as the inequality 
$L(f_{\mathcal{S}},\mathcal S) <
L(f_*,\mathcal S) $, which follows since $f_{\mathcal{S}}$ minimizes the training loss. Note that, for another model $f$, which may optimize the training loss only approximately, the last inequality in \eqref{gen bound} would not hold, and one should account for the contribution to the optimality gap caused by a non-ideal optimizer.  Using \eqref{uniform deviation} and \eqref{gen bound} in \eqref{eq:decomposition},  we have the following upper bound on the optimality gap of the learnt function $f_{\mathcal{S}}$, 
\begin{align}
    \mathcal{E}(f_{\mathcal{S}}) \leq 2 \mathcal{D}(\mathcal{F},\mathcal{S}). \label{eq:excessrisk_uniformdeviation}
\end{align}

The relation \eqref{eq:excessrisk_uniformdeviation} suggests that the optimality gap can be controlled by ensuring that the uniform deviation $\mathcal{D}(\mathcal F,\mathcal S)$ is sufficiently small. In the rest of this section, we describe a statistical complexity measure, the Rademacher complexity, that provides a way to quantify the uniform deviation.

The 
\emph{Rademacher complexity} $R_P(\mathcal F)$ of a given function class $\mathcal F$ under the true data distribution $P(x,y)$ is defined as
\begin{align}
	R_P(\mathcal F) &= \ave_{\mathcal S}[R(\mathcal F,\mathcal S)], &
	R(\mathcal F,\mathcal S) &= \ave_{\bs \sigma}\left[ \sup_{f\in \mathcal F} \frac1N\sum_{n=1}^N
	\sigma_n \ell(f,x_n,y_n)\right],
	\label{rademacher}
\end{align}
where the first expectation is taken over the data $\mathcal{S}$, whose each entry is  drawn from distribution $P(x,y)$; and the second expectation is over independent  Rademacher variables $\sigma_j$ that take values $\pm1$ with equal probability. The quantity   
$R(\mathcal F,\mathcal S)$ is known as the \emph{empirical Rademacher complexity}, and is a function solely of the model class $\mathcal{F}$ and of the training set $\mathcal{S}$. 

While the empirical Rademacher  complexity $R(\mathcal F,\mathcal S)$ can be evaluated on the basis of the available training data, the Rademacher complexity  $R_P(\mathcal F)$ requires  averaging over the distribution $P(x,y)$, and, as such, it 
cannot be computed. However, in the large training data set limit, i.e., with large $N$, the two quantities become increasingly close. In fact,  
with probability higher than $1-\delta$ for $\delta\in[0,1]$, when the loss is in $[-1,1]$ we have the inequality 
\begin{equation}
	|R_P(\mathcal F) - R(\mathcal F,\mathcal S)| \leq \mathcal O\left(\sqrt{\frac{\log(1/\delta)}{N}}\right).
	\label{rada vs emp rada}
\end{equation}
Similar bounds hold for more general losses. Furthermore, replacing the Rademacher variables with normally distributed 
variables $\sigma_j\in\mathcal N(0,1)$ yields the Gaussian Rademacher  
complexity
\cite{bartlett2002rademacher}, which shares similar properties with $R_P(\mathcal F)$.

Although faster rates are known for specific 
instances\cite{bartlett2021deep},  the Rademacher complexity for a model class $\mathcal{F}$ typically scales as  $\mathcal O(B(\mathcal{F})/\sqrt N)$, where $B(\mathcal{F})$ is some constant that depends on the model class $\mathcal{F}$. Using this result and (\ref{rada vs emp rada}), we can assume the approximate equality  $R(\mathcal F,\mathcal S) \simeq R_P(\mathcal F)  
	\simeq \mathcal O(\sqrt{B(\mathcal F)/N})$, which is increasingly accurate for large values of $N$.
Note that the differences between $R(\mathcal F,\mathcal S)$ and $ R_P(\mathcal F) $ result in 
$\mathcal O(1)$ corrections to $B(\mathcal F)$ that are not significant for large $N$ and will be 
omitted here. See Appendix~\ref{a:math} for more precise definitions.

The Rademacher complexity provides tight upper and lower bounds on the uniform deviation,
 as summarized in  Theorem~\ref{thm:rada} in the Appendix. Specifically, using Theorem~\ref{thm:rada} and the approximate equality between Rademacher complexity and empirical Rademacher complexity discussed above, we obtain that, 
with high probability, we have the approximate equalities  
\begin{equation}
	\mathcal D(\mathcal F,\mathcal S) \simeq R(\mathcal F,\mathcal S) \simeq R_P(\mathcal F)  
	\simeq \mathcal O\left(\sqrt{\frac{B(\mathcal F)}{N}}\right)
	\label{rada simeq}
\end{equation} for sufficiently large $N$.  Using \eqref{rada simeq} into \eqref{uniform deviation} and \eqref{eq:excessrisk_uniformdeviation}, we can finally  conclude that the generalization error and optimality gap scale as 
\begin{align}
  \mathcal{G}(f,\mathcal{S}) \leq \mathcal D(\mathcal F,\mathcal S) \simeq 
\mathcal O\left(\sqrt{\frac{B(\mathcal F)}{N}}\right)
\label{bound train test}, \quad  \mathcal{E}(f_{\mathcal{S}}) &\leq 2 \mathcal D(\mathcal F,\mathcal S) \simeq 
\mathcal O\left(\sqrt{\frac{B(\mathcal F)}{N}}\right).
\end{align}
The above inequalities 
show that the difference between the training and testing errors can be bounded via the same function $B(\mathcal{F})$ of the model class $\mathcal{F}$
up to $\mathcal O(1)$ constant factors.

While, thanks to Theorem~\ref{thm:rada},  the bounds on the uniform deviation are tight, 
the bounds in \eqref{bound train test} may be loose due to the maximization over all functions in the model class that underlies the definition of the uniform deviation (\ref{uniform deviation}). 
For instance, in neural networks it is known that the Rademacher complexity, via function  $B(\mathcal F)$, grows polynomially with 
the number of network parameters. So, for deep networks with trillions of parameters, one generally has the strong inequality  
$B(\mathcal F) \gg N$, making the bounds \eqref{bound train test}  vacuous. Indeed, deep learning is well known to contradict the common statistical wisdom based on Occam's razor, which stipulates that, all else being equal, 
simpler models compatible with the data  are preferred, since models with trillions of parameters are able to 
routinely get with little efforts both small training and testing errors.

This discussion points to some of the limitations of the capacity-based analysis presented in this work. In particular, this type of investigation leaves out the impact of the training algorithm and of the data distribution \cite{bartlett2002rademacher,kawaguchi_bengio_kaelbling_2022,simeone2022machine}.  
Although the function class $\mathcal F$ is hugely complex for deep networks, when using gradient descent
for training, the set of explored functions is a very small fraction of the entire 
function space, since only functions in the neighbourhood of the starting point are considered by 
gradient descent. Therefore, in practice, the size of the overall model class may not matter as much as the quality of solutions obtained in the vicinity of randomized initializations \cite{jacot2018neural}.

\subsection{Model Selection via Structural Risk Minimization}\label{s:structural risk minimization}
The capacity-based analysis outlined in the previous subsection can be leveraged for \emph{model selection}. This approach is also known as {\it structural risk minimization}, and is the subject of this subsection. 

To elaborate, fix a hierarchy of function classes
$\{\mathcal F_r\}$
 with some notion of ordering such that the inequality $r_1 \leq r_2$ implies the inclusion $\mathcal{F}_{r_1} \subseteq \mathcal{F}_{r_2}$.  For instance, we may define $\mathcal F_r$ as a set of neural networks  
with the norm of parameter vector bounded by $r$. Such constraints are common in the machine learning literature, e.g., in support vector machines (SVMs) or Lasso regression. For each constrained function class $\mathcal{F}_r$,  as discussed in this section,  the learning problem is typically formulated as the minimization \begin{equation}f_{\mathcal{S}}^r = \argmin_{f\in\mathcal F_r} L(f,\mathcal S).\end{equation} 

This constrained optimization problem is typically reformulated as the minimization of the \emph{regularized loss} $L(f,\mathcal{S}) + \mu_rg(f)$ where $\mu_r>0$ is a Lagrange multiplier and $g(f)$ is a suitable penalty function. As explained in the next paragraph, this reformulation is exact  when strong duality holds, e.g., for convex loss functions and constraints \cite{boyd2004convex}. 

As a notable example, consider a linear regression problem with model defined by a vector of parameters $\theta$ and the squared loss. The family $\mathcal{F}_r$ is defined by the constraint $g(f)= \|\theta\|_2^2\leq r$. Accordingly, the training loss can be expressed as $L(f,\mathcal S) =
\|X\theta -y\|_2^2/N$, where $X$ is a matrix with rows given by the inputs $x_j$ and $y$ is the vector of outputs 
$y_j$. The regularized training problem is given as \begin{equation}
	\theta_\lambda = \argmin_\theta L(f,\mathcal S) + \lambda \|\theta\|_2^2.
\end{equation} By the Karush–Kuhn–Tucker theorem, a suitable choice of the Lagrange multiplier $\lambda$ ensures the satisfaction of constraint $g(f)= \|\theta\|_2^2\leq r$ \cite{boyd2004convex}. In the limit $\lambda\to 0^+$, the solution of this problem converges to  $\theta_*=X^\sharp y$, where $X^\sharp$ is the pseudo-inverse of $X$.

 In structural risk minimization, one optimizes the choice of the parameter $r$ of the function class $\mathcal{F}_r$ by considering the following optimization problem  \cite{bartlett2002rademacher}
\begin{align}
	r_* & = \argmin_{r} L(f_{\mathcal{S}}^r,\mathcal S) + p_r,
\end{align} which aims to find an optimal tradeoff between the empirical risk and a complexity penalty term $p_r$. The final optimal function is then taken as $f_{\mathcal{S}} = f_{\mathcal{S}}^{r_*}$. The upper bound on the generalization error in \eqref{bound train test} gives a natural choice of the penalty term as $p_r = B(\mathcal F_r)$. For example, if the function class $\mathcal{F}_r$ consists of $L$-layered deep neural networks with $r$-bounded Frobenius norm of weight matrices in each layer, then the penalty term  $B(\mathcal F_r)$ is known to behave as $r^L \sqrt{L/N}$ \cite{golowich2020size}.
Finally, in the zero noise limit, namely when the mapping between $x$ and $y$ 
is deterministic, overparameterized neural networks are complex enough to ensure zero training error,
while zero generalization error is possible as long as the complexity $r$ grows sufficiently slowly 
so that it converges to zero for large number of samples. 

\subsection{Unsupervised Learning} \label{sec:unsupervisedlearning}
In this section, we study unsupervised learning problems in which each data instance contains a single vector $x$, assumed to be drawn from an underlying, unknown, distribution $P(x)$. While the class of unsupervised learning problems encompasses tasks as diverse as clustering and generative modeling, in this paper we focus on the fundamental task of estimating the distribution $P(x)$. Accordingly, the \emph{model class} $\mathcal{F}$ includes a set of candidate probability distributions for vector $x$.  We use $f \in \mathcal{F}$ to denote a probability distribution $f(x)$ in the model class.

Model classes $\mathcal{F}$ may encompass explicit or implicit models $f(x)$. \emph{Explicit} models provide a  function $f(x)$ that can be evaluated for any $x$. Examples of such models include normalized flows or quantum models with a classical likelihood based on expected values of observables (see, e.g., \cite{simeone2022machine,simeone2022introduction}). In contrast, \emph{implicit} models do not enable the evaluation of function $f(x)$, but they only support \emph{sampling} of vectors $x$ from distribution $f(x)$. Examples include generative adversarial networks, diffusion models \cite{murphy2022probabilistic}, and single/ multi-shot quantum models \cite{park2023quantum}. In both cases, as for supervised learning, models are typically functions of a vector of trainable parameters. 

In this paper, given the focus on quantum models, we will assume \emph{implicit} models that produce samples $x\sim f(x)$. We observe that explicit models are obtained in the limit in which one can produce an arbitrarily large number of samples from the model. In fact, in this case, such samples can be used to obtain an arbitrarily accurate estimate of the distribution $f(x)$.

To evaluate how far a candidate distribution $f(x)$ is from the true distribution $P(x)$, we need to introduce a measure of \emph{divergence} between probability distributions. Let $D(\cdot,\cdot)$ denote such a divergence measure, which may be selected as, among others, the Jensen-Shannon divergence, the Wasserstein distance, or the \emph{maximum mean discrepancy} (MMD). 

For instance, the \emph{total variation distance}, \begin{align}\label{eq:TV}
    D_{\rm TV}(P,f) = \sup_{A \subseteq \mathcal{X}} \Bigl| P(A)- f(A)\Bigr|,
\end{align} is the maximum absolute difference between the probabilities that the distributions $P$ and $f$ can assign to  subsets $A$ of the space $\mathcal{X}$ of  vectors $x$. As another example, the class of \emph{integral probability metrics} (IPMs) contains divergences of the form 
\begin{align}
     D_{\rm IPM}(P,f) =  \sup_{h \in \mathcal{H}} \Bigl |\ave_{x \sim P(x)}[h(x)] -\ave_{x \sim f(x)}[h(x)] \Bigr | \label{IPM},
\end{align} where $\mathcal{H}$ is an appropriately chosen class of functions. The function $h(\cdot)$ optimized in \eqref{IPM} is known as the \emph{discriminator}. This is because the optimal function $h(\cdot)$  in \eqref{IPM} should output different values for samples $x\sim P(x)$ generated from the true distribution and for samples $x\sim f(x)$ produced by the model, hence discriminating between the two distributions. 

Different classes $\mathcal{H}$ determine distinct IPMs.  For instance,  when the discriminator class $\mathcal{H}$ consists of the set of all $1$-norm bounded and $1$-Lipschitz functions, the IPM in \eqref{IPM} recovers the $1$-Wasserstein distance between true and model-generated distributions \cite{arjovsky2017wasserstein}. When the discriminator class is the set of neural networks, the IPM \eqref{IPM} equals the  {\it neural-net distance}\cite{arora2017generalization}. Finally, when the set $\mathcal{H}$ is the set of functions within a unit ball in the reproducing kernel Hilbert space, the IPM equals the MMD.

Having specified a divergence measure, the ultimate goal of unsupervised learning problems aimed at distribution estimation is to find the distribution in the function class $\mathcal{F}$ that is closest to the true distribution $P(x)$ in divergence measure $D(\cdot,\cdot)$, i.e.,
\begin{align}
    f_*=\arg \min_{f\in \mathcal{F}} D(P,f). \label{risk_unsupervised}
\end{align}
The criterion $D(P,f)$ can be considered as the counterpart of the average loss $L_{\mathcal{P}}(f)$  in Section \ref{sec:summary} or of the average loss (\ref{risk}) in supervised learning, in that its minimization represents the ultimate aim of the learning problem. Accordingly, we will refer to it as test divergence. Note, however, that, while the average loss (\ref{risk}) depends on the data-generating distribution only through an expectation, the divergence  $D(P,f)$ has a more general functional dependence on $P(x)$, as seen, e.g., in the total variational distance (\ref{eq:TV}).

As for the average loss in supervised learning, finding the minimizer of the test divergence in \eqref{risk_unsupervised} is impossible without access to the true distribution  $P(x)$. In fact, in a learning problem, one only has access to a finite number, $N$, of samples generated i.i.d. according to the true distribution $P(x)$, which are included in  the {\it training set} $\mathcal{S}=\{x_n\}_{n=1}^N$. In addition, in unsupervised learning with implicit models, the distribution $f(x)$ itself is not directly accessible, and the learner can only produce a {\it generated set} $\mathcal{S}_f=\{x^f_m\}_{m=1}^M$ of $M$ examples sampled i.i.d. from the distribution $f(x)$.

 Using training and generated sample data sets, one can estimate the divergence measure  $D({P},{f})$ in different ways. A  general approach is given by \emph{plug-in} estimators that first obtain estimates $\hat{P}(x)$ and $\hat{f}(x)$ of distributions $P(x)$ and $f(x)$, respectively, and then plug these estimates into the divergence metric to obtain the estimate $D(\hat{P},\hat{f})$. Alternatively, one could directly use the samples $\mathcal{S}$ and $\mathcal{S}_f$ to evaluate an estimate $\hat{D}(\mathcal{S},\mathcal{S}_f)$ of the test divergence $D(P,f)$. For example, an empirical estimate of the  IPM (\ref{IPM}) can be obtained using the available samples as
\begin{align}
\hat{D}_{\rm IPM}(\mathcal{S}, \mathcal{S}_f) =  \sup_{h \in \mathcal{H}} \Bigl |\frac{1}{N}\sum_{n=1}^N h(x_n) -\frac{1}{M}\sum_{m=1}^M h(x^f_m) \Bigr | \label{emp IPM}.
\end{align} 
In the following, we write $\hat{D}(\mathcal{S},\mathcal{S}_f)$ to denote either of these two types of estimates, which is referred to as \emph{training divergence}.

The learning problem can be formulated as the minimization of the training divergence
\begin{align}
    f_{\mathcal{S}, \mathcal{S}_f}=\arg \min_{f\in \mathcal{F}} \hat{D}(\mathcal{S},\mathcal{S}_f).\label{empiricalrisk_unsupervised}
\end{align} 
In the limiting case in which one can generate an arbitrarily large number of samples from the model, we recover, as mentioned, the setting with explicit model classes. In this case, the learner can directly leverage the distribution $f(x)$, and we write the corresponding training divergence as $\hat{D}(\mathcal{S},f)$. 
 The  minimizer of the training divergence for explicit models -- obtained equivalently in the limit of large $M$ for implicit models -- is accordingly defined as
\begin{align}
    f_{\mathcal{S}}=\arg \min_{f\in \mathcal{F}} \hat{D}(\mathcal{S},f)\label{empiricalrisk_unsupervised_1}.
\end{align}

\subsection{Generalization Error in Unsupervised Learning}\label{sec:generror_unsupervisedlearning}
As explained in the previous subsection, we can interpret the empirical divergence $\hat{D}(\mathcal{S},\mathcal{S}_{f})$ as the training loss accrued with model $f$, and the  divergence $D(P,f)$ as the average loss. Accordingly,
as for supervised learning, we are interested in analyzing  the  {\it optimality gap},
\begin{align}
\mathcal{E}_D(f_{\mathcal{S}, \mathcal{S}_f}) = D(P,f_{\mathcal{S}, \mathcal{S}_f}) -  D(P,f_*), \label{eq:optimalitygap_unsup}\end{align} of the learnt distribution $f_{\mathcal{S}, \mathcal{S}_f}$ with respect to the divergence measure $D(\cdot,\cdot)$. This is defined as the difference in test divergences of the learnt distribution and of the optimal distribution $f_*$ in \eqref{risk_unsupervised} from the true data distribution $P(x)$.

Following \eqref{eq:decomposition}, the above optimality gap can be written as the sum of the {\it generalization error} of the learnt distribution $f_{\mathcal{S}, \mathcal{S}_f}$ with respect to divergence $D(\cdot,\cdot)$, denoted here as
\begin{align}
    \mathcal{G}_D(f_{\mathcal{S},\mathcal{S}_f},\mathcal{S},\mathcal{S}_{f_{\mathcal{S},\mathcal{S}_f}})= D(P,f_{\mathcal{S}, \mathcal{S}_f})-\hat{D}(\mathcal{S},\mathcal{S}_{f_{\mathcal{S}, \mathcal{S}_f}}), \label{eq:generror_unsupervised}
\end{align} 
and of the {\it excess empirical error} of the learnt distribution $f_{\mathcal{S}, \mathcal{S}_f}$, denoted as
\begin{align}
    \mathcal{A}_D(f_{\mathcal{S},\mathcal{S}_f},\mathcal{S},\mathcal{S}_{f_{\mathcal{S},\mathcal{S}_f}})= \hat{D}(\mathcal{S},\mathcal{S}_{f_{\mathcal{S}, \mathcal{S}_f}})-D(P,f_*). \label{eq:excessempiricalerror_unsupervised}
\end{align}
Note that the generalization error is now a function also of the samples $\mathcal{S}_{f}$ generated by model $f$.

As in the inequalities \eqref{uniform deviation} and \eqref{gen bound} for supervised learning, the generalization error \eqref{eq:generror_unsupervised} and excess empirical error \eqref{eq:excessempiricalerror_unsupervised} can be upper bounded via   the \emph{uniform deviation}
\begin{equation}\label{eq:unidevunsup}\mathcal{D}_D(\mathcal{F}, \mathcal{S},\mathcal{S}_{\mathcal{F}})=\sup_{f \in \mathcal{F}} |\mathcal{G}_D(f,\mathcal{S},\mathcal{S}_{f}) | \end{equation} of the function class $\mathcal{F}$ with respect to data sets $\mathcal{S}$ and $\mathcal{S}_{\mathcal{F}}=\cup_{f \in \mathcal{F}}\mathcal{S}_f$.
Overall, this results in the following upper bound on the optimality gap
\begin{align} \mathcal{E}_D(f_{\mathcal{S}, \mathcal{S}_f}) & = \mathcal{G}_D(f_{\mathcal{S},\mathcal{S}_f},\mathcal{S},\mathcal{S}_{f_{\mathcal{S},\mathcal{S}_f}})+\mathcal{A}_D(f_{\mathcal{S},\mathcal{S}_f},\mathcal{S},\mathcal{S}_{f_{\mathcal{S},\mathcal{S}_f}}) \leq 2 \mathcal{D}_D(\mathcal{F}, \mathcal{S},\mathcal{S}_{\mathcal{F}}).
\end{align}  In general, the analysis of the uniform deviation depends on the choice of divergence measure, as well as on the specific unsupervised learning model considered.

As shown in Appendix~\ref{app:genboundunsup}, for any IPM $D_{\rm IPM}(\cdot,\cdot)$ in \eqref{IPM}, under suitable assumption on functions $h \in \mathcal{H}$, as in Theorem~\ref{thm:rada}, we can apply the uniform deviation relation \eqref{rada simeq} to obtain the bound 
\begin{align}\label{eq:optgapunsup}
    \mathcal{E}_{\rm IPM}(f_{\mathcal{S}, \mathcal{S}_f}) \leq 2 \mathcal{D}_{\rm IPM}(\mathcal{F},\mathcal{S},\mathcal{S}_{\mathcal{F}}) \simeq \mathcal O\left(\sqrt{\frac{B(\mathcal H)}{N}}\right) + \mathcal O\left(\sqrt{\frac{B(\mathcal F \times \mathcal{H})}{M}}\right), 
\end{align}where 
$B(\cdot)$ is defined in Section \ref{sec:Rad} as the function determining the dependence of the Rademacher complexity on the argument class (see (\ref{rada simeq})), and $\mathcal{F} \times \mathcal{H}=\{(f,h):h \in \mathcal{H}, f \in \mathcal{F}\}$ denotes the combined function space of discriminators and models. This bound relates the optimality gap to the statistical complexity of  the chosen class $\mathcal{H}$ of discriminators, as well as to the class of models $\mathcal{F}$. 
 
In the limit of a large $M$, and in particular for explicit models, the dominant term in (\ref{eq:optgapunsup}) is the first one, which depends only on the discriminator class $\mathcal{H}$. This observation provides useful guidelines for the choice of the divergence to be used for training. For instance, as remarked in \cite{arora2017generalization, zhang2017gendiscr},  since the set of all $1$-norm bounded and $1$-Lipschitz class of functions is larger than the set of parameterized neural networks, the bound motivates the use of the neural-net distance, as opposed to the widely adopted 1-Wasserstein distance.

\section{Quantum State Discrimination} \label{sec:qsd}

Quantum state discrimination (QSD) is the task of deciding which state a certain test quantum system is in, given knowledge of the finite set of possible states.  This corresponds to a special case of the quantum learning problem in which the generation mechanism $\mathcal{P}$ is fully known, and the output $y$ is a label that determines the identity of the test state. We specifically focus on the case of binary QSD, in which the test state may be equal to one of two known states: $\rho_+$, identified with label $y=+$, or $\rho_-$, identified with label $y=-$. Furthermore, it is known that the test state is equally likely to be either $\rho_+$ or $\rho_-$. Note that in this case, there is no classical input $x$ (see Section~\ref{sec:summary}), and there is no need for training data, since the data-generation mechanism $\mathcal{P}$ is available to the learner.

QSD underlies several applications of quantum information. For instance, in quantum cryptography and key distribution, one needs to assess whether the received state corresponds 
to the known state encoding bit 0 or bit 1\cite{pirandola2020advances}. As other examples, in quantum illumination \cite{tan2008quantum} and quantum radars 
\cite{maccone2020quantum}, a probe light is sent to illuminate an object and, based on the scattered 
quantum state received by the detector, the task is to decide whether a target was there or not. 
Extensions beyond the binary case were considered for barcode reading and pattern classification 
with quantum light \cite{banchi2020quantum} and for channel position finding \cite{zhuang2020ultimate}. 

The quantum state discrimination routine may be summarised as follows 

\begin{enumerate}[leftmargin=45pt,align=left,labelwidth=45pt,labelsep=0pt]
	\item[QSD.1] Obtain a classical description of the density matrices,
		$\rho_+$ and $\rho_-$, describing the two classes of states, e.g., by using 
		a theoretical model. 
	\item[QSD.2] Construct optimal or almost optimal measurement strategies, e.g., based 
		on Holevo-Helstrom measurements \cite{holevo1973statistical,helstrom1969quantum} for binary classes 
		or pretty good measurements for multiple classes \cite{PGM2,barnum2002reversing,montanaro2019pretty}. 
	\item[QSD.3] Apply such a measurement strategy to discriminate an unknown
		state, assuming that it is in either state
		$\rho_+$ or state $\rho_-$. 
\end{enumerate}

Fundamental works in the theory of state discrimination were performed by Helstrom \cite{helstrom1969quantum} and Holevo \cite{holevo1973statistical}, which considered the most general set of measurements. However, there are certain cases, e.g., in multi-qubit systems or in more general many-particle settings, where the most general operator can be highly non-local and difficult to implement. In these cases, it makes sense to constrain the available set of measurements and operations
\cite{matthews2009distinguishability}. Another reason is that, as we will show, unconstrained measurements can lead to large generalization errors when the physical problem is described by many degrees of freedom. From the discussion in the previous section we know that in all of these cases it is useful to constrain the function class to improve the generalization performances.

The fact that the states are known a priori 
makes QSD different from a learning problem, while forming the conceptual basis, as well as a key benchmark, for many learning algorithms. In this section, we focus on QSD for settings with two possible density matrices $\rho_+$ and $\rho_-$ in order to set the necessary background for the learning problems studied in the following sections.

\subsection{Single-Shot Discrimination with Fixed Measurements}\label{sec:fixed}

We focus on known states, in the sense that a
classical description of the density matrices $\rho_+$ and $\rho_-$ is available to the discriminator.  In
this subsection, we fix an arbitrary measurement described by a 
positive operator-valued measure (POVM), i.e.,  
by a set of positive semi-definite operators $\Pi_k$ satisfying the equality $\sum_k \Pi_k=\openone$. We recall that projection matrices $\Pi_k$ define a specific subclass of POVMs known as projective measurements. Given a fixed POVM, we study the problem of optimizing the binary decision of whether the test system is in state $\rho_+$ or $\rho_-$ on the basis of a single observation, or shot, of the test state. In later subsections, we will address the case in which more copies of the test state are available, allowing multiple measurements to be made on the test system.

By Born's rule, a measurement $\mathcal M = \{\Pi_k\}$ maps a density matrix $\rho$ to a random classical outcome $k$ 
with probability 
\begin{equation}
	p_\rho(k) = \Tr[\Pi_k \rho]. 
	\label{measurement}
\end{equation} In principle, as assumed in this subsection, the number of possible output values $k$ is arbitrary, although, as we will see in the next subsection, for QSD with two possible states, it can be taken to be two without loss of generality. 

Given the random observation $k\sim p_\rho(k)$ output of a single-shot measurement, a decision is made via a  classical post-processing, i.e.~a stochastic distribution  $f(y|k)$ that maps the measurement 
outcome $k$ into the predicted class $y \in \{+,-\}$, indicating a decision that the measured density matrix was $\rho_+$ or $\rho_-$ for $y=+$ or $y=-$, respectively.  By following the notation used in the previous sections, we will write $f$ to denote either probabilistic or deterministic mappings from $k$ to $y$.

For a fixed measurement $\mathcal{M}$, the optimal design of the mapping $f(y|k)$  amounts to the problem of distinguishing the two probability distributions $p_{\rho_\pm}(k) = p_{\pm}(k)= \Tr(\rho_\pm \Pi_k)$ based on measurement output $k\sim p_\pm(k)$. The corresponding optimization problem is defined in terms of the minimization of the average loss as \begin{equation}
 	L(f,\mathcal M) = \frac12 \sum_{k,y} \ell(f,k,y) \Tr[\Pi_k \rho_y]
 	= \frac12 \sum_{k,y} \ell(f,k,y) p_y(k),
 	\label{qsd loss}
\end{equation} where we have assumed that the possible states $\rho_+$ and $\rho_-$ are equally probable. A natural loss function $\ell(f,k,y)$  is the probability of error,  also known as 0-1 loss, 
$\ell_{01}(f,k,y)$, which equals 0 when $y=f(k)$ and 1 otherwise. For such a loss function, the optimal decision, also known as Bayes decision rule, is deterministic, and it sets 
$f(k)=+$ if $p_+(k)\geq p_-(k)$ and $f(k)=-$ otherwise.  Accordingly, the Bayes decision rule can be written as \begin{equation}\label{eq:Bayesdr}f_*(k) = {\rm sign}[p_+(k)-p_-(k)]. \end{equation}

A loss is said to be Bayes consistent if the optimal decision 
function $f_* = \argmin_f L(f,\mathcal M)$ equals the Bayes decision rule (\ref{eq:Bayesdr}). Bayes consistent losses with 
more desirable properties for numerical optimization include the hinge loss 
$\ell_h(f,x,y) = \max\{0,1-yf(x)\}$, used in support 
vector machines, as well as the smooth approximation 
provided by the logistic loss $\ell_{\gamma}(f,x,y) = \gamma^{-1}\log(1+e^{-\gamma yf(x)})$ 
with  margin parameter $\gamma>0$.

\subsection{Single-Shot Discrimination with Optimized Measurements} \label{sec:Helstrom}

In the previous subsection, we fixed the POVM and optimized over the classical post-processing map $f(y|k)$ with the goal of minimizing an average loss criterion. In this section, we address the problem of optimizing the POVM.

We start by observing that when the POVM is optimized, there is no need for a post-processing map $f(y|k)$, since the latter can be effectively integrated into the POVM.
Given any POVM $\mathcal M = \{\Pi_k\}$ and any classical stochastic mapping $f(y|k)$, we can define a new POVM $\{\Pi_+,\Pi_-\}$ that returns the same probabilities  $p_\rho(y)=\sum_k f(y|k)p_\rho(k)$ of decisions $y\in \{+,-\}$ via the matrices $\Pi_y=\sum_k f(y|k)\Pi_k$ for $y=\{+,-\}$. Therefore, one can implement an optimized POVM $\{\Pi_+,\Pi_-\}$ and use the output of the measurement as a decision without loss of generality. As a note, one way to implement such a binary-valued POVM on an $n$-qubit system is to apply a 
unitary operation $U$ on the entire system followed by a Pauli measurement on a single qubit, the binary outcome of which
provides the predicted class. We write this measurement as $\mathcal M_U$ where 
$\Pi_y = U(\ket {k_y}\!\bra {k_y}\otimes \openone_{n-1})U^\dagger$, where $k_+=0$ and $k_-=1$.

The optimal choice of POVM depends on the objective function that we want to minimize.
For instance, 
when we consider the single-shot probability of error $L_{01}(f,\mathcal{M})$ as a loss and optimize over all possible 
measurements, the solution is given by 
\begin{align}
	\inf_{\mathcal M} L_{01}(\mathcal M) &= \frac12 - \frac14\|\rho_+-\rho_-\|_1,
&
	\argmin_{\mathcal M} L_{01}(\mathcal M) &= \{\Pi^{\rm HH}_{\pm}\}, 
		&
	\Pi^{\rm HH}_\pm &= \frac{\openone \pm {\rm sign}(\rho_+-\rho_-)}2,
	\label{helstrom}
\end{align}
where the optimal measurement $\mathcal M_{\rm HH} =\{\Pi^{\rm HH}_{\pm}\}$ is known as 
Holevo-Helstrom measurement.
As another celebrated example, if one maximizes the  mutual information between the 
true class index and the predicted outcome, the best POVM is not 
known in closed form, but the maximum mutual information can be upper bounded via the accessible 
information $S(\frac{\rho_++\rho_-}2)-\sum_{y=\pm} S(\rho_y)/2$, where $S=-\Tr[\rho\log_2\rho]$
is the von Neumann entropy.

\subsection{Multiple Shot Discrimination via the Majority Rule}\label{sec:majority rule}

In the previous subsections, we have studied the case in which a single observation $k$ is made on a copy of the system in state $\rho_+$ or $\rho_-$. In this subsection, we study the case in which $V$ copies of the quantum system, all in the same state $\rho_+$ or $\rho_-$, are available. Accordingly, the system is in either of the  many-copy states $\rho_+^{\otimes V}$ or $\rho_-^{\otimes V}$.

The measurement postulates of quantum mechanics stipulate that a projective measurement on a quantum system places it in an eigenstate of the measured observable, and so it is not possible to obtain more information by carrying out multiple, independent, projective measurements on a single copy of a quantum state \cite{holevo1973bounds}. Moreover, due to the no-cloning theorem \cite{wootters1982single}, one cannot produce copies of an unknown state and so it is not possible to copy the system in order to obtain more measurements of the same state. That said, in spite of the no-cloning theorem, different copies of the state can be made if the {\it recipe} to  build such state is known, e.g., by lowering the temperature or applying some external control onto a system.


When multiple copies $V$ are available, by the discussion in the previous subsection, the optimal measurement in terms of probability of error, or 0-1 loss, is given by the Helstrom measurement. Using Fuchs-van de Graaf inequalities and 
properties of the fidelity function $F(\rho_+,\rho_-)=\|\sqrt{\rho_+}\sqrt{\rho_-}\|_1^2$, the resulting minimum probability of error can be upper bounded as
\begin{equation}
	\inf_{f,\mathcal M} L^V_{01}(f,\mathcal M) = \frac12 - 
	\frac14\|\rho_+^{\otimes V}-\rho_-^{\otimes V}\|_1 
	\leq \frac{F(\rho_+,\rho_-)^{\frac V2}}2.
	\label{fidelity decay}
\end{equation}
Therefore, the probability of error with an optimal measurement decreases 
exponentially with the number of copies, as long as the two states are not identical, i.e., as long as we have  $F(\rho_+,\rho_-)\neq 1$. 

However,
the resulting optimal Helstrom measurement in \eqref{helstrom} 
is in general highly non-local, 
requiring the application of coherent measurements over all $V$ copies of the state.  
Simpler adaptive strategies are only known to be optimal for discriminating certain classes of states, such as pure states 
\cite{acin2005multiple}. In the rest of this subsection, we explore a simple, suboptimal, measurement strategy based on independent, local, measurements of each copy followed by a \emph{majority 
vote}.

Accordingly, we consider performing  independent measurements, possibly at different times, of the different copies, without having 
to physically build many copies of the state in parallel. To elaborate, suppose that each local measurement applies some arbitrary binary measurement $\mathcal M = \{\Pi_\pm\}$. 
Based on local measurement $\mathcal M$, the majority vote over $V$ copies can be described by a binary POVM $\mathcal M_{\rm maj}^{\mathcal M,V} 
	= \{\Pi^V_\pm\}$ applied on the $V$ copies. Specifically, POVM $\mathcal M_{\rm maj}^{\mathcal M,V} $ applies the local measurement $\mathcal M$ onto all copies of the state, obtaining the random, i.i.d. outcomes $y_1,\dots,y_V \sim p_\rho(y)$,
with $y_j\in\{\pm\}$, and then outputs the class $\pm$ depending on the majority of outcomes. To avoid the possibility of ending in a draw, we assume that $V$ is odd for simplicity. 

Let us denote as $p_+ = \Tr[\Pi_+ \rho_+]$  the probability that a local measurement returns the correct decision $+$ when the true state is $\rho_+$. The following discussion would equally apply to the case in which the true state is $\rho_-$ by swapping the signs. The probability of outputting the correct decision $+$ when the true state is $\rho_+$ is given by the binomial distribution 
\begin{equation}
	p_{\rho_+}^V(+) = \Tr[\Pi_+^V \rho_+^{\otimes V}] =  
	\sum_{v > V/2} \binom{V}{v} p_+^s(1-p_+)^{V-v} = 1-C^V_{V/2}(p_+), \label{eq:probcorrectcond}
\end{equation}
where  we have defined $C^V_k(p) = \sum_{v\leq k} \binom{V}{v} p^v (1-p)^{V-v}$
as the cumulative distribution function of the binomial distribution. We now
show that, under mild conditions on the probability $p_+$ of correct detection
of each local measurement, the majority rule also yields a probability of error
that decreases exponentially with the number of copies, $V$, as for the
corresponding probability (\ref{fidelity decay}) of the optimal, global,
measurement.

To do this, we observe that, for large $V$, we have the approximation \cite{ash1990information}
$C_k(p) \approx \exp(-V D_{\rm KL}(k/V,p))$ for $k/V<p$,
where $D_{\rm KL}(a,p) = a\log(a/p)+(1-a)\log((1-a)/(1-p))$ is the binary Kullback-Liebler (KL) divergence. Therefore, as long as the condition 
\begin{equation}
	p_+ > \frac{v+1}{V} =  \frac 12 + \mathcal O(V^{-1})
\label{p+ V}
\end{equation} holds, 
where $v$ is the integer such that $V=2v+1$, then the KL divergence term is positive, and the  probability of  correct detection (\ref{eq:probcorrectcond}) converges exponentially quickly to 1 as a function of the number 
of measurement shots $V$. Note that, as compared to the probability~\eqref{fidelity decay} for optimal global measurements, the exponent of the probability of error is generally suboptimal. 

We conclude that, even without the optimal  global measurements \eqref{helstrom}, 
provided that the local measurement is able to distinguish the state with probability \eqref{p+ V} above chance, i.e., larger than 1/2, 
perfect discrimination can be obtained in the limit of many measurement shots $V$.

\subsection{Multiple Shot Discrimination via Expected Value of an Observable}

In the previous subsections, we have assumed discrimination models based on POVMs. In this subsection, we consider a conceptually distinct family of discriminators that are based on the expected values of an observable. Hence, instead of optimizing a POVM, such schemes optimize over observables. As we will see, for some specific penalties this optimization can be formalized via the representer theorem, creating a link between quantum discriminative models for QSD and kernel-based methods.

\subsubsection{Problem Formulation}\label{s:measurement problem formulation}

To elaborate, consider an observable defined by a Hermitian operator $A$ and denote the expected value of observable $A$ over state $\rho$ with $\rho\in\{\rho_+,\rho_-\}$ as $\langle A\rangle_{\rho} = \Tr[A\rho]$. We study the class of decision functions of the form \begin{equation}
	y = {\rm sign} (\langle A\rangle_{\rho_y}).
	\label{y predicted A}
\end{equation} 

In order to implement and optimize the observable $A$, one typically relies on a linear decomposition of observable $A$ into operators that are easier to realize. A first approach is to  decompose the observable in terms of a POVM $\{\Pi_k\}$ as  $A = \sum_k a_k \Pi_k$, where $a_k$ are real numbers. This way, one can implement an observable-based predictor in the same way as for the POVM-based predictors studied earlier in this section. To this end, applies the POVM $\{\Pi_k\}$, and considers the random variable $a_k \sim \mathrm{Tr}[\Pi_k \rho]$ as the output of a measurement on the system. By averaging this random variable one obtains the expectation in \eqref{y predicted A}. A natural decomposition of this type is obtained via the spectral decomposition of operator $A$. However, this becomes impractical for large Hilbert spaces, since the eigenprojectors of an operator $A$ are generally non-local.

Alternative decompositions can rely on more convenient bases for the operator space that consist of operators with locality properties. As a notable example, with the basis of multi-qubit Pauli operators $P_\alpha$, an observable $A$ can be  decomposed as
$A = \sum_\alpha a_\alpha P_\alpha$
with real coefficients $a_\alpha$. Each operator $P_\alpha$ is the tensor 
product of single-qubit Pauli operators. Such single-qubit observables can be efficiently measured, since each single-qubit Pauli operator can be written as a local rotation followed 
 by a  local projective POVM in the computational basis $\{\ket0\!\!\bra0,\ket1\!\!\bra1\}$.  
Therefore, the expected value $\langle A \rangle_\rho$ can be efficiently evaluated as the sum $\langle A \rangle_\rho=  \sum_\alpha a_\alpha \mathrm{Tr}[P_\alpha \rho],$ where each expectation $\mathrm{Tr}[P_\alpha \rho]$ can be evaluated separately via local operations.

While it is convenient to assume that decisions are made on the basis of the expectation \eqref{y predicted A}, in practice this decision function  cannot be directly evaluated, but only estimated on the basis of measurements over $V$ copies of the state. Accordingly, one  replaces the expected value $\langle A\rangle_{\rho_y}$ with an empirical average using $V$ independent measurements of the observable $A$. 
The corresponding average quadratic  
error is given by $\Delta_y/\sqrt V$, where $\Delta_y^2 = \Tr[A^2\rho_y]-\Tr[A\rho_y]^2$ is the single-shot 
variance. This error 
should be made sufficiently small to avoid wrong predictions. 
Informally, one should ensure the condition  
\begin{equation}
	\frac{\Delta_y}{\sqrt V} \ll |\langle A\rangle_{\rho_y}|.
	\label{Delta condition}
\end{equation}
When this is not the case, the model \eqref{y predicted A} uses an unreliable approximation of the actual empirical average to produce a decision. Furthermore, in this regime, the classifier is known to be  vulnerable to adversarial attacks 
\cite{ren2022experimental,banchi2022robust}, since tiny perturbations in the inputs can alter 
the prediction of the classifier. 

More precisely, when the average is estimated
using $V$ samples, assuming a zero-error predictor \eqref{y predicted A} and that the magnitude of the expected value of the observable is the same for both states, so that $|\langle A\rangle_{\rho_+}|=|\langle A\rangle_{\rho_-}|$, one can bound the residual probability of error caused by the use of only $V$ samples, using
Hoeffding's inequality (Appendix~\ref{a:math}), as
\begin{equation}
	L_{01}^{V}(A) \leq \exp\left(-\frac{V \langle A\rangle_{\rho_{\pm}}^2}{2\|A\|^2_\infty}\right).
	\label{perr observable}
\end{equation} 

In the next subsections we study different ways of optimizing over a set of observables. We start from the simplest case in which the operator structure is fixed in terms of a linear decomposition based on a POVM, and optimization is done only over the linear coefficient of the decomposition. Then, we address the more challenging scenario in which constraints on the operator $A$ do not limit the optimization space to a specific linear decomposition. Finally, we observe that a specific formulation of such constraints enables the application of the representer theorem, which yields a convenient parametrization of the solution 
to the optimization problem as a combination of the states to be distinguished.

\subsubsection{Decision Observables with Fixed Operator Structure} 

Let us fix a POVM $\{\Pi_k\}$, and write the decision observable as the linear combination $A = \sum_k a_k \Pi_k$ with trainable parameters $a_k$. In this case, the problem amounts to the classical detection of two probability distributions, namely $p_+(k) = \Tr[\Pi_k \rho_+]$ and $p_-(k) = \Tr[\Pi_k \rho_-]$, based on an average of the observations $a_k \sim \Tr[\Pi_k \rho]$. Therefore, it can be addressed via a ``quantum 
data collection'' phase followed by classical post-processing as per the decision function (\ref{y predicted A}). 

We focus here on common  losses that take the form
\begin{equation}
    \ell(f,x,y) = \Lambda[yf(x)],\label{eq: convex loss function}
\end{equation}
where $\Lambda(\cdot)$ is a \emph{convex} function. Examples include the hinge loss with $\Lambda(z)=\max\{0,1-z\}$, and the logistic loss with $\Lambda(z)=\log(1+\exp(-z))$ (see Table 6.1 in \cite{simeone2022machine}). With this choice, the average loss can be written as
\begin{equation}
	L(f,\mathcal M_A) =  \sum_{k,y} \ell(f,k,y) \Tr[\Pi_k \rho_y] = 
	\sum_{k,y} \Lambda[ya_k] \Tr[\Pi_k \rho_y], 
	\label{observable loss}
\end{equation}which amounts to a classical binary classification problem over the coefficients $\{a_k\}$.

\subsubsection{Norm-Constrained Decision Observables}\label{sec:norm constrained}

We now focus on optimizing the entire operator $A$. In this case, the problem of designing operator $A$ does not reduce to the classical problem of detecting two classical probability distributions, since the structure of the operator determines the distributions of the measurement outputs. 

Focusing again on losses of the form in Eq.~(\ref{eq: convex loss function}), the design problem amounts to the minimization of the loss  \begin{equation} \label{observable loss inequality1}
	L(A) =  
 \sum_{y} \Lambda\left(y\langle A\rangle_{\rho_y}\right). 
\end{equation} This optimization is in principle feasible, since the problem is convex as long as the domain of matrix $A$ is a convex set.  However, for large systems  with many qubits, the size of the optimization variable $A$ becomes unmanageable without imposing some restrictions on the optimization domain. 

Before addressing this problem, we observe that, given an eigendecomposition $A=\sum_k a_k \Pi_k$, for any operator $A$, the loss $L(A)$ in (\ref{observable loss inequality1}) is no larger than that in (\ref{observable loss}). In fact, by Jensen's inequality, we have \begin{equation}
	L(A) =
	\sum_{y} \Lambda\left(\sum_k ya_k \Tr[\Pi_k \rho_y]\right)  
 \leq L(f,\mathcal M_A).
	\label{observable loss inequality}
\end{equation} This result will be useful in the next sections.

As discussed, in order to address the minimization of function $L(A)$, one needs to impose some constraints on matrix $A$, while not fixing its structure, as we did in the previous subsection. A typical approach is to assume the form $A= U Z_1 U^\dagger$, where 
$U$ is a trainable unitary, e.g., via a parametric quantum circuit followed by a Pauli-Z measurement 
on the first qubit. With this choice, however, the minimization problem is no longer convex, making algorithmic solutions and analysis more problematic.

Inspired by structural risk minimization (Section~\ref{s:structural risk minimization}),
we will take a different route and map the constraints on the decision observable as a penalty 
term. As we elaborate next with an example, the penalty should ideally reflect an underlying constraint on the structure of the operator $A$. 

As an example, the assumption  
$A= U Z_1 U^\dagger$ discussed above satisfies the constraint  $A^2=\openone$,  and hence this structure can be approximately imposed by adding  
a penalty term of the form $\Tr[(A^2-\openone)]$. This choice, however, is not unique. For instance, the observable also  satisfies the constraint $\|A\|_\infty = 1$, since its eigenvalues are $\pm1$. Therefore, one could also add a penalty based on the norm $\|A\|_\infty$.

Generalizing the example, we consider imposing a constraint on the operator $A$ based on the value of a norm $\|A\|_\sharp $. Accordingly, the optimization domain is defined as  $\mathcal A_{\sharp,p} = \{A : \|A\|_\sharp \leq p\}$. Introducing a positive convex barrier function $g(\cdot)$, one can formulate the problem in an unconstrained form, as  discussed in Section~\ref{s:structural risk minimization}, yielding the optimized observable \begin{equation}
	A_{*,\lambda} = \argmin_{A}[ L(A) +  \lambda g(\|A\|_\sharp)].
	\label{eq:criterionrepr}
\end{equation} In the rest of this subsection we discuss some physically motivated operator norms that 
can be used to construct penalties terms, while the next subsection shows how problem (\ref{eq:criterionrepr}) can be addressed for the norm $\|A\|_2$.

Classical shadows represent a powerful technique for predicting the expectation values 
of many observables without doing full tomography\cite{huang2020predicting}. 
The prediction error with classical shadows is quantified by the shadow norm $\|A\|_{\rm shadow}$,
whose definition depends on the choice of the shadow representation. In particular, for Clifford and local operations 
it was found that the following bounds hold, respectively, 
\begin{align}
	\|A\|_2^2 &\leq	\|A\|^2_{\rm shadow, Clifford} \leq 3 \|A\|_2^2, & 
		\|A\|^2_{\rm shadow,local} \leq 4^k \|A\|_\infty^2,
		\label{shadow norm ineqa}
\end{align}
where $k$ is the number of qubits on which $A$ acts non-trivially. As a result, 
a penalty dependent on the norm $\|A\|_2$ can model observables that can be estimated efficiently using 
classical shadows with Clifford circuits; while a penalty on $\|A\|_\infty$ can model 
observables that can be estimated efficiently using local operations. 
Therefore, for instance, if two states can be distinguished using local observables, a penalty based on norm $\|A\|_\infty$ 
may be more appropriate. 

Another notion of locality is at the heart of the quantum Wasserstein distance of 
order 1\cite{de2021quantum}. For a traceless operator $X$, the $W_1$ norm is defined as 
$\|X\|_{W_1} = \max_{A : \|A\|_L\leq 1} \Tr[X A]$
where $\|A\|_L$ is the quantum Lipschitz constant of the traceless observable $A$, which is the dual norm 
of $\|\cdot\|_{W_1}$ 
\cite{de2021quantum}. 
The $W_1$ distance provides a quantum version of the Hamming distance, 
since two quantum states $\rho_\pm$ that coincide after discarding $k$ qudits satisfy the inequality  
$\|\rho_+-\rho_-\|_{W_1} \leq 2k$. We can summarize the properties of these two norms 
for traceless observables\cite{de2021quantum} acting on $n$ qudits as 
\begin{align}
	\|A\|_1 &\leq	\|A\|_{W_1} \leq  k \|A\|_1,
					&
	\|A\|_L &\leq \max_{1\leq i\leq n} \|A_i\|_\infty.
\end{align}
where $k$ is the number of qudits on which $A$ act non-trivially, 
while $A_i$ is the sum of terms in the operator expansion of $A$ that 
act non-trivially on qudit $i$. These norms can be used to define tight 
bounds for parametric quantum circuits \cite{de2023limitations}. 
As such, these penalties may represent a natural choice to enforce constraints 
in the depth of the quantum circuit used to classify the two states.

\subsubsection{Decision Observables from the Representer Theorem } \label{sec:representer}

In general, obtaining the optimized observable (\ref{eq:criterionrepr}) is  intractable when the dimension 
of the Hilbert space is large. However, an application of a classical result from statistical learning, namely 
the representer theorem \cite{shalev2014understanding}, can be leveraged to simplify the problem when the average loss function being optimized is suitably regularized. 

Specifically, assume that we are interested in minimizing the criterion 
\eqref{eq:criterionrepr}
with the choice $g(\|A\|_\sharp) = \Tr[A^2]$.
For this particular choice, by the representer theorem, the optimal solution of problem (\ref{eq:criterionrepr}) can be always expressed as a linear combination  of the two possible states, i.e.,  
\begin{equation}
	A = \alpha_+ \rho_+ + \alpha_- \rho_-,
	\label{representer qsd}
\end{equation}
where the real coefficients $\alpha_{\pm}$ must be optimized to minimize the objective function in  \eqref{eq:criterionrepr} \cite{schuld2021supervised}.

As a specific example, consider the hinge loss function. Then using \eqref{representer qsd} and 
\eqref{observable loss inequality1}
we can write the average loss and regularizer as  
\begin{align}
	L_{\rm hinge}(A) &= \frac12 \left(\max\{0,1-\alpha_+P_+-\alpha_- F\} +  \max\{0,1+\alpha_+F+\alpha_- P_-\}\right),
	\\
	\Tr[A^2] &= \alpha^2_+ P_+ + \alpha_-^2 P_- + 2\alpha_+\alpha_- F,
\end{align}
where $P_\pm = \Tr[\rho_\pm^2]$ is the purity of each of the two possible states and $F=\Tr[\rho_+\rho_-]$,
which represents the fidelity between two quantum states when at least one of the two states $\rho_+$ and $\rho_-$ is pure. With this choice, problem (\ref{eq:criterionrepr}) has the solution 
\begin{equation}
	A_* =\frac{ (P_-+F)\rho_+ - (P_++F)\rho_-}{P_+P_--F^2}  
	\stackrel{\rm pure}{=} \frac{ \rho_+ - \rho_-}{1-F},
	\label{observable optimal}
\end{equation}
where the second equality holds when both states are pure, and hence we have the equalities $P_\pm=1$. Accordingly, in the special case of pure states, for an unknown true state $\rho\in\{\rho_+,\rho_-\}$, the optimal predictor (\ref{y predicted A}) computes the fidelities, or overlaps,  
$\Tr[\rho \rho_+]$ and $ \Tr[\rho \rho_-]$, and outputs $y=+1$ or $y=-1$ depending on whether the first or the second is larger than the other.
Note that this discriminator was called the ``fidelity'' classifier in \cite{lloyd2020quantum}.

As mentioned, in practice, the predictor \eqref{y predicted A} cannot be evaluated exactly based on the availability of $V$ copies of the state.
The test copy complexity should be large enough to enforce the condition \eqref{Delta condition} and make the probability of error \eqref{perr observable} sufficiently small. 
In the case of the ``fidelity'' classifier with optimal observable (\ref{observable optimal}), when both possible states are pure, we have  $\langle A_*\rangle_{\rho_\pm}=\pm1$, and $\|A\|_\infty=1/(1-F)$ so the conditions \eqref{Delta condition} and \eqref{perr observable} result in 
\begin{align}
	\Delta^2 &= \frac{F}{1-F} \ll  V,
					 &
	L_{01}^{V}(A) &\leq \exp(-V(1-F)^2/2).
\end{align}
Accordingly, for larger overlaps $F$ between the two possible states, one needs more shots $V$ to resolve the differences between the two states with sufficiently high probability.

\section{Learning To Discriminate Unknown Quantum States}\label{sec:learning discriminate}

When knowledge of the states  to be distinguished is not available, the state discrimination problem is a proper learning problem as described in Section 1. To formalize this problem, assume that, as in the previous section, the system of interest is equally likely to be in one of two possible states $\rho_+$ and $\rho_-$. Unlike in the previous section, however, the states $\rho_+$ and $\rho_-$ are now assumed to be unknown. The only information available at the learner about the states $\rho_+$ and $\rho_-$ is in the form of a training data set composed of $S$ quantum systems in state $\rho_+$ and of $S$ quantum systems in state $\rho_-$, on which the learner can act with measurements. 
Accordingly, the training set can be described as the single composite state 
$\rho_{+}^{\otimes S} \otimes \rho_{-}^{\otimes S}$.
Note that the positions of the states in the ordering implied by the state description
implicitly determines the label of each state as being $y=+$ for the first $S$ systems and $y=-$ for the last $S$ systems.

As per the framework in Section 1, the learner is also given $V$ additional unlabelled test systems which, unbeknownst to the learner, are all in either state $\rho_+$ or state $\rho_-$. The goal is to determine whether  the test systems are in state $\rho_+$ or $\rho_-$. The test systems are hence collectively in either state $\rho_+^{\otimes V}$ or state $\rho_-^{\otimes V}$.

Overall, the learner has access to a composite quantum system that is in state 
\begin{equation}\label{eq:overallstate}
	\underbrace{\rho_{+}^{\otimes S} \otimes \rho_{-}^{\otimes S}}_{\rm training}
\otimes\rho_{+}^{\otimes V} \textrm{~~ or ~~} 
	\underbrace{\rho_{+}^{\otimes S} \otimes \rho_{-}^{\otimes S}}_{\rm training}
\otimes\rho_{-}^{\otimes V}.
\end{equation} 
However, unlike the discrimination problem of Section~\ref{sec:qsd}, the individual states $\rho_+$ and $\rho_-$ are unknown, and hence it is not possible to design an optimal measurement using the procedures described in the previous section. In fact, as we will discuss in Appendix \ref{sec:averagevswc}, there are even different ways to define optimality in this case.

The most general strategy is to apply a \emph{joint} binary measurement on both training and test systems. The goal of the measurement is to determine if the test systems are more ``similar'' to the first $S$ systems in the training data, in which case the detector outputs the label $y=+$ as its decision; or rather if the test systems are more ``similar'' to the last $S$ systems in the training data, in which case the detector outputs the label $y=-$ as its decision. Following the definitions given in Section 1, the general class of joint measurements implements a \emph{transductive learning} strategy. As seen, in transductive learning, training and test data are jointly processed to produce a decision on the test inputs.

In classical machine learning, transductive learning strategies have the disadvantage that, when new test inputs are presented to the learner, training and test inputs would have to be jointly processed anew in order to produce a decision \cite{derbeko2004explicit}. In contrast, for more conventional \emph{inductive learning}, as described in Section 1, the learner obtains a general rule $f$ from the training data set, which is then used to make decisions on any new test input. In the case of quantum machine learning, the limitations of transductive learning approaches are compounded by the fact that, once the training data are operated on via a measurement, they are no longer available to be jointly measured with  new test input states. Therefore, with quantum transductive learning, training data can be used only once. That said, as also hinted at in Section 1, transductive learning may have advantages in terms of learning performance.

In the next subsections we introduce different inductive and transductive state discrimination strategies, 
and analyze the performance of the inferred predictors with respect to the average loss via the decomposition  \eqref{eq:error optimality gap_1}. We start by specializing the average loss decompositions introduced in Section 1 to the problem at hand in the next subsection.

\subsection{Generalization Analysis}
We now review and specialize the framework introduced in Section 1 to analyze the performance of inductive and transductive learning algorithms.

   \textbf{Inductive learning:} As discussed
 in Section 1, the inductive learning strategy adopts a two-step procedure. In the first step, the available $S$-copy training set is used to extract some classical knowledge, which we denote as $\mathcal{S}_o$. This can for example be a set of observations generated via measurements on the $S$-copy training set, or  some classical description of the unknown states. We can then define an empirical training loss $L(f,\mathcal{S}_o)$ on this set of observations (see Section~\ref{sec:optimality}), which can be minimized to obtain the optimized inference function $f_{\mathcal{S}_o}$. Since $f_{\mathcal{S}_o}$ is our finite-$S$ approximation of $f_{\mathcal{S}}$, we also call it $f^S_{\mathcal{S}}$.

    One can then study the average loss $L_{\mathcal{P}}(f^S_{\mathcal{S}})$ of the inferred function using \eqref{eq:error optimality gap_1}. For the setting under study here, with only two possible states, the abstract dataset loss $L(f,\mathcal{S})$, which uses knowledge of the unknown quantum states, coincides with the average loss $L_{\mathcal{P}}(f)$. This in turn results in the respective minimizers being equal, i.e., $f_{\mathcal{S}}=f_*$. Using this equality in \eqref{eq:error optimality gap_1}, we get the following equivalent decompositions,
    \begin{align}
        L_{\mathcal{P}}(f^S_{\mathcal{S}}) &= L_{\mathcal{P}}(f_*)+E^S_{\rm test} 
        =L_{\mathcal{P}}(f_*)+E^S_{\mathcal{S}} 
        =L_{\mathcal{P}}(f_*)+\mathcal{E}(f^S_{\mathcal{S}}).\label{eq:sec4_2}
    \end{align} Consequently, for inductive learning schemes, a small optimality gap, or equivalently knowledge gap, results in smaller average loss, thereby ensuring generalization of the classification strategy  built using partial information to the general case in which an arbitrary number of observations (or copies) are available. 
    
    In the next sub-sections, we will reserve analysis via optimality gap $\mathcal{E}(f^S_{\mathcal{S}})$ for scenarios when $f^S_{\mathcal{S}}$ is the result of minimizing an empirical training function and we want to emphasize the connection to the Rademacher complexity.

     \textbf{Transductive learning:} In contrast, a transductive learning strategy is a one-step implementation that eliminates the need to extract classical knowledge to define a training loss as in the inductive learning strategy. As such, the transductive scheme aims to directly approximate $f_*$. It does this by applying a single joint measurement on the training and test states that enacts the classifier $f^S_{\mathcal{S}}$ on the test state.
    In this setting, as seen in Section 1, the average loss can be decomposed as $L_{\mathcal{P}}(f_{\mathcal{S}}^S)= L_{\mathcal{P}}(f_*)+E^{S}_{\mathcal{S}}$. Accordingly, the joint measurement must be chosen such that for any pair of states $\rho_{\pm}$, in the asymptotic limit of $S\to\infty$, $E^{S}_{\mathcal{S}}\to 0$. As in the inductive scheme, a small $E^{S}_{\mathcal{S}}$ is an indication that the transductive strategy learned using partial information about unknown states generalizes to the case when arbitrarily large number of copies are available.

We now discuss different inductive and transductive learning strategies.

\subsection{Tomography-based Quantum State Classification}\label{sec:tomography classification}
The first naive approach is based on full state tomography. In this case the state discrimination routine 
proceeds as in Section~\ref{sec:qsd}, with the only difference that the initial step QSD.1 is done 
empirically by reconstructing the state with state tomography.
It is known that \cite{haah2016sample} 
full state tomography requires $S=\mathcal O(d^2/\epsilon^2)$ copies of a $d$-dimensional state to obtain an approximate classical description with precision $\epsilon$ in the trace distance. This can be reduced 
to $S=\mathcal O(dr/\epsilon^2)$ if the rank $r$ is known. 
For mathematical simplicity, we assume that both $\rho_\pm$ can be reconstructed up to the desired 
precision $\epsilon$ with the same number of copies $S$, since the extension to the general case 
is straighforward. 

Let us call $\rho^{\rm emp}_y$ the approximate reconstructions of the true states $\rho_y$ and define $\Pi^{\mathrm{HH}}_{\pm}$ and $\Pi^{\mathrm{HH},\mathrm{emp}}_{\pm}$ as the Helstrom POVMs constructed using the true and empirical states respectively (per Eq.~\ref{helstrom}). Finally, let us define $\eta = \|\rho_+-\rho_-\|_1$ and $\eta^{\mathrm{emp}} = \|\rho^{\mathrm{emp}}_+-\rho^{\mathrm{emp}}_-\|_1$. Note that, when carrying out the measurement, we have direct access to $\epsilon$ (which is set by $S$) and $\eta^{\mathrm{emp}}$ (which we can calculate from our approximate states), but not to $\eta$ (which depends on the unknown true states). However, using the triangle inequality, we can write the bound $|\eta - \eta^{\mathrm{emp}}|<2\epsilon$.

The average loss for our empirically constructed measurement is
$(1-\Tr[\Pi^{\mathrm{HH},\mathrm{emp}}_{+}(\rho_+ - \rho_-)])/2$, so as long as
$\Tr[\Pi^{\mathrm{HH},\mathrm{emp}}_{+}(\rho_+ - \rho_-)]>0$, it is better than
random guessing. This is guaranteed to be the case as long as
$\eta^{\mathrm{emp}}>2\epsilon$. Conversely, if
$\eta^{\mathrm{emp}}<2\epsilon$, the empirically constructed measurement can
be the worst, rather than the best, choice on the true states.
Indeed, for a given pair of empirical states, the true
states could be the linear combinations $\rho_\pm =
(1-\epsilon/\eta^{\mathrm{emp}})\rho^{\mathrm{emp}}_\pm +
\epsilon/\eta^{\mathrm{emp}}\rho^{\mathrm{emp}}_\mp$, so, for
$\eta^{\mathrm{emp}}<\epsilon$, we could have $\rho_\pm =
\rho^{\mathrm{emp}}_\mp$. It is also intuitively understandable that if the
states are closer together, we need to know them to greater precision in order
to discriminate between them.

Now let us bound the knowledge gap in terms of $\epsilon$, and hence the training copy complexity, $S$. Note that 
 $\Pi^{\mathrm{HH}}_{\pm}$ and $\Pi^{\mathrm{HH},\mathrm{emp}}_{\pm}$ are, respectively, the  
 optimal measurements for the average and empirical losses, so the knowledge gap, $E^{S}_{\mathcal{S}}$, is the only quantity of interest. 
Using the 0-1 loss \eqref{helstrom}, the knowledge gap is
\begin{equation}
    E^{S}_{\mathcal{S}} = \frac{1}{2}\Tr[(\Pi^{\mathrm{HH}}_{+}-\Pi^{\mathrm{HH},\mathrm{emp}}_{+})(\rho_+ - \rho_-)] = \frac{1}{4}\eta - \frac{1}{2}\Tr[\Pi^{\mathrm{HH},\mathrm{emp}}_{+}(\rho_+ - \rho_-)].
\end{equation}
Using the linearity of the trace, we can rewrite the last term as
\begin{equation}
    \Tr[\Pi^{\mathrm{HH},\mathrm{emp}}_{+}(\rho_+ - \rho_-)] = \Tr[\Pi^{\mathrm{HH},\mathrm{emp}}_{+}(\rho^{\mathrm{emp}}_+ - \rho^{\mathrm{emp}}_-)] + \Tr[\Pi^{\mathrm{HH},\mathrm{emp}}_{+}(\rho_+ - \rho^{\mathrm{emp}}_+)] + \Tr[\Pi^{\mathrm{HH},\mathrm{emp}}_{+}(\rho^{\mathrm{emp}}_- - \rho_-)].
\end{equation}
The first term is exactly $\eta^{\mathrm{emp}}/2$, whilst the last two terms are lower bounded by $-\epsilon/2$, so the knowledge gap is upper bounded by
\begin{equation}
  E^{S}_{\mathcal{S}}  \leq \frac{1}{4}(\eta - \eta^{\mathrm{emp}} + 2\epsilon) \leq 2\epsilon = \mathcal O\left(\frac{d}{\sqrt{S}}\right).\label{eq:tomography optimality}
\end{equation}

Note that in this approach we have built up a set of measurement outcomes, $\mathcal{S}_o$, and then minimized the corresponding empirical loss $L(f,\mathcal{S}_o)$. We could therefore, in theory, bound the optimality gap $\mathcal{E}(f_{\mathcal{S}_o}) = \mathcal{E}(f^S_{\mathcal{S}})$ using the Rademacher complexity. Whether we do so depends on which approach yields tighter bounds.

Suppose $V>1$, so we have more than one copy of the test state. If our empirically constructed measurement is better than a random guess for a single copy of the test state, then the error will decay exponentially with $V$. We know that this is the case if we carry out the (suboptimal) majority vote. The Rademacher complexity bound on the knowledge gap increases with $V$, but only sublinearly as $\mathcal{O}(\sqrt{V})$. See Section~\ref{sec:rademacher majority} for more details.

A similar analysis can be done using decision observables with loss \eqref{observable loss inequality1}.
More precisely, given a $\lambda$-Lipschitz function $\Lambda$ and a set of decision observables $\mathcal A$
we can use the inequality in 
\eqref{eq:excessrisk_uniformdeviation} to write 
\begin{equation}
    E^{S}_{\mathcal{S}} 
\leq2\sup_{A\in\mathcal A}| L(A)-L_{\mathcal S}(A)| 
\leq 3\lambda \sum_y \sup_{A\in\mathcal A} \left|\Tr[ A(\rho_y-\rho_y^{\rm emp})]\right|.
\label{empirical risk A}
\end{equation}
If the observables satisfy $\|A\|_\infty \leq B$, then by H\"older's inequality 
$E^{S}_{\mathcal{S}}
\leq \mathcal O(\lambda B \epsilon) =\mathcal O(\lambda B d/\sqrt{S})$. Therefore, if $S\gg d^2$,
the knowledge gap will go to zero. This comes at the price of performing a large amount of measurements 
(exponentially many, in the case of many-qubits).

A different approach consists in using the classical shadows formalism\cite{huang2020predicting}, 
which does not provide a tomographic reconstruction of the density matrices $\rho_\pm$ but 
rather some estimators $\rho_y^{\rm emp}$, which are possibly quite different in trace norm from $
\rho_y$, but such that $\Tr[A\rho_y^{\rm emp}]\approx \Tr[A\rho_y]$ for a large number of 
observables. More precisely, given a set of $M$ traceless observables $A_i$,
using the classical shadow estimator it was found that
	$|\Tr[A_i\rho_y^{\rm emp}]- \Tr[A_i\rho_y]|\leq \epsilon$,
	as long as 
\begin{align}
	S\geq \mathcal O\left(\frac{\log(M)}{\epsilon^2}\max_i \|A_i\|^2_{\rm shadow}\right),
\end{align}
where the ``shadow norm'' was already introduced in Eq.~\eqref{shadow norm ineqa}. 
For $M\approx e^m$ it is hence possibly to predict exponentially many observables with a number 
of copies that scale linearly with $m$.
Assuming that the states $\rho_\pm$ can be distinguished with linear combinations of 
the observables $A_i$, we may define the decision class $\mathcal A$ as the set 
of observables $A=\sum_{i=1}^M \alpha_i A_i$, then using 
\eqref{empirical risk A}, we get 
\begin{equation}
    E^{S}_{\mathcal{S}} 
	\leq 3\lambda \sum_{i=1}^M |\alpha_i|\epsilon 
	\lesssim 3\lambda \frac{\sqrt{\log(M)}\sum_{i=1}^M |\alpha_i|}{\sqrt S} \max_i \|A_i\|_{\rm shadow}.
\end{equation}
Therefore, although the shadow tomography is able to accurately reconstruct exponentially many observables
$A_i$, because of the sum  $\sum_{i=1}^M |\alpha_i|$, it is unclear whether this advantage persists over {\it families} of observables. As such, it is unclear whether the shadow tomography approach 
can provide an exponential advantage for state discrimination.

\subsection{Discrimination with fixed measurements}\label{sec:discrimination fixed measurements}

Another simple inductive strategy is to first get {\it some} classical information 
about $\rho_\pm$ using a fixed measurement strategy $\mathcal M = \{\Pi_k\}$,
-- e.g.,those compatible with the experimental platform -- 
and then use a purely classical learning approach.

Suppose for instance that we have performed $S$ measurements 
on $\rho_+$ and $S$ measurements on $\rho_-$ with outcomes $k_n$. We may 
group these outcomes and define a purely classical training set $\mathcal{S}_o = \{ (k_n, y_n) \}$ 
where $y_n=\pm1$ defines whether $k_n$ was obtained by performing the measurement 
on either $\rho_\pm$. Suppose that the information contained in $\mathcal{S}_o$ is not enough
to perform quantum state tomography to the desired precision, either because $\mathcal M$ is not 
tomographically complete or because the number of shots is not enough to 
reconstruct $\rho_\pm$ with the desired precision, or both. 
Since the probabilities $p_\pm(k)$ are unknown, we cannot explicitly optimize
the loss \eqref{qsd loss}. The only thing we can train is the empirical loss 
$L(f,\mathcal{S}_o)$  of Eq.~\eqref{emp risk}, with $N=2S$, 
which only depends on the set of outcomes, $\mathcal{S}_o$, and not explicitly on the choice of measurements used to obtain them. The latter is a purely classical optimization problem
and the only quantum part is in getting the experimental data contained in the original training set.
As such, errors can be studied via the optimality gap $\mathcal E(f^S_{\mathcal S})$
\eqref{eq:excessrisk_uniformdeviation},
see also Figure~\ref{fig:gaps}, which can be bounded by 
the (empirical) Rademacher complexity \eqref{rademacher}, with a purely classical decision function.

Different bounds on the Rademacher complexity for ``classical'' function classes $\mathcal F$,
such as neural networks or support vector machines, 
are known in the literature \cite{shalev2014understanding,bartlett2021deep}. 
For functions parameterized by real parameters $\bs w$, 
these bounds typically scale as 
\begin{equation}
	R_N \lesssim \ave_{\mathcal S}\left[\frac{ \|\bs w\|\|\bs k\|}{\sqrt S}\right],
	\label{bound svm}
\end{equation}
where $\bs k$ is the vector of measurement outcomes and the choice of the norm depends on the function class. However, from the physical point of view 
these bounds are not particularly informative, other than saying that POVMs with many outcomes 
perform more poorly as $\|\bs k\|$ gets bigger. 
When the set of outcomes is continuous, e.g.,when using homodyne or heterodyne measurements, this shows 
that models with spread outcomes are penalized. However, for discrete outcomes, such dependence 
cannot be justified. 

In order to get a better bound we focus on the 0-1 loss (probability of error), with dictionary-type 
decision functions that map a measurement outcome $k_n\in\{1,\dots,K\}$ to an output $y_n=\pm1$. 
In that setting,
using Theorem~\ref{thm: rada finite} from Appendix~\ref{a:math} we get the empirical Rademacher 
complexity as 
\begin{equation}
	R^{01}(\mathcal M,\mathcal{S}_o) \leq \sum_{k=1}^K \frac{\sqrt{N_k}}{2N} 
	\label{fixed measurement S}
\end{equation}
where $N=2S=\sum_k N_k$ and $N_k=N_k^++N_k^-$ is the number of
outcomes $k$ that we get from POVM $\mathcal M = \{\Pi_k\}$,
applied on either $\rho_\pm$, 
which follows a multinomial distribution.
Using Jensen's inequality, we then obtain (in Corollary~\ref{coro: entropy})  an $\mathcal M$-dependent bound on the true Rademacher 
complexity 
\begin{equation}
	R^{01}_N(\mathcal M)
	\leq
	\sum_{k=1}^K \frac{\sqrt{\ave_{\mathcal S}[N_k]}}{2N} =\frac14 
	\sum_{k=1}^K \sqrt{\frac{\Tr[\Pi_k\rho_+] + \Tr[\Pi_k\rho_-]}{S}} = \sqrt\frac{2^{H_{1/2}(\mathcal M)}}{8S},
	\label{fixed measurement}
\end{equation}
where we have introduced the 
R\'enyi entropy $H_\alpha(\mathcal M) = \log_2(\sum_k p_k^\alpha)/(1-\alpha)$ 
and $p_k=\Tr[\Pi_k\frac{\rho_++\rho_-}2]$. 
From the above inequalities, it appears that measurements with less entropy on the 
average state $\frac{\rho_++\rho_-}2$ may result in a lower optimality gap. 
Moreover, since $2^{H_{1/2}(\mathcal M)}\leq K$, POVMs with fewer outcomes are 
preferable, in line with a similar bound that we get from support 
vector machines and neural networks \eqref{bound svm}.

On the other hand, we still require our measurement to provide information about which state we have. More informative measurements may result in a larger value of $H_{1/2}(\mathcal M)$ (in fact, $H_{1/2}(\mathcal M)$ provides a trivial upper bound on the classical mutual information between the measurement outcomes and the class), so a measurement with a smaller value of $H_{1/2}(\mathcal M)$ may have a larger training error.


\subsection{Learning with decision observables}\label{sec:learning decision observables}  
We now focus on learning to classify states according to the sign of an expectation value of an observable $A$
\eqref{y predicted A}. When the quantum states are unknown, finding the optimal  observable $A$ for a given loss
is non-trivial because changing $A$ may also change the probability distribution of the measurement outcomes,
making the problem quite different from the classical supervised learning paradigm. 
Nonetheless, expectation values over quantum states 
can be measured in different ways, depending on the choice of the resolution $A=\sum_k a_k \Pi_k$, 
where $a_k$ are real numbers and $\Pi_k$ form a POVM. For any such resolution, the
measurement outcomes $a_k$ have associated probabilities $\Tr[\rho \Pi_k]$. 
In the particular case where the above resolution is provided by the spectral decomposition, 
when we optimize over $A$, we also optimize over 
$\Pi_k$, and the resulting probability distribution is not fixed. 
This setting is more similar to reinforcement, rather than supervised, learning.

There are different ways of simplifying this problem, for instance using a
fixed basis of operators constructed from Pauli measurements,
as discussed in Section~\ref{s:measurement problem formulation}. In this case the problem can be 
formulated using the classical learning framework on Section~\ref{sec:discrimination fixed measurements}.
Let $A = \sum_{j=1}^M \alpha_j A_j$ be a decomposition of $A$, 
where $\alpha_j$ are real trainable parameters, with $\bs\alpha\in\mathcal A$, and $A_j$ are fixed 
observables that can be measured efficiently. 
By performing $MS'$ different measurements, where $S'=S/M$ (so that the overall copy complexity is still $S$), we can build a classical training set $\mathcal{S}_o = (a_1^{(s)},\dots,a_M^{(s)},y^{(s)})$ 
where $s=1,\dots,S'$ and $a_j^{(s)}$ denotes the measurement outcome of observable $A_j$ at the $s$th shot,
while $y^{(s)}$ depends on whether we took the measurements on $\rho_{\pm}$. 
By construction 
\begin{equation}
	A(S,y) =	\frac{1}{S'} \sum_{s=1}^{S'} \delta_{y^{(s)},y}\sum_{j=1}^M \alpha_j a_j^{(s)},
	\label{eq:Asy}
\end{equation}
provides an unbiased estimator of the observable, so $A(S,y) \to \langle A\rangle_{\rho_y}$ for $S\to\infty$. 
In the machine learning framework, we can treat the outcomes $a_j^{(s)}$ as 
the inputs of classical learning model with linear function class $f(\bs a) = \sum_{j=1}^M \alpha_j a_j$,
and then optimize the empirical loss with a suitable regularization to obtain the 
empirical parameters $\alpha_j^{\rm emp}$. The latter allows us to construct the optimal empirical 
decision observable $A^{\rm emp} = \sum_{j} \alpha_j^{\rm emp} A_j$. Note that the machine 
learning model is trained to predict $y$ given a single measurement outcome from all observables, while 
in Eq.~\eqref{eq:Asy} the decision is based on the average over all the measurement shots. 
According to Eq.~\eqref{observable loss inequality}, the average loss in the latter case can only be smaller. 
In the machine learning framework the the error can be studied via the 
optimality gap $\mathcal E(f^S_{\mathcal S})$, 
see Figure~\ref{fig:gaps}, 
which can be bounded by the Rademacher complexity, thanks to 
\eqref{eq:excessrisk_uniformdeviation} and \eqref{rada simeq}.
Using Corollary \ref{coro: loss finite} and a $\lambda$-Lipschitz loss 
of the form in Eq.~(\ref{eq: convex loss function}) we get 
\begin{equation}
	\mathcal E(f_{\mathcal S}) 
	 \lesssim \mathcal O\left(\sup_{\bs\alpha\in\mathcal A}
		 \frac{\lambda{\sqrt M} a_{\rm max} \|\bs\alpha\|_p}{\sqrt S}\right),
	\label{excess risk observables}
\end{equation}
where $a_{\rm max}\geq |a_j^{(s)}|$ and the choice of $p$ depends on the regularization using for training. 
The error depends on 
how big can $\|\bs\alpha\|_p$ can be. We will consider an example in Section~\ref{sec:learning representer} 
where $\|\bs\alpha\|\approx (1-F)^{-1}$ and 
$F$ is the overlap between $\rho_+$ and $\rho_-$. So these bounds are useful to quantify 
how, for a given family of observables, specific properties of the quantum states can affect the 
number of measurements that are necessary to obtain the desired classification accuracy.

\begin{figure}[t]
	\centering
	\begin{tikzpicture}
		\node at (-9,2) { (a)};
		\node at (0,2) { (b)};
		\node at (-9,-0.25) {
				\begin{quantikz}[row sep=7.5mm]
					\rho~		 &  && \ \ldots\ & \swap[partial swap={~t~},partial position=0.81]{3} &  \ground{\rm trace~out} \\
					\wave[black!10]&&&&& \\
					\rho~ &  \swap[partial swap={~t~},partial position=0.5]{1} & \ground{} \\
					\sigma~		 & \targX{} && \ \ldots\ & \targX{} & & ~~\approx e^{imt \rho}\sigma e^{-imt \rho}
				\end{quantikz}
			};

		\draw[very thick,red] (-4.65,-2) -- +(0,4);

		\node at (0,0) {
				\begin{quantikz}[row sep=5mm]
					\ket0 ~ & \gate{H} & \ctrl{3} & \ \ldots\ & &  \gate[3]{{\rm QFT}^\dagger} & \meter[1]{b_1}\\
					\wave[black!10]&&&&\\
					\ket 0~		 & \gate{H} & & \ \ldots\ & \ctrl{1} & & \meter[1]{b_m}\\
					\ket\psi		 & & \gate{U} & \ \ldots\ & \gate{U^{2^{m-1}}} & & ~ \ket{\lambda_k}
				\end{quantikz}
			};
	\end{tikzpicture}
	\caption{(a) The state exponentiation algorithm uses a target state $\sigma$ and $m$ copies of $\rho$ 
		to act on $\sigma$ with an approximate unitary $U=e^{imt \rho}$, up to an error $\mathcal O(m t^2)$. 
		It only uses partial SWAP gates, as in Eq.~\eqref{state exponentiation}.
		(b) The phase estimation algorithm uses two registers, one initialized in $\ket 0^{\otimes m}$ and 
		the other initialized in $\ket\psi$. It applies Hadamard gates, powers of controlled-$U$ gates 
		For a generic unitary in some diagonal (possibly unknown) basis  
		$U= \sum_k e^{2\pi i \phi_k} \ket{\lambda_k}\!\!\bra{\lambda_k}$, this algorithm transforms a generic 
		input $\ket\psi = \sum_k \psi_k \ket{\lambda_k}$ into $\sum_{k} \psi_k \ket{b_1,\dots,b_m} 
		\ket{\lambda_k}$, where the measurement of the first register provides a bitstring approximation 
		of the phase as $\phi_k \approx b_1/2 + \dots + b_m /2^m$ and
		prepares the second register in the eigenvector $\ket{\lambda_k}$ -- see \cite{Nielsen_Chuang} for 
		an extended discussion on the precision as a function of $m$. 
	}
	\label{fig:helstrom circuits}
\end{figure}
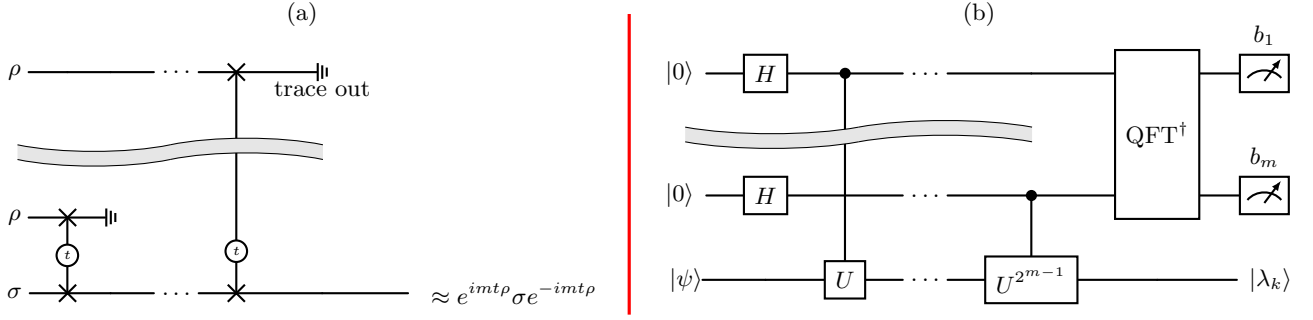

\subsection{Learning with Helstrom measurements}\label{sec:learning with helstrom} 
We now introduce the first transductive approach, which is based on the 
implementation of the Helstrom measurement \eqref{helstrom} when the states $\rho_\pm$ are unknown.
This approach was developed in Refs.~\cite{lloyd2020quantum,kimmel2017hamiltonian} using 
the state exponentiation (SE) algorithm \cite{lloyd2014quantum} followed by phase estimation (PE). The state exponentiation algorithm, shown in Figure~\ref{fig:helstrom circuits}(a), uses a target state $\sigma$ 
and many copies of another arbitrary state $\rho$, and 
acts on $\sigma$ with a unitary $U=e^{it \rho}$ 
that depends on $\rho$. It is based on the observation that
\begin{equation}
	e^{-i\rho t} \sigma e^{i\rho t} = \sigma -i[\rho,\sigma]t + ... = 
	\Tr_1 \left[e^{-it\, {\rm SWAP} } \left(\rho\otimes\sigma\right) e^{it\, {\rm SWAP} }\right]  + O(t^2),
	\label{state exponentiation}
\end{equation}
where SWAP is the swap operator and the $\Tr_i$ is the partial trace over the $i$th subsystem.  
This is the first operation shown in Figure~\ref{fig:helstrom circuits}(a).
This routine was generalized in \cite{kimmel2017hamiltonian} to use 
$S$ copies of $\rho_+$ and $S$ copies of 
$\rho_-$, with $S=\mathcal O(t^2/\delta)$ to simulate $e^{i H t}$, with $H=(\rho_+-\rho_-)/2$ 
up a precision $\delta$ in the trace norm. The number of 
operations scales as $\mathcal O(S \log(d))$, where $d$ is the dimension of the Hilbert spaces 
of $\rho_\pm$, so it is efficient as the dimension increases. For instance, in multi-qubit 
systems, it scales linearly with the number of qubits. 

This simulation method was employed in \cite{lloyd2020quantum} to obtain ${\rm sign}(H)$, and hence the Helstrom measurement \eqref{helstrom}. The starting point is that the exponentiated operator $e^{2\pi i H}$ is a unitary, $U$, with the same eigenvectors as $H$ and eigenvalues $e^{2\pi i \phi_j}$, where the $\phi_j$ are the eigenvalues of $H$ (which range between $-1/2$ and $1/2$). Since the $2\pi\phi_j$ are phases, we can equivalently say that the eigenvalues of $U$ are $e^{2\pi i \phi'_j}$, where $\phi'_j=\phi_j$ when $\phi_j>0$ and $\phi'_j=1+\phi_j$ when $\phi_j<0$ (assuming none of the $\phi_j=0$, with a simple extension if this is not the case).

Given a target state $\sigma$ (which is either $\rho_\pm$), we can carry out the phase estimation algorithm (shown in Figure~\ref{fig:helstrom circuits}(b)) with unitary $U=e^{2\pi i H}$ and initial state $\sigma$. If the algorithm succeeds, this results in the creation of a bitstring approximation $b_1\dots b_m$ of $\phi'_j$, up to a desired precision $m$, with probability $\Tr[\Pi_j \sigma]$. Since $\phi'_j>1/2$ iff $\phi_j$ (the corresponding eigenvalue of $H$) is negative, the value of the first bit $b_1$, tells us ${\rm sign}(\phi_j)$ (specifically, it is $0$ iff $\phi_j>0$). The probability of $b_1$ being $0$ is therefore $\sum_{j:\phi_j>0}\Tr[\Pi_j \sigma]$. Note that $\sum_{j:\phi_j>0}\Pi_j$ is precisely the Helstrom operator.

Calling $Z_1$ the Pauli operator on the first qubit, we can summarise 
\begin{equation}
	\sigma \otimes \rho_+^{\otimes S} \otimes \rho_-^{\otimes S} 
~~~~~~~
\xrightarrow{~~~\rm SE+PE~~~}
~~~~~~~
\langle Z_1\rangle  = \sum_j {\rm sign}(\phi_j) \Tr[\Pi_j \sigma] = 
 \Tr[{\rm sign}(\rho_+-\rho_-) \sigma],
\end{equation}
where, as already mentioned, $\sigma$ is either $\rho_\pm$.

We now present an analysis of the failure probability of this approximate Helstrom measurement in terms of the training copy complexity, $S$. The phase estimation algorithm has a chance of failing (i.e., incorrectly approximating $\phi'_j$). If we want the algorithm to correctly give the first $k$ digits of the bitstring approximation with probability at least $1-\epsilon$, we require $m = k +\log_2(2+1/(2\epsilon))$. Consequently, to carry out the Helstrom measurement with a probability of failure no more than $\epsilon$, we must set $m \geq 1 + \log_2(2+1/(2\epsilon))$.

To carry out the approximate Helstrom measurement, we must therefore simulate $U$, $U^2$,..., $U^{2^{m-1}}$. To understand the error scaling of the total operation, we can assume that each unitary has the same error, $\delta$, so that we need $O(2^0/\delta)$ copies of $\rho_+$ and $\rho_-$ to simulate the first unitary, $O(2^2/\delta)$ copies for the second unitary, etc., up to $O(2^{2(m-1)}/\delta)$ copies for the last unitary. With $O((\sum_{x=0}^{m-1}2^{2x})/\delta) = O(2^{2m}/\delta)$ copies of $\rho_+$ and $\rho_-$, we have an additional error (to be added to the failure probability of the phase estimation algorithm) of at most $\mathcal{O}(m\delta)$ due to the imperfect simulations of $U$. The phase estimation algorithm then has a failure probability of at most $\epsilon=\mathcal{O}(2^{-m})$. Therefore, the total failure probability is upper bounded by $\mathcal{O}(2^{-m})+ \mathcal{O}(m\delta)$, and so, setting $\delta = \mathcal{O}(2^{-m}/m)$, we can carry out the Helstrom measurement with a probability of failure (i.e., an additional probability of error above the minimum error probability afforded by the exact Helstrom measurement) of at most $\mathcal{O}(2^{-m})$ using $S=O(2^{3m} m)$ copies. Finally, we can say that we can enact the Helstrom measurement with a failure probability of at most $\epsilon$ using $S=\mathcal{O}(-\log(\epsilon)\epsilon^{-3})$ copies of $\rho_+$ and $\rho_-$.

Recalling that full-state tomography requires $\mathcal O(d^2/\epsilon)$ copies to achieve the same error, we see that the scaling is much worse in terms of the error. However, for the approximate Helstrom measurement scheme, the number of copies does not depend on the dimension of the states at all, so it can be more efficient for high-dimensional states. This efficiency comes at the price of having to manipulate all copies coherently. This strategy is also transductive, so for $V>1$, in order to achieve the same error as tomography, we require approximately $V$ times as many copies (i.e., we can split our training set into $V$ subsets and use one for each test state, then carry out a majority vote, or we could use all of the copies coherently to enact the $V$-copy Helstrom POVM, but now considering $\rho_\pm^{\otimes V}$ to be a single copy for the purposes of our circuit).

\subsection{Learning with the representer theorem}\label{sec:learning representer}
We now focus on a different transductive strategy based on the representer theorem \eqref{representer qsd},
which tells us that, for a given loss with a particular penalty, the optimal observable to discriminate 
two states can be written as a linear combination of the training states $\rho_\pm$. 
This strategy is hence transductive as, after training, we still need to use copies of $\rho_\pm$ 
to discriminate a new state $\sigma$.
Consequently, whilst we still refer to our overall copy complexity as $S$, we also define $S=S'+S''$, where $S'$ are the states used during the learning phase and $S''$ are the states used at test time.

Using the representer theorem, the discrimination is based on the sign of the following expectation value
\begin{equation}
	\langle A \rangle_\sigma = \alpha_+ \Tr[\sigma \rho_+] + \alpha_- \Tr[\sigma \rho_-].
	\label{eq:A representer}
\end{equation}
For known $\rho_\pm$, the optimal coefficients $\alpha_\pm$ were found explicitly and the resulting 
optimal observable was analytically constructed in \eqref{observable optimal}. Both the coefficients 
and the final expectation values can be obtained from the overlap $\Tr[\rho\sigma]$ between two states 
$\rho$ and $\sigma$ (possibly equal). 
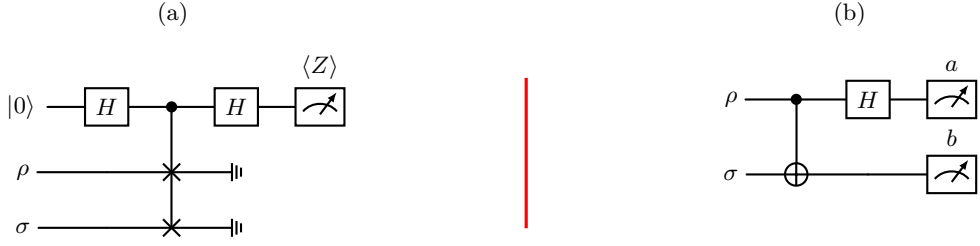
\begin{figure}[t]
	\centering
	\begin{tikzpicture}
		\node at (-9,1.5) { (a)};
		\node at (0,1.5) { (b)};
		\node at (-9,-0.25) {
				\begin{quantikz}
					\ket{0}~ & \gate{H} & \ctrl{1} & \gate H &\meter{\langle Z\rangle}  \\
					{\rho}~ & & \swap{1}& \ground{}\\
					{\sigma}~ & &\targX{}&\ground{}
				\end{quantikz}
			};
		\draw[very thick,red] (-4.3,-1.35) -- +(0,2); 
		\node at (0,0) {
				\begin{quantikz}
					\rho~ & \ctrl{1}  & \gate{H} &\meter{a} \\
					\sigma~ & \targ{} & &\meter{b}
				\end{quantikz}
			};
	\end{tikzpicture}
	\caption{Two different algorithms for computing the overlap $\Tr[\rho\sigma]$, 
		the swap test (a) and the swap measurement (b). The swap test (a) uses an ancillary 
		qubit, two Hadamard gates and a Fredkin gate (control-SWAP), followed by a measurement 
		on the ancilla. Measurement on the ancilla results in the desired result as 
		$\langle Z\rangle = \Tr[\rho\sigma]$. 
		The swap measurement (b) uses the fact that $\Tr[\rho\sigma] = \Tr[\rho\otimes\sigma {\rm SWAP}]$ 
		and that the single-qubit  swap operator can be diagonalized via a CNOT
		gate and Hadamard gate, with diagonal form
		$\sum_{a,b=0}^1 (-1)^{ab}\ket{ab}\!\!\bra{ab}$. Performing these operations in each pairs of qubits 
		from either $\rho$ and $\sigma$ we can then estimate the result as 
		$\Tr[\rho\sigma] = \mathbb[\prod_{i=1}^n (-1)^{a_i b_i}]$ where $n$ is the number of qubits in $\rho$ and 
		$\sigma$. 
	}
	\label{fig:swap}
\end{figure}
Two strategies to measure such overlap are presented in Figure~\ref{fig:swap}, using either 
the swap test\cite{barenco1997stabilization} or the swap measurement.
In the swap test Figure~\ref{fig:swap}(a),
prediction is done from the measurement of a Pauli-Z observable on an ancillary qubit, coupled 
to $\rho$ and $\sigma$, whose outcome at each 
measurement shot can be either $\pm1$. Therefore, we can describe this procedure as a POVM 
$M^\rho_\pm$, where $\pm$ denotes the measurement outcome. 
Something similar can be done using the swap measurement Figure~\ref{fig:swap}(b), by 
defining the resulting POVM $M^\rho_\pm$ depending on the sign of $(-1)^{\sum_i a_i b_i}$.
Since any measurement of the overlap at test time consumes both a test and a training state, we can assume that $S''=V/2$ and $S'=S-V/2$. For fixed $V$, the states consumed during testing therefore only constitute a constant correction to $S$, which does not affect the error scaling.

When the overlaps are estimated with $S'$ measurement shots
via either algorithm from Figure~\ref{fig:swap}, the parameters $P_\pm$ and $F$ 
that enter into Eq.~\eqref{observable optimal} have a variance 
\begin{equation}
	\Delta^2 = \frac{\langle Z^2\rangle - \langle Z\rangle^2}{S'} = \frac{1-\Tr[\rho\sigma]^2}{S'}.
	\label{precision swap}
\end{equation}
The approximate observable $A^{S'}$ is constructed by replacing the 
coefficients from Eq.~\eqref{observable optimal} with the ones estimated with $S'$ shots. 
As such, for any given $\lambda$-Lipschitz loss \eqref{observable loss inequality1} we can
bound the component of the training error that arises as a result of our imperfect knowledge of the overlap as 
\begin{align}
	E^{S'}_{\mathcal{S}} &\leq \lambda \sum_y |\langle A\rangle_{\rho_y} - \langle A^{S'}\rangle_{\rho_y}| 
	\leq \lambda\left|\frac{1}{1-F}-\frac{1}{1-F^{S'}}\right|\sum_y |\Tr[\rho_+\rho_y]-\Tr[\rho_-\rho_y]|
	\nonumber
	\\ &	\leq 2\lambda\left|1-\frac{1-F}{1-F^{S'}}\right| = 
	2\lambda \frac{|F-F^{S'}|}{1-F^{S'}} \approx2\lambda \sqrt{\left(\frac{1+F}{1-F}\right)\frac{1}{S'}},
		\label{excess risk representer}
\end{align}
where for simplicity we have restricted the analysis to pure states, for which $P_\pm=1$,
and in the last expression we assume $F-F^{S'}\approx \Delta$. As expected, more 
measurements are required to reduce the error when $F\approx 1$, namely when $\rho_\pm$ are 
less distinguishable.
We must note that there is a second component of the training error, which arises due to our imperfect estimation of the constructed observable at test time. This component is bounded by Eq.~(\ref{perr observable}).

More generally, given \eqref{eq:A representer} and the above expression,
\begin{align}
	E^{S'}_{\mathcal{S}} \leq \lambda \sum_y |\langle A\rangle_{\rho_y} - \langle A^{S'}\rangle_{\rho_y}| 
	\leq 2\lambda \|A- A^{S'}\|_\infty \leq  2\lambda\sum_{y=\pm} (\alpha_y-\alpha_y^{S'})
	\simeq 2\lambda \frac{f(\Delta P_+,\Delta P_-,\Delta F)}{\sqrt{S'}(P_+P_--F^2)^2},
	\label{excess risk representer general}
\end{align}
where for each estimated quantity $X$, we have assumed that $X-X^{S'}\approx \delta X/\sqrt{S'}$.
When the states are less distinguishable, $F^2\approx P_-P_+$ and a larger $S$ is required.

The analytic solution from Eq.~\eqref{observable optimal} is optimal for a particular loss and penalty. 
In general though analytic solutions are not available and the coefficients $\alpha_\pm$ must be 
optimized numerically given a few measurement results. In this case the optimization process fits 
into the framework discussed in Section~\ref{sec:learning decision observables}. Accordingly, 
the scaling of the knowledge gap is given by Eq.~\eqref{excess risk observables} with $a_{\rm max}=1$,
since the circuits 
in Figure~\ref{fig:swap} output estimators with outcomes $\pm1$. In the examples discussed before, 
we have shown that the optimal parameters satisfy $\|\bs \alpha\|\approx (1-F)^{-1}$. 
Although the behaviour as a function of $S$ shown in Eq.~\eqref{excess risk observables} and 
\eqref{excess risk representer} is the same, the bound in Eq.~\eqref{excess risk representer} 
may be smaller when $\rho_\pm$ are less distinguishable, i.e., when $1-F=\epsilon$. Indeed, 
in this case \eqref{excess risk representer} is small whenever $S\gg \epsilon^{-1}$, 
while \eqref{excess risk observables} is small whenever $S\gg \epsilon^{-2}$. 
Therefore, exploiting the analytic solution \eqref{observable optimal} is expected to provide an advantage 
to get the same accuracy with fewer copies of the two states.

\subsection{Inductive vs. Transductive Strategies}\label{sec:inductive vs transductive}

A key difference between inductive and transductive learning processes can be seen in how the error of a process of each type scales with $V$. Suppose we pick a scheme, calculate the error for $V=1$, and then increase $V$ to see how the error changes for that scheme.

For an inductive learning process, there is an obvious way of extending the process to multiple copies of the test state. Since we have an intermediate result that tells us which measurement to enact on the test state, we can carry out the same initial measurement on the training set as for the $V=1$ case and then repeat the second part of the measurement on as many copies of the test states as we like, with the same probability of error. We are therefore at least able to use the scaling of the majority vote, which is exponential in the number of copies ($\mathcal{O}(e^{-V})$). In other words, we have learned a rule that we can then apply to any number of unknown states.

On the other hand, for a transductive process, the measurement is not divided into two separate parts, so the extension to $V>1$ is not necessarily obvious. We do not learn any ``reusable" information, and since the training set is used up when measuring it, it must be shared amongst the $V$ test states in some way. One option would be to divide our training set into subsets of $S/V$ elements, use each subset to carry out the process for $V=1$ separately on each test state, and carry out the majority vote. However, since we are decreasing the number of training states per test state, and 
the number of training states determines the excess error, the (single-copy) probability of error for each test state will be worse than in the $V=1$ case (if $S$ is fixed). As long as the single-copy error probability remains better than that of a random guess, the total error probability for the majority vote will continue to decrease as $V$ increases, but for large enough $V$ (and fixed $S$), this will eventually no longer be the case. The scaling with $V$ will therefore be worse than for an inductive process -- how much worse depends on how the single-copy error depends on $S$.

As an example, we compare the performance of two approaches, one inductive and one transductive, based on the Helstrom classifier, with the goal of classifying $V$ copies of a test state using $S$ copies of the training states. 
In the inductive approach, we can use state tomography to achieve a single-copy error of $\mathcal O(d/\sqrt{S})$, per Eq.~(\ref{eq:tomography optimality}), which is independent of $V$. In the transductive framework, 
the copies have to increase as $\mathcal O(V)$ to use a majority rule.
%
We can also consider a case in which we must find the identity of $M$ different test states, i.e., our test set comprises of multiple unknown states and we must find the identity of all of them. 
In this scenario, the number of training samples, $S$, required in order to achieve a given single-copy error, $\epsilon$ in the two cases is
\begin{equation}
	{\rm inductive{:}}~~\mathcal O\left(\frac{d^2}{\epsilon^2}\right),  ~~~~~~~ {\rm vs.} ~~~~~~~ 
	{\rm transductive{:}}~~\mathcal O\left(VM\epsilon^{-3}(-\log\epsilon)\right).
\end{equation}
Although the scaling with $V$ is worse in the transductive case than in the inductive case (linear vs $\mathcal{O}(1)$), the scaling with dimension can be much more important. Hence, the most suitable approach depends on the dimension of the Hilbert space $d$ and on the number of test states. In multi-qubit systems, where $d=2^n$ and $n$ is the number of qubits, if $V$ is at most polynomial
in $n$ and $\epsilon$ is independent on $n$, the transductive strategy may provide an exponential (in $n$) advantage compared with tomography based methods. On the other hand, in the ``big data'' regime, where $V$ is large, the inductive strategy may be preferable.


The difference between inductive and transductive measurements has also been formulated in terms of a non-signalling condition between different test instances, in Ref.~\cite{monras_inductive_2017}, where the definition of a separation between training and test states is also used.

As a final note, for methods based on the expectation value of an observable, it makes more sense to consider how the training error scales as a whole, rather than the single-copy error probability, since we do not base our classification decision on the majority vote over single copies. For mixed strategies that consume copies of the training states both to learn parameters and then again at test time, such as in Section~\ref{sec:learning representer}, the training error has two additive components. If we define $S=S'+V/2$, one component only decreases with $S'$ (for the strategy based on the representer theorem, this component is given by Eq.~(\ref{excess risk representer general})) and one only decreases with $V$ (for the strategy based on the representer theorem, this component is bounded by Eq.~(\ref{perr observable}).

\section{Supervised Learning for Quantum Classification} \label{sec:supervised learning}

In the previous two sections, we have first studied the case in which the data-generation mechanism $\mathcal{P}$ is known (Section~\ref{sec:qsd}), and we have then addressed the new challenges arising from lack of knowledge about the possible states the test input may be in (Section~\ref{sec:learning discriminate}). With reference to the general framework introduced in Section~\ref{sec:summary}, in the settings considered so far, there is no classical input $x$, and hence the joint distribution $P(x,y)$ of classical input $x$ and classical output $y$ plays no role. As a result, the performance of the learner depends solely on the number of copies $S$ and $V$ available during training and testing, respectively (with training being irrelevant for the case of known $\mathcal{P}$). In this section, we investigate a more complex scenario in which data consists of classical input $x$, quantum embedding $\rho(x)$, and classical output $y$, with the latter taken to be binary as in the previous sections.   In this situation, the number $N$ of training data points determines the information that the learner can extract about the joint distribution $P(x,y)$. Therefore, the optimality gap is a function not only of the number of copies $S$ and $V$, but also of the size of the training set, $N$.

The training set $\mathcal S^S =\{ (x_n, y_n, \rho(x_n)^{\otimes S})\}_{n=1,\dots,N}$ consists 
of triples containing the classical inputs $x_n$, $S$ copies of the corresponding quantum state 
$\rho(x_n)$, and the true class $y_n=\pm1$. We provide some discussion on how one can generate $S$ copies of state $\rho(x)$ in practice at the end of this subsection. For both training and test data, the classical input-output pairs are generated via a generally \emph{unknown} joint distribution $P(x,y)$.

In this setting, the inference operation $f$ introduced in Section~\ref{sec:summary} is a quantum measurement applied to the test input $\rho(x)$ to extract the label $y$. The optimal inference operation $f_*$ is obtained by assuming knowledge of the joint distribution $P(x,y)$ and of embedding mapping $\rho(x)$.  The learner optimizes operation $f$ on the basis of training data $\mathcal{S}^S$. We will first consider the case in which the abstract training set $\mathcal{S}$ is available to the learner, and hence the states $\rho(x)$ in the training set are known. In this case, the learner can compute and optimize the training loss.   Then, we will tackle the case in which the states $\rho(x)$ are unknown, and hence the inference function must be optimized using the available 
$S$-copy training set $\mathcal S^{S}$.

As mentioned, in the case of unknown quantum states $\rho(x)$, we assume that it is possible to create $S$ copies of the states $\rho(x)$, either by 
repeating the embedding procedure or by repeating the experiment multiple times. 
In real experiments, where $\rho(x)$ are typically mixed states, one may distinguish the following situations:
\begin{enumerate}
	\item The  states $\rho(x)$  account for a statistical description of the 
		experimental uncertainties. Physically, this means that the timescales on which the environment 
		acts on the system are faster than the measurement times. In other terms, equilibration happens before
		the measurements are carried out, so  each single measurement effectively ``sees'' the same
		mixed state $\rho(x)$. In this regime, by repeating the experiment $S$ times, we effectively 
		act on the product state $\rho(x)^{\otimes S}$. 
	\item In contrast, with a ``slow'' environment, the states $\rho(x)$ will always be different,  
		and it is impossible to create perfect copies. Without loss of generality, we may describe the 
		uncertainty about the states before measurement by introducing a random variable $e$ that models all the imperfections that the environment could apply to the states. This yields the mixed state as the statistical mixture 
		\begin{equation}
			\rho(x) = \int de \, p_{\rm env}(e|x) \rho(x,e),
			\label{eq:environment dep}
		\end{equation}
		where $\rho(x,e)$ model the joint distribution of the classical input $x$
		and of the environment disturbance $e$. When trying to create copies $S$, 
		the environment effectively creates $\rho(x,e_1)\otimes\dots,\rho(x,e_S)$, where
	  $e_s\sim p_{\rm env}(e|x)$ are possibly unknown. Nonetheless, since in the general 
		framework of Section~\ref{sec:summary} the inputs $x$ are possibly unknown to the learner, 
		we can theoretically describe this limit by introducing a new training set 
		$\mathcal S^S_{\rm env} = \{(x_n,e_s,y_n,\rho(x_n,e_s)\}_{n=1,\dots,N; s=1,\dots,S}$,
		with $NS$ different states $\rho(x_n,e_s)$ and a larger input variable, modelled by 
	  $x$ and $e$.
\end{enumerate}

\subsection{Known Quantum State and Unconstrained Complexity} \label{sec:unconstrainedPOVMs}

Let us first study the case in which the states $\rho(x)$ are known, and hence the learner has access to the abstract training set $\mathcal{S}$. Therefore, assuming an equal probability for the two classes, the only element that is unknown about the data generation process $\mathcal{P}$ is given by the probability distributions $p_\pm(x)$ of the classical input.  Furthermore, we focus here on unconstrained-complexity operations.

Adopting the 01 loss, the average loss can be written as 
\begin{equation}
	\ell_{01}(\mathcal M,x,y) = \sum_{\bar y} \Tr[ \Pi_{\bar y} \rho(x)]= 
	1-\Tr[ \Pi_{y} \rho(x)].
	\label{single shot linear loss}
\end{equation}
Given a state $\rho(x)$ with true class $y$, 
the above loss quantifies the probability that the outcome of $\mathcal M$ is different from $y$. 
Due to the linearity $\ell_{01}$, the average loss~\eqref{risk} depends on the average 
states as 
\begin{align}
	L_{01}(\mathcal M) &= 1-\frac{1}{2}\sum_{y=\pm}\Tr[\Pi_y\bar\rho_y],
				 &
	\bar\rho_y &= \int dx\, p_y(x)\, \rho(x),
	\label{average loss general} 
\end{align}
where we used the rule of conditional probabilities $P(x,y) = p_y(x)/2$. Accordingly, 
the optimal classifier $f_*$, see Figure~\ref{fig:gaps},
is given by 
the Helstrom measurement \eqref{helstrom} between the two average states $\bar\rho_\pm$. However, 
this optimal inference operation is not accessible by the learner since the probabilities  $p_\pm(x)$ are unknown.

 Given that the states $\rho(x)$ are assumed to be known, the training loss can be evaluated by the learner as \eqref{emp risk} 
\begin{align}
	L_{01}(\mathcal M,\mathcal S) &= 1-\sum_{y=\pm}\frac{N_y}{N}\Tr[\Pi_y\bar\rho^{\mathcal S}_y],
				 &
	\bar\rho^{\mathcal S}_y &= \frac{1}{N_y} \sum_{n=1}^N \delta_{y,y_n}\rho(x_n),
	\label{training loss 01}
\end{align}
where $N_\pm$ is the number of samples with $y_n=\pm$ and 	$\bar\rho^{\mathcal S}_y $ 
is the ensemble average over all states from the training set $\mathcal S$ with class $y$.
For simplicity, from now on we will assume an equal 
number of samples per class, namely $N_\pm/N=1/2$. The optimal data-driven classifier $f_{\mathcal S}$,
one minimizing the dataset loss \eqref{emp risk}, is now given by 
the Helstrom measurement $\mathcal M^{\rm emp} = \{\Pi_+^{\rm emp},\Pi_-^{\rm emp}\}$ over the empirical averages 
\begin{align}
	\Pi_\pm^{\rm emp} &= \frac{\openone\pm {\rm sign}\left(\bar\rho^{\mathcal S}_+ -\bar\rho^{\mathcal S}_-\right)}2, 
										&
	L_{01}(\mathcal M^{\rm emp},\mathcal S) &= \frac{1}{2}-\frac14\big\|\bar\rho^{\mathcal S}_+ -\bar\rho^{\mathcal S}_-\big\|_1,
    \label{empirical helstrom}
\end{align}
where the second expression provides the analytical single-shot training loss. 
As long as the two average states are distinguishable, namely 
$\bar\rho^{\mathcal S}_+ \neq \bar\rho^{\mathcal S}_-$, arbitrarily small training loss can be obtained 
by using multiple shots, as we will discuss in a following section. 

We now study generalization via the Rademacher complexity \eqref{rademacher}. As we derive 
in Appendix~\ref{app:unconstrained povm}, both the empirical and Rademacher complexities scale as  in 
Eq.~\eqref{rada simeq}, namely as $R\leq \sqrt{B/N}$ with 
\begin{align}
	B(\mathcal M) &\leq \left(\Tr\sqrt{\int dx\, p(x)\, \rho(x)^2}\right)^2,
								&
	B(\mathcal M,\mathcal S) &\leq \left(\Tr\sqrt{\frac1N\sum_{n=1}^N \rho(x_n)^2}\right)^2,
	\label{B 01 loss}
\end{align}
where the second expression is the empirical approximation of the Rademacher complexity,
which can be explicitly computed for given data. On the other hand, $B(\mathcal M) $ is purely 
formal, since the distribution $p(x)$ is unknown. Nonetheless, asymptotically the 
two quantities differ at most by $\mathcal O(1)$ factors.

\subsubsection{Lowering the Training Error via the Majority Rule}\label{sec:rademacher majority}

We now extend the result of the previous section by employing $V$ copies 
of the test state. We repeat a local POVM classifier on each copy, with 
binary outcomes $\mathcal M = \{\Pi_+,\Pi_-\}$, and then use a majority rule. 
We set $V=2v+1$ as an odd integer to avoid ending up in a draw.
The training error is given by the average probability of error 
\begin{equation}
	L^{V}_{01}(\mathcal M,\mathcal S)  
	= 1-\frac1N \sum_{n=1}^Np^V(x_n,y_n) = \frac1N \sum_{n=1}^Nc_v\left(\Tr[\rho(x_n)\Pi_{y_n}]\right),
	\label{loss majority vote}
\end{equation}
where $p^V(x, y) = 1-c_v\left(\Tr[\rho(x)\Pi_y]\right)$ is the probability that
the majority vote over the state $\rho(x)$ 
returns the correct class $y$, as in Eq.~\eqref{eq:probcorrectcond}, 
$\Tr[\rho(x)\Pi_y]$ is the probability that a single vote provides the correct answer $y$, 
and $c_v(p) = C^{2v+1}_v(p) = \sum_{n=0}^v \binom{2v+1}{n} p^n(1-p)^{2v+1-n}$ is the probability 
that a majority vote provides an incorrect answer, given a probability of success $p$ for 
a single vote. 
As shown in Section~\ref{sec:majority rule}, as long as $p=1/2 + \mathcal
O(V^{-1})$, $c_v(p)$ decreases exponentially with 
$V$, so that the training error goes to zero for $V\to\infty$.

In order to study the generalization error, 
in Appendix~\ref{app:majority rule}
we show that the Rademacher complexity behaves as in Eq.~\eqref{rada simeq} with 
\begin{align}
	B^V(\mathcal M) &\leq \sqrt{V+1}\left(\Tr\sqrt{\int dx\, p(x)\, \rho(x)^2}\right)^2,
								&
	B^V (\mathcal M,\mathcal S) &\leq \sqrt{V+1}\left(\Tr\sqrt{\frac1N\sum_{n=1}^N \rho(x_n)^2}\right)^2,
	\label{B 01 loss V}
\end{align}
thus providing a proof of the conjecture from \cite{banchi2021generalization}, namely that 
the generalization error only slightly increases with the number of copies used 
in the majority vote. Since the training error decreases exponentially with $V$,
using more shots is expected to be  beneficial,
as long as we have enough data to make \eqref{B 01 loss V} small.

\subsection{Unknown Quantum States and Unconstrained Complexity}\label{section: scaling with N and S}

In the previous section we consider the error with the optimal unconstrained POVM, which for the 0-1 loss is given by the Helstrom classifier. We now consider the error in implementing such a measurement when 
the states $\bar{\rho}_\pm^{\mathcal S}$ in \eqref{empirical helstrom} are unknown, and only a finite number of copies of the states $\rho(x)$ in the training set are available. We specify the problem as follows: instead of having classical knowledge of the $N$ different states in the abstract training set, $\mathcal{S}=\{\rho(x_i)\}$ for $i$ ranging from $1$ to $N$, we instead have $S_i$ copies of the $i$th state. The training set can therefore be expressed as $\bigotimes_{i+1}^N \rho(x_i)^{\otimes S_i}$. We constrain the number of copies by demanding that $\sum S_i = NS$, namely that $S$ is the average of $S_i$. In other words, we can only draw $NS$ states in total from $N$ different options.

This setting makes sense when it is computationally expensive to make copies of the states in the training set but less so than adding new states to the training set. I.e., there are different costs associated with drawing a new state from the true distribution and making copies of that state, so that it makes sense to distinguish between $N$ and $S$. This is the case for many of the examples in Section.~\ref{sec: examples supervised unsupervised}. For instance if we are classifying phases of matter, there may be a cost associated with producing copies of a state that we know to be in one phase or the other, but the cost of finding the settings to produce a new state in a known phase might be higher. For quantum classification of quantum sensed data, there will be some cost associated with probing the same sample multiple times, but a different cost associated with producing a new sample, so it again makes sense to make a distinction between $N$ and $S$.

Note that we could have specified the problem in a slightly different way. Instead of simply requiring that a total of $NS$ states are drawn from $\mathcal{S}$, we could have required that we have precisely $S$ copies of each of the $N$ different states in $\mathcal{S}$. However, in many scenarios of interest, there is no reason why we would have this extra constraint and we can simplify the mathematical analysis by only constraining the total number of states in the training set.

Let us consider the two possible extremes that $S$ could take. If there is no cost associated with $S$, so that we can take $S$ to infinity, we once again have classical knowledge of the states, so that the only source of error is the generalization error (with no excess testing error or knowledge gap). In the language of Figure~\ref{fig:gaps}, we will always find the minimum of the dataset loss (the red curve) and the only source of error is the difference between the dataset loss and the average loss (the blue curve). As $N$ becomes larger, the two curves get closer together, with the difference between them scaling with $N^{-1/2}$, per Eq.~(\ref{bound train test}). This is essentially a classical learning problem implemented using quantum states.

On the other hand, if it is as costly to make the same sample multiple times as it is to produce a new sample, we may have $S=1$, so that we have a single copy of each training state. In this particular, special case, we have $N$ samples drawn from the true average states for each class, $\bar{\rho}_{\pm}$. We can then treat the problem as one of state discrimination between the unknown states $\bar{\rho}_{+}$ and $\bar{\rho}_{-}$, as per Section~\ref{sec:learning discriminate}. We have no generalization error, since the average states $\bar{\rho}_{+}$ and $\bar{\rho}_{-}$ are precisely what we want to discriminate between, and so the only source of error is the knowledge gap. In the language of Figure~\ref{fig:gaps}, the dataset loss is the same as the average loss, but we are not at their minimum value. In this case, we can directly replace $S$ with $N$ in our previous (Section~\ref{sec:learning discriminate}) expressions for the knowledge gap, which are different for each protocol. As $N$ increases, the two curves remain the same (and identical to each other) but we get closer to the minimum.

Now let us return to the intermediate setting, where $S$ is finite, but $S\gg 1$. We will have a knowledge gap, coming from our finite knowledge of the states in the training set, and a generalization error, coming from the difference between the training set and the true average states.

We define the difference between the empirical average states as
\begin{equation}
    H^{\mathcal S} = \bar \rho_+^{\mathcal S}-\bar \rho_-^{\mathcal S} = \frac{2}{N} \sum_n {y_n} \rho(x_n),
    \label{H ope}
\end{equation}
and the difference between the true average states as
\begin{equation}
    H = \bar \rho_+-\bar \rho_- = \int dx dy \, P(x,y)\, y \rho(x).
    \label{H abstract}
\end{equation}
Then, the knowledge gap quantifies how different the final measurement we perform on the test state is from the empirical Helstrom measurement, $\mathcal M^{\rm emp} = \{\Pi_+^{\rm emp},\Pi_-^{\rm emp}\}$, where $\Pi_{\pm}^{\rm emp}=(\openone\pm {\rm sign}(H^{\mathcal S}))/2$, as per Eq.~(\ref{empirical helstrom}). The optimality gap $\mathcal{E}(f_{\mathcal{S}})$, and the generalization error quantify how different $\mathcal M^{\rm emp}$ is from the true Helstrom measurement, $\mathcal M = \{\Pi_+,\Pi_-\}$, where $\Pi_{\pm}=(\openone\pm {\rm sign}(H))/2$, as per Eq.~(\ref{helstrom}).

To understand how the knowledge gap and generalization errors scale with $N$ and $S$, we must clarify exactly how we generalize the learning strategies from Section~\ref{sec:learning discriminate} to this new setting where we have multiple different states in each class. We draw $NS$ random samples from the set $\mathcal{S}$ and apply the protocols from Section~\ref{sec:learning discriminate} to learn to discriminate between the empirical average states $\rho_+^{\mathcal S}$ and $\rho_-^{\mathcal S}$. The knowledge gap can therefore be found using the expressions from Section~\ref{sec:learning discriminate}, but replacing $S$ with $NS$ (since we have $NS$ samples), and the generalisation error comes from the difference between the empirical average states and the true average states, and follows Eq.~(\ref{bound train test}).

To show how this works, and the interplay between the two types of errors, let us consider the tomography-based (inductive) approach, from Section~\ref{sec:tomography classification}, and the (transductive) approach based on applying the approximate Helstrom measurement via state exponentiation, from Section~\ref{sec:learning with helstrom}.

Per Section~\ref{sec:tomography classification}, the knowledge gap, $\epsilon$, for tomography-based quantum state classification scales with the total number of training states, $NS$, as $\epsilon = \mathcal{O}(d/\sqrt{NS})$. The generalization error, $\delta$, scales with the number of different training states, $N$, as $\delta = \mathcal{O}(\sqrt{B/N})$, with $B\leq d$. Thus, increasing $S$ decreases only the knowledge gap, whilst increasing $N$ decreases both the knowledge gap and the generalization error. For large $d$, the knowledge gap will dominate. 
However, assuming the states from the training set have rank at most $r$ and the rank of the average 
states is at most $Nr\ll d$, we can write $\epsilon \leq \mathcal O(\sqrt{d Nr}/\sqrt{NS}) = \mathcal O(\sqrt{dr/S})$, which is independent of $N$. In the worst case $B\approx d$, assuming $r=\mathcal O(1)$ 
we need $S$ as large as $N$ to make 
$\epsilon$ and $\delta$ of the same order.

For the approximate Helstrom measurement, following Kimmel et al.\cite{kimmel2017hamiltonian} we can apply the state exponentiation algorithm with samples from either 
$\bar\rho_y^{\mathcal S}$ or $\bar \rho_y$ to approximate either $e^{it H^{\mathcal S}}$ or $e^{it H}$ to the desired precision, and hence $\mathcal M^{\rm emp}$ or $\mathcal M$.
The latter is the $S=1$ case, whilst the former is the extension to $S\gg 1$ (but still finite).
In the former case, the total number of training states required to achieve a knowledge gap of $\epsilon$ is $NS=\mathcal{O}(-\log(\epsilon)\epsilon^{-3})$ (per Section~\ref{sec:learning with helstrom}), whilst again the generalization error scales with $N$ as $\delta = \mathcal{O}(\sqrt{B/N})$. Again, increasing $S$ decreases the knowledge gap and increasing $N$ decreases both the knowledge gap and the generalization error. If $S$ is kept constant and $N$ is increased, the generalisation error will decrease much more quickly than the knowledge gap. If we want the two errors to be of the same order in $N$, we require $\epsilon = \mathcal{O}(\delta) = \mathcal{O}(\sqrt{B/N})$. To do this, we must set $S=\mathcal{O}(S_{N,B})$, with $S_{N,B}=\log(N/B)\sqrt{N/B^3}$. Thus, if we do not want either the knowledge gap or the generalization error to dominate (for large enough $N$), we must also increase $S$.
	As an example, suppose that the model complexity $B$ is known, and that the learner has chosen 
	a dataset with $N=\mathcal O(B\epsilon^{-2})$ elements to reach a maximum generalization error $\epsilon$. 
	Then, a knowledge gap of order $\epsilon$ requires $S=\mathcal O(B\epsilon^{-1}(-\log\epsilon))$.

If $S$ is constant or increases slower than $\mathcal{O}(S_{N,B})$, the knowledge gap will become dominant compared to the generalization error. If $S$ increases more quickly than $\mathcal{O}(S_{N,B})$, the generalization error will dominate.
We can understand this behaviour intuitively. We know that for $S\to\infty$ we only have a generalization error and for $S=1$ we only have a knowledge gap. Thus, if $S$ is large (compared to $N$), the generalization error will dominate, since we are close to the $S\to\infty$ case, whilst for small $S$, we are close to the $S=1$ case, so the knowledge gap dominates.

\subsection{Information theoretic understanding of training/testing errors}\label{sec:informationtheoretic}

The errors introduced in Section~\ref{sec:unconstrainedPOVMs} can be understood using information 
theoretic quantities, as summarized in Section~\ref{sec:summary}. Here we present an extended 
derivation. 

We start by defining an abstract space 
which contains all the possible knowledge of the problem. In our data we have inputs $x$, 
outputs $y$ and quantum states $\rho(x)$, we may model all of this together by 
defining the abstract and empirical average states
\begin{align}
	\rho_{XYQ} &= \sum_y \int dx P(x,y) \ket{xy}\!\!\bra{xy}\otimes \rho(x),
						 &
	\rho^{\mathcal S}_{XYQ} &= \frac1N \sum_{n=1}^N \ket{x_ny_n}\!\!\bra{x_ny_n}\otimes \rho(x_n),
	\label{rho XYQ}
\end{align}
where $X$ is the space of inputs (possibly continuous), $Y$ is the space of outputs,
and $Q$ is the space of the  quantum states $\rho(x)$. 
In $\rho^{\mathcal S}_{XYQ}$ the average over the unknown $P(x,y)$ is replaced with the empirical 
	average over the $N$ data in $\mathcal S$. 
Technically both $\rho_{XYQ}$ and $	\rho^{\mathcal S }_{XYQ} $
describe a classical-classical-quantum state, as both $X$ and $Y$ are classical. 
We remark that states in the extended XYQ space should never be computed in applications, 
this is just a mathematical framework to rephrase some quantities in a information-theoretic language.

It was shown \cite{banchi2021generalization} that both the training 
error $L_{01}(\mathcal M^{\rm emp},\mathcal S)$ and the testing error  
$L_{01}(\mathcal M^{\rm emp})$ can be expressed via information theoretic quantities, 
computed with respect to the states in Eq.~\eqref{rho XYQ}. 
In particular, as already discussed in Section~\ref{sec:summary} 
the average loss with the optimal Helstrom classifier $\mathcal M^{\rm HH}$ 
can be written as 
\begin{align}
	L_{01}(\mathcal M^{HH}) &= 1-2^{-H_{\rm min}(Y|Q)} \leq 1-2^{-H(Y|Q)},
													&
	L_{01}(\mathcal M^{\rm emp},\mathcal S) &  \leq 1-2^{-H^{\rm emp}(Y|Q)},
	\label{HQY}
\end{align}
where $H(Y|Q)$ is the quantum conditional entropy and $H_{\rm min}$ is a similar quantity 
with a different notion of entropy, see \cite{banchi2021generalization,konig2009operational} for details,
while $H^{\rm emp}(Y|Q)$ is the quantum conditional entropy, but computed over the empirical
average state $\rho^{\mathcal S }_{XYQ}$. 
The average loss is zero when the conditional entropy is zero. This happens when $Y$ 
is completely determined by $Q$, namely when for a given quantum 
state there is a direct mapping to find its class. 
When, given a test state, the information about its class is imperfect, 
the conditional entropy is greater than zero, and so is the loss. 

As for the generalization error, as shown in Section~\ref{sec:summary}, we can express 
the bounds \eqref{B 01 loss} using information theoretic quantities to get the bounds 
\begin{equation}
	\mathcal G_{01} = 	L_{01}(\mathcal M^{\rm emp}) -	L_{01}(\mathcal M^{\rm emp},\mathcal S) 
	\leq \mathcal O\left(\sqrt{\frac{B(\mathcal M)}{N}}\right) = 
	\mathcal O\left(\sqrt{\frac{2^{I_{1/2}(X{:}Q)}}N}\right) = 
	\mathcal O\left(\sqrt{\frac{2^{I^{\rm emp}_{1/2}(X{:}Q)}}N}\right),
	\label{generalization info}
\end{equation}
where $I_{1/2}(X{:}Q)$ is the R\'enyi quantum mutual information between $X$ and $Q$, 
see also Eq.~\eqref{eq:mutual info},
which quantifies the amount of information that the knowledge of $X$ provides to the 
knowledge of $Q$, and vice versa. On the other hand, 
$I^{\rm emp}_{1/2}$ is the same quantity, but computed over $\mathcal \rho_{XYQ}^{\mathcal S}$, as 
in \eqref{B 01 loss}. 
In the last approximate equality in \eqref{generalization info}
we replace $\rho_{XYQ}$ with $\rho_{XYQ}^{\mathcal S}$, 
since the difference between the Rademacher and empirical Rademacher complexities go
to zero for large $N$ -- see Eq.~\eqref{rada vs emp rada}.

Since $\rho_{XYQ}^{\mathcal S}$ are classical-classical-quantum states,
the R\'enyi quantum mutual information satisfies some simple properties, 
$I^{\rm emp}_{1/2}(X{:}Q)\leq \min\{H_{1/2}(Q),\log_2 N\}$,
	where the first term is due to the fact that 
the space $X$ describes classical information represented as $N$ orthogonal vectors. 
When the mutual information becomes comparable with $\log_2 N$
the bound \eqref{generalization info} becomes trivial. 
Only when the R\'enyi entropy of the $Q$ subsystem is much smaller than $\log_2N$, 
we can expect a small generalization error, even in the worst case.
In other words, good generalization is possible when the quantum embedding $x\mapsto\rho(x)$ 
effectively discards ``irrelevant'' information from the input $X$ that is 
		unnecessary to predict the output $Y$. However, if too much information is discarded, then 
		$H(Y|Q)$ gets larger, and so does the training error. 

The above information-theoretic interpretation of generalization error can be extended to a multi-task, transfer learning framework as in \cite{jose2023transfer} or to adversarial learning \cite{georgiou2024adversarial}.

\subsubsection{Learning with imperfect copies} 

We finally comment on what happens when copies of an unknown quantum state are used 
for learning. Thanks to the analysis of Section~\ref{sec:unknown states many copies}
the $S$ copies of each state (on average) used during training, do not 
enter into the generalization error. On the other hand, applying a majority 
rule over more copies of the same test state only slightly increases the 
generalization error, as shown in \eqref{B 01 loss V}. 

In this section we discuss another scenario, introduced in Section~\ref{sec:supervised learning},
namely when it is impossible to create 
perfect copies of the training states. As shown in that section, we have to replace 
$\rho(x_n)^{\otimes S}$ with $\rho(x_n,e_1)\otimes\dots\otimes \rho(x_n,e_S)$, where 
the auxiliary classical inputs $e_s$ describe the unknown action of an environment. 
If these copies are processed at different times, we may model the learning process 
as  a single-shot processing the $NS$ states $\rho(x_n,e_s)$. 

At test stage, thanks to \eqref{eq:environment dep} and the linearity of the 01 loss, 
we may simply focus on $V$ copies of the average state $\rho(x)^{\otimes V}$ and use 
the results of the majority vote. 
Calling $E$ the space of possible unknown actions performed by the environment, 
we get a generalization error
\begin{equation}
	\mathcal G^{E}_{01} \leq  
	O\left(\sqrt{\frac{V+1}{S N} 2^{I^{\rm emp}_{1/2}(X,E{:}Q_1)}}\right).
	\label{generalization info }
\end{equation}
Therefore, if $V = \mathcal O(S)$ the dependence on either $S$ and $V$ disappears from 
the generalization error, making it independent on the number of copies employed either 
during training or testing. However, the price to pay is modeled by a different mutual information 
$I^{\rm emp}_{1/2}(X,E{:}Q_1)$, which takes into account both the effect of the classical inputs and of 
the environment. If the environment can significantly alter the states, 
than this mutual information can be higher, and hence the number of samples 
$N$ to get a small generalization error will increase. On the other hand, when the perturbations 
inflicted by the environment are negligible, it is reasonable to expect that 
$I^{\rm emp}_{1/2}(X,E{:}Q_1) \simeq I^{\rm emp}_{1/2}(X{:}Q_1) $. In this regime, 
the learner can basically ignore the presence of an environment.

\subsection{Learning with Observables}\label{sec:learning_observables} 

We now focus on discriminating quantum states according to the expectation value 
of an observable $A$. Given a state $\rho(x)$ we will predict its class as in 
Eq.~\eqref{y predicted A}, namely as 
\begin{equation}
	y_{\rm predicted} = {\rm sign}\left(\langle A\rangle_{\rm \rho(x)}\right).
	\label{eq:y predicted} 
\end{equation}
The average loss and the training error are constructed from the cost function of  
Eq.~\eqref{observable loss inequality1} as 
\begin{align}
	L(A) &= \sum_y \int dx \, P(x,y) \Lambda\left(y \langle A\rangle_{\rm \rho(x)}\right), &
	L(A,\mathcal S) &= \frac1N \sum_{n=1}^N \Lambda\left(y_n \langle A\rangle_{\rm \rho(x_n)}\right).
	\label{eq:A loss leaning}
\end{align}
where $\Lambda$ is a $\lambda$-Lipschitz convex function and $\mathcal A$ is a set of observables. 
The optimal empirical observable 
is the one minimizing $L(A,\mathcal S)$ under suitable constraints. As discussed 
in Section~\ref{sec:norm constrained}, many practical restrictions can be rephrased 
using norm constraints. Let $\mathcal G_\alpha$ be the generalization error 
when the observables are constrained to have $\|A\|_\alpha\leq c_\alpha$, 
for a certain norm and constant $c_\alpha$. 

In Appendix.~\ref{app:rade observables} we study different constraints. In particular, we find 
\begin{align}
	\mathcal G_\infty &\leq  O\left( \lambda c_\infty 
	\sqrt{\frac{2^{I^{\rm emp}_{1/2}(X{:}Q)}}N}\right),
										&
		\mathcal G_2 &\leq  O\left( \frac{\lambda c_2}{\sqrt N}\right),
								 &
		\mathcal G_1 &\leq  O\left(\lambda c_1 \sqrt{\frac{\log d}{N}}\right).
	\label{eq:generalization bound inf12}
\end{align}
For observables with maximum eigenvalue equal to 1, we can focus on $\mathcal G_\infty$ 
with $c_\infty=1$ and the generalization bound 
that we get by optimizing the loss \eqref{eq:A loss leaning}, aside from the constant $\lambda$,
is equivalent to that obtained by optimizing over POVM, namely Eq.~\eqref{generalization info}. 
For the hinge loss, where $\lambda=1$, the bounds are exactly equal. 
For observables with bounded 2-norm, generalization 
does not explicitly depend on the properties of the quantum state. 
Something similar is obtained for observables with bounded 1-norm, but with an
extra factor due to the logarithm of the dimension $d$ of the quantum states. 

As an example, we consider a system with $n$ qubits, and we focus on 
observables constructed as a linear combination of Pauli measurements, 
\begin{equation}
	A = \sum_j \alpha_j P_j,
\end{equation}
where $\alpha_j$ are real trainable coefficients and $P_j$ are Pauli matrices. Since 
these satisfy $\Tr[P_iP_j]=2^n\delta_{ij}$,
we get $\Tr[A^2] = 2^{2n}\|\bs \alpha\|_2^2$, 
so any bound $c_2$ on the 2-norm of the observable can be rewritten in terms of the norm of the coefficients as $c_2\geq 2^n\|\bs\alpha\|_2$.
Therefore, unless the coefficients $\alpha_j$ are rescaled with an exponentially small quantity, 
the bound \eqref{eq:generalization bound inf12} grows exponentially with the number of qubits. 
A similar problem happens for $\mathcal A_1$, as $\|A\|_1\geq \|A\|_2$.
On the other hand, since the eigenvalues of $P_j$ are $\pm1$ we get $\|P_j\|_\infty = 1$
and $c_\infty \geq \sum_j |\alpha_j| = \|\bs\alpha\|_1$, which has better scaling as long 
as $A$ is constructed from few Pauli observables.

\subsection{Learning with Kernels} \label{sec:supervised kernel}

As discussed in Section~\ref{sec:representer}, 
the optimal observables with a $\ell_2$ penalty can be expressed as 
a linear combination of the training data 
\begin{equation}
	A = \sum_{n=1}^N \alpha_n \rho(x_n).
	\label{representer extended}
\end{equation}
Therefore, for computing the expectation value of such quantity with respect to some state $\rho(x)$ 
we need to be able to compute all possible overlaps $\Tr[\rho(x)\rho(x_n)]$. Algorithmically, this can be 
done using the techniques presented in Section~\ref{sec:learning representer}, e.g., using the 
swap test of swap measurements from Figure~\ref{fig:swap}. Such overlaps define the kernel 
\begin{equation}
	k(x,x') = \Tr[\rho(x)\rho(x')].
	\label{kernel} 
\end{equation}
According to Eq.~\eqref{precision swap}, if we want to estimate the kernel with precision $\epsilon$ 
we need 
\begin{equation}
	S \simeq \frac{1-k(x,x')^2}{\epsilon^2},
	\label{copies kernel}
\end{equation}
copies of $\rho(x)$ and $\rho(x')$. 
Training consists in minimizing the loss \eqref{eq:A loss leaning} with a $\ell_2$ penalty  
\begin{align}
	L(\bs\alpha,S) &= \frac1N\sum_{n=1}^N \Lambda\left(y_n\langle A\rangle_{\rho(x_n)}\right) + \mu \Tr[A^2]
	\\
								 &= \frac1N\sum_{n=1}^N \Lambda\left(y_n\sum_{m=1}^N \alpha_m k(x_n,x_m)\right) + \mu \sum_{m,n=1}^N 
	\alpha_m\alpha_n k(x_n,x_m). 
	\label{kernel loss}
\end{align}
which, once the kernel matrix has been estimated, becomes a convex problem in the parameters $\alpha_n$. 
Popular machine learning libraries can be used to efficiently solve the above problem numerically, 
given the kernel matrix. After training, classification of a new state is done as Eq.~\eqref{eq:y predicted} 
as 
\begin{equation}
	y_{\rm predicted} = {\rm sign}\langle A\rangle_{\rho(x)} = 
	{\rm sign}\left(\sum_{n=1}^N \alpha_n k(x_n,x)\right).
	\label{y predicted kernel}
\end{equation}
Notice that for prediction the evaluation of new kernels $k(x_n,x)$ is necessary, which requires 
further copies of the training states $\rho(x_n)$, as discussed in Section~\ref{sec:summary}. This strategy is therefore transductive.
Using the language of Section~\ref{sec:learning representer}, we split $S$ into $S'$ and $S''$, where $S''$ is now given by $S''=V/N$, since $V/N$ copies of each training state will be consumed at test time.
Moreover, we have to ensure that the additional error in the estimation of the new kernels at test time, which can be 
controlled via the test copy complexity, does not significantly alter 
the classification accuracy.  

Generalization can be studied thanks to \eqref{eq:generalization bound inf12}. In particular, from 
Eq.~\eqref{representer extended} we get $\Tr[A^2] = \bs\alpha^T K\bs \alpha$, where 
$K_{nm} = K(x_n,x_m)$ is the kernel matrix. Hence 
\begin{equation}
	\mathcal G \leq \mathcal O\left(\lambda \sqrt{\frac{\sup_{\alpha_j} \bs\alpha^T K \bs\alpha}{N}}\right)
	\simeq \mathcal O\left(\lambda B \sqrt{\frac{k_{\rm max}}{N}}\right),
\end{equation}
where $k_{\rm max}$ is the largest eigenvalue of $K$ and $B\geq \|\bs \alpha\|_2$. 
In numerical simulations, this term can be constrained by using a larger value of $\mu$, which penalises solutions with 
large $\Tr[A]^2$.

Finally, we conclude this section by studying what happens when the kernel entries are 
estimated using the techniques of Section~\ref{sec:learning representer}, and assuming large $V$, so that the additional error at test time, due to estimating the new kernels, is not the dominant source of error. Getting analytical 
results with the loss \eqref{kernel loss} is complicated, so we focus on a lower bound.
Applying Jensen's inequality we can lower bound the problem as the discrimination 
of the two average states $\bar \rho_y^{\mathcal S}$, which 
was treated in Section~\ref{sec:learning representer}. Indeed,
\begin{equation}
	L(A,\mathcal S) \geq \frac12 \left[\Lambda(\langle A\rangle_{\bar \rho_+^{\mathcal S}})
	+ \Lambda(-\langle A\rangle_{\bar \rho_-^{\mathcal S}})\right] + \mu\Tr[A]^2.
\end{equation}
In this way, we can use Eq.~\eqref{excess risk representer general}
to show that  the error using $S$ copies scales as $\mathcal O(S^{-1/2})$, but with a prefactor 
that diverges when $\rho_+^{\mathcal S}\simeq \bar \rho_-^{\mathcal S}$.
Accordingly, if the states are less distinguishable 
more shots are required.

\subsection{Parameterized Quantum Circuits as Classifiers}\label{sec:pqc}
So far in this section, we have discussed the problem of classifying a quantum state $\rho(x)$, indexed by a classical variable $x$, into one of the two possible class labels $y \in \{\pm 1\}$, via an optimal POVM (Section~\ref{sec:unconstrainedPOVMs}) or an optimal observable  (Section~\ref{sec:learning_observables}). These optimal solutions generally entail the implementation of complex circuits that may not be compatible with noisy intermediate scale quantum (NISQ) computers. In this subsection, we discuss the pragmatic approach of constraining the optimization of the classifying quantum circuit to architectures, or {\it ans\"{a}tze} that can be efficiently implemented on NISQ hardware.

A \emph{parameterized quantum circuit} (PQC), also referred to as a \emph{quantum neural network} (QNN), consists of a sequence of quantum gates that  can be efficiently implemented on a given hardware, while possibly allowing for optimization via the tuning of some real-valued parameters. Typical examples include parameterized single-qubit rotations and fixed two-qubit gates such as CNOT (see, e.g., \cite{simeone2022introduction}). Accordingly, a PQC implements unitary transformations of the form  \begin{equation}\label{eq:PCQ1}U(\theta)=\prod_{i=1}^{N_g} U_l(\theta_l),\end{equation} where each of the $N_g$ unitary matrices $U_l(\theta_l)$ is described by a number of fixed gates and by one parameterized gate with real-valued parameter $\theta_l$. 

In practice, quantum gates are subject to quantum noise. Writing as $\mathcal{U}_{\theta_l}(\cdot)$ the operator defined by the parameterized unitary $U_l(\theta_l)$ -- i.e., $\mathcal{U}_{\theta_l}(\rho)=U_l(\theta_l)\rho U_l(\theta_l)^\dagger$ for any input density $\rho$ --, the actual operation of each term in (\ref{eq:PCQ1}) is described by a quantum channel $\mathcal{N}_{\theta_l}(\cdot) = \bar{\mathcal{N}} \circ  \mathcal{U}_{\theta_l}(\cdot)$, where $\bar{\mathcal{N}}(\cdot)$ is quantum channel describing gate noise, e.g., depolarizing noise. The notation $\circ$ denotes a composition of channels. Accordingly, the PQC implements the overall quantum channel $\bar{\mathcal{N}}\circ \mathcal{U}_{\theta_{N_g}} \circ \hdots \circ \bar{\mathcal{N}} \circ \mathcal{U}_{\theta_1}(\cdot)$.

Given a classical input $x$, PQCs operate on a quantum state $\rho(x)$ produced using a fixed {\it quantum encoding}. In general, one can apply a unitary gate $U(x)$, parameterized by index $x$, to act on an initial fiducial state to obtain the quantum embedding \begin{equation}\label{eq:emb}\rho(x)=U(x)\vert0\rangle\langle 0\vert U(x)^{\dag}.\end{equation} It is also possible to consider strategies in which the input $x$ is ``reloaded'' multiple times by interleaving input-dependent unitaries of the form $U(x)$  and parameterized quantum gates of the form $U_l(\theta_l)$ as in (\ref{eq:PCQ1}).

The PQC produces the quantum state \begin{equation}\rho_{\theta}(x)=\mathcal{N}_{\theta}(\rho(x)).\end{equation}  A classification decision can then be made by measuring the output using a fixed projective measurement. This is typically implemented by applying a standard measurement on one of the qubits. The classification function is then of the form \begin{align}
f_{\theta}(x) = {\rm Tr}(\Pi\rho_{\theta}(x)). \label{eq:f_theta_0}
\end{align} where $\Pi$ is a fixed projection matrix. With input (\ref{eq:emb}), and a noiseless PQC (\ref{eq:PCQ1}), this output can be expressed as \begin{align}
f_{\theta}(x) = {\rm Tr}(\Pi U(\theta) U(x) |0\rangle \langle 0| U(x)^\dagger U(\theta)^\dagger). \label{eq:f_theta}
\end{align} This expression shows that one can think of the optimization of the PQC $U(\theta)$ as the design of the encoding circuit $U(\theta)U(x)$ for a fixed observable $\Pi$, or as the design of the observable $U(\theta)^\dagger\Pi U(\theta)$ for a fixed encoding circuit $U(x)$,

We now discuss the generalization error of the PQC-based classifiers. To this end, we consider the $01$ loss $\ell_{01}(f,x,y)$ as in the previous subsections.
The generalization error of the learnt PQC-based classifier $f_{\theta,\mathcal{S}}$ can be studied via the Rademacher complexity of model class $\mathcal{F}=\{f_{\theta}(\cdot): \theta \in \Theta \}$, where $\Theta$ is the domain of the parameters to be learnt. 

While upper bounds on Rademacher complexity in the form of $B(\mathcal{F})$ can be evaluated directly for simple model classes, for PQC-based classifiers we have to leverage additional tools to evaluate $B(\mathcal{F})$. One such important tool is the Dudley entropy integral bound \cite{dudley1967sizes} that bounds Rademacher complexity of $\mathcal{F}$ via the {\it covering number} of $\mathcal{F}$. In Table~\ref{tab:PQC}, we summarize some recent results on the generalization error of PQC-based classifiers that leverage covering number based bounds on the Rademacher complexity.  Besides covering number-based analysis, Ref.~\cite{bu2022statistical} studies a characterization of $B(\mathcal{F})$ in terms of the \textit{magic resource} of quantum channels; the latter quantified using the $(p,q)$-group norm of the representation matrix of quantum channels. The magic resource has also been used to analyze the Rademacher complexity of noisy quantum circuits \cite{bu2023effects}, which is found to be non-increasing with increasing strength of gate noise.
\begin{table}[h!]
  \centering 
  \begin{threeparttable}
    \begin{tabular}{ccc}
     Source & Model Complexity  & Generalization Error Scaling\\
     \midrule\midrule
Caro \emph{et al.} \cite{caro2022generalization}   &   
Covering Number     &   $\mathcal{O}\biggl(\sqrt{\frac{N_g\log (N_g)}{N}}\biggr)$                     \\
& & $N_g$: Trainable gates in the  PQC \\
    \cmidrule(l  r ){1-3}
    Du \emph{et al.} \cite{du2022efficient} & Covering number &  ${\mathcal{O}}\biggl( 2^k \sqrt{\frac{N_g}{N}} \log \Bigl(N_g(1-p)^{\frac{N_g}{N_g2^{2k}}}\Bigr)\biggr)$
     \\
     & & $k$= largest number of qubits operated  \\
     & & on by a single parameterized quantum gate \\
     & & $p$ = strength of the local depolarizing noise\\
      \cmidrule(l  r ){1-3}
      Caro \emph{et al.} \cite{caro2021encoding} & Rademacher complexity of generalized  & $\mathcal{O}\biggl(\sqrt{\frac{|\Omega|}{N}}\biggr)$ \\
      & trignometric polynomial (GTP)-based & $|\Omega|=$ cardinality of the set of accessible frequencies \\
      & function class $\mathcal{F}$ & of the quantum encoding strategy \\ 
    \midrule\midrule
    \end{tabular}
\end{threeparttable}
\caption{Generalization error bounds for PQC-based Classifiers} \label{tab:PQC}
  \end{table}

Table \ref{tab:PQC} shows that the generalization error of PQC-based classifiers depends on $(i)$ the architecture of the PQC, accounted for by the number of trainable gates $N_g$, as well as the largest number $k$ of qubits operated on by a single parameterized quantum gate in the PQC; $(ii)$ on the strength $p$ of the quantum gate noise; and $(iii)$ on the quantum encoding strategy. 

While the bound of Caro \emph{et al} is tighter than Du \emph{et al}, both bounds show that PQCs with large number of trainable gates, $N_g$, incur a larger generalization error when trained on a fixed set of $N$ examples.  In particular, the bound of Du \emph{et al} also demonstrates, akin to the observation in Ref.~\cite{bu2023effects}, that  the quantum gate noise, which is inherent to currently available NISQ devices, may not be detrimental to generalization and can help prevent over-fitting.  None of these works account for the impact of the quantum encoding strategy on the generalization error. 

This aspect is addressed in \cite{caro2021encoding}. Specifically,  the authors of \cite{caro2021encoding} consider the quantum encoding in (\ref{eq:emb}), with the unitary $U(x)=\exp(-ixH)$, where $H$ is a Hermitian matrix known as the data-encoding Hamiltonian $H$. With this choice, the function $f_{\theta}(x)$ in \eqref{eq:f_theta} can be written 
as the \emph{generalized trigonometric polynomial} (GTP),
\begin{align}
f_{\theta}(x) =\sum_{\omega \in \Omega} c_{\omega}(\theta,\Pi) \exp(i\omega x),
\end{align} where the coefficents $c_{\omega}$ depend on the tunable parameters $\theta$ and observable $\Pi$, while $\Omega$ denotes the set of accessible frequencies determined by the spectrum of the data encoding Hamiltonian $H$. The key idea in  \cite{caro2021encoding} is to consider the function class \begin{equation}\label{eq:caro}\mathcal{F}_{\Omega}=\left\{f(x)=\sum_{\omega \in \Omega} c_{\omega} \exp(i\omega x): \{c_{\omega}\}_{\omega \in \Omega} \hspace{0.1cm} \mbox{ensures that } \Vert f \Vert_{\infty} \leq M \right\},\end{equation} that encompasses the original model class $\mathcal{F}$, i.e, $\mathcal{F} \subseteq \mathcal{F}_{\Omega}$. In fact, the model class (\ref{eq:caro}) allows for a larger class of coefficients $c_{\omega}$ that need not depend on tunable PQC parameters $\theta$ and observable $\Pi$.

This way, an upper bound on Rademacher complexity of the PQC-based classifiers $\mathcal{F}$ follows as $R_P(\mathcal{F}) \leq R_P(\mathcal{F}_{\Omega})$ via the Rademacher complexity of the GTP-based model class $\mathcal{F}_{\Omega}.$ As seen from Table \ref{tab:PQC}, the resulting generalization error depends only on the set of accessible frequencies determined by the quantum encoding strategy, and not on the trainable part of the PQC. Furthermore, set of accessible frequencies can be further upper bounded in terms of  the number of encoding gates used.  

Note that the Rademacher complexity-based generalization bounds of Caro \emph{et al.} \cite{caro2022generalization} and Du \emph{et al,} \cite{du2022efficient}, where the training data scales with the number of trainable parameters $N_g$, cannot explain the low generalization error of QNNs observed in the over-parameterized \footnote{A QNN with $N_g$ trainable parameters is said to be overparameterized if the quantum Fischer information (QFI) matrices, evaluated simultaneously for all data in the training set, saturates (i.e, has maximal rank) on at least one point in the parameter space \cite{larocca2023theory}.} regime. In this regime, adopting an encoding-based generalization bound as in \cite{caro2021encoding} can help explain the generalization error. Alternatively, as in classical deep learning, one must move from the capacity-based analysis to a general analysis that accounts for the training algorithm as well as the data distribution. In this regard, Ref.~\cite{du2022demystify} adopts an \textit{algorithmic robustness}-based analysis to quantify the generalization error of an ERM-classifier that scales as $\mathcal{O}(\sqrt{4^kN_{ge}\log N_{ge}/N})$, where $N_{ge}$ denotes the number of data encoding gates each acting on at most $k$ qubits.

\section{Unsupervised Learning for Quantum Generative Modelling}\label{sec:unsupervised learning}
In this section, we turn our attention to the less studied problem of  quantum unsupervised learning. As reviewed in Section~\ref{sec:unsupervisedlearning},  classical unsupervised learning  amounts to the general problems of estimating properties of an unknown probability distribution $P(x)$, or of sampling from an unknown probability distribution $P(x)$, by observing a data set drawn from $P(x)$. In particular, \emph{generative models} implement a model class $\mathcal{F}$ of candidate probability distributions $f(x)$ from which sampling can be efficiently carried out on a classical computer. Functions $f(x)$ typically consist of neural networks with classical sources of randomness.  In  quantum unsupervised learning for generative modelling, {\it quantum generative  models}  implement PQCs, as well as quantum measurements and classical post-processing.

\subsection{Learning Tasks}
Quantum generative models can be used to approximately sample from an unknown classical distribution $P(x)$, hence addressing the same problem as classical generative models,  or 
to generate a quantum state $\hat{\rho}$ that approximates an unknown quantum state $\rho$. For the former case, the learner leverages information available in the form of an training set $\mathcal{S}$ of $N$ classical examples sampled i.i.d. from the unknown distribution $P(x)$, while  in the latter case the available information is in the form of an $S$-copy training set  $\mathcal{S}^S$ consisting of $S$ copies of the unknown state $\rho$. Note that the $S$-copy training set consists of copies of a single ($N=1$) unknown quantum state, and that there is no classical input $x$. Not that there may be also situations in which the generation of the quantum state is conditional on some classical input $x$ \cite{zeng2023conditional}, but we will not elaborate on such \emph{conditional quantum generative models} here.  Depending on the type of observed {\it input data} and on the type of generated {\it target output}, 
we may distinguish the following situations.

\begin{enumerate}
    \item  {\it Classical input-classical target}: In this setting, the goal is the same as for classical generative modelling. Given a training set $\mathcal{S}=\{x_1,\hdots,x_N\}$ of examples sampled i.i.d. from the classical unknown distribution $P(x)$, we wish to optimize an implicit model that can generate samples from a distribution $f(x)$ that is a close approximation of $P(x)$. Unlike the classical case, here the model class consists of PQCs and the output is obtained via quantum measurements \cite{benedetti2019generative, situ2020quantum,romero2021variational}.

    \item {\it Classical input-quantum target}: In this setting, the goal is to load an unknown data distribution $P(x)$ into a quantum state $\hat{\rho}$ by only observing a training set $\mathcal{S}=\{x_1,\hdots,x_N\}$ of examples sampled i.i.d. from $P(x)$ \cite{zoufal2019quantum}. This setting finds application in {\it quantum state preparation} \cite{benedetti2019generative}.

    \item {\it Quantum input-quantum target}: In this setting, the goal is to obtain an approximation $\hat{\rho}$ of  an unknown quantum state $\rho$. The only information available is in the form of $S$ copies of the unknown quantum state $\rho$,  which constitutes the input data set $\mathcal{S}^{S} = \rho^{\otimes S}$. This class of problems is also termed \textit{quantum state compilation} \cite{ezzell2023quantum}, and QGLMs have been employed to approximate unknown pure states \cite{benedetti2019adversarial} as well as mixed states \cite{ezzell2023quantum}.
\end{enumerate}

In the following, we will focus on the classical input-classical target and quantum input-quantum target cases.

\subsection{Quantum Generative Learning Models}
Quantum generative learning models (QGLMs) are PQC-based models used for the purpose of generating synthetic data samples or for approximating an unknown quantum state. At their core, QGLMs consist of a  quantum channel $\mathcal{N}_{\theta}(\cdot)$, parameterized by tunable classical parameters $\theta \in \Theta$, that maps an input quantum state $\rho_{\rm in}$ to an output quantum state $\mathcal{N}_{\theta}(\rho_{\rm in})$. We now review  \emph{quantum circuit Born machines} (QCBMs) and \emph{variational quantum generators} (VQGs) as notable representatives of QGLMs.

A QCBM describes a parameterized quantum state \begin{equation}\label{eq:qcbm}\hat{\rho}_{\theta} =\mathcal{N}_{\theta}(\vert 0\rangle \langle 0 \vert )\end{equation} obtained via the operation of the channel $\mathcal{N}_{\theta}(\cdot)$ on a fiducial input state $|0\rangle$. For the \emph{classical input-classical target} case, QCBM leverages the intrinsic randomness of quantum measurements to generate discrete classical data, while for the \emph{quantum input-quantum target} case, the state $\hat{\rho}_{\theta}$ in (\ref{eq:qcbm}) can directly serve as the learned approximation. 
 
To elaborate further on the classical input-classical target case, according to  Born's rule, a projective measurement $\Pi_i=\vert i \rangle \langle i \vert$, for $i \in \{0,1, \hdots,2^{n-1}\}$, of the quantum state $\hat{\rho}_{\theta}$ onto the $i$th computational basis generates discrete samples $i\in \{0,1, \hdots,2^{n-1}\}$ with probability $f_{\theta}(i)= {\rm Tr}(\Pi_i \rho_{\theta})$. Note that QCBMs are implicit models, since they produce samples $i$, and not the probabilities  $f_{\theta}(i)$.

While QCBM can generate classical discrete data, VQGs address the classical input-classical output case  to generate real-valued data samples \cite{romero2021variational}. To do this, VQG leverages an external source of randomness, and evaluate expected values of observables, rather than relying on single-shot measurements as QCBMs.
 
A VQG takes as input to the  PQC $\mathcal{N}_{\theta}(\cdot)$ a quantum state $\rho_{\rm in }(z)$ that encodes a classical variable $z \sim Q(z)$ sampled randomly from a fixed prior distribution $Q(z)$. Accordingly, the VQG prepares the random quantum state $\hat{\rho}_{\theta}(z)=\mathcal{N}_{\theta}(\rho_{\rm in }(z))$. To generate a $d$-dimensional real-valued output sample $x \in \mathbb{R}^d$, VQG evaluates the expected value of  $d$ observables $A_1, \hdots,A_d$ as \begin{equation}\label{eq:VQG} x=[\langle A_1\rangle_{\rho_{\theta}(z)}, \hdots \langle A_d\rangle_{\rho_{\theta}(z)}].\end{equation} Assuming the possibility to accurately estimate the expectations in (\ref{eq:VQG}), the only randomness in $x$ is due to the classical random variable $z\sim Q(z)$.

\subsection{Excess Risk of Quantum Generative Modelling} 
In contrast to extensive recent empirical research on the use of quantum models for generative learning,  the generalization analysis of quantum generative models has been studied in very few existing works. In this section, we review some known results, and highlight some open problems in this field.

To begin with, consider the classical input-classical target setting, where QGLMs are used to generate classical samples from an unknown distribution $P(x)$. In this setting, one can think of QGLM as an implicit model generating samples according to an abstract distribution $f(x)$. Note that, unlike classical unsupervised learning, the distribution $f(x)$ is determined by PQCs as well as quantum measurements. Using a divergence measure $D(P,f)$ to quantify how far the true distribution $P(x)$ is from the generated distribution $f(x)$, we can proceed to define excess risk as in Section~\ref{sec:unsupervisedlearning}. As discussed before, bounds on the excess risk depends on the choice of divergence measure and the specific QGLM used.  When IPM is used as the divergence measure, the upper bound on excess risk in \eqref{eq:optgapunsup} directly applies. For VQGs,  recent Ref.~\cite{du2022theory} derived bounds on the excess risk with the squared maximum mean discrepancy (MMD) as the divergence measure. Under the assumption of Lipschitz continuous kernel function $\kappa(\cdot,\cdot)$, the resulting upper bound solely depends on the model complexity of VQGs, which is then quantified via the covering number of the considered class of PQCs in a manner similar to the results in Table \ref{tab:PQC}. Apart from Ref.~\cite{du2022efficient}, very less is known about the generalization performance of other QGLMs under different divergence measures.

More challenging is the problem of quantifying the excess risk of QGLMS used to approximate an unknown quantum state $\rho$.  To define excess risk in this setting, one can use appropriate distance measures $D(\rho,\hat{\rho}_{\theta})$ between the unknown quantum state $\rho$ and the approximated quantum state $\hat{\rho}_{\theta}$. Popular distance measures include trace distance \cite{benedetti2019adversarial}, fidelity, quantum-Wasserstein semi-metric \cite{chakrabarti2019quantum}, or the quantum relative entropy \cite{dallaire2018quantum}. However, the distance measure $D(\rho,\hat{\rho}_{\theta})$ cannot be evaluated  since the true quantum state $\rho$ is unknown. Instead, we have access only to partial information about it in the form of $S$ copies $\rho^{\otimes S}$. Since quantum measurements consume copies, the available partial information is not sufficient to search for the optimal approximation, thereby yielding a sub-optimal approximation. As such, the optimality gap in this setting will be a consequence of the   availability of limited number $S$ copies of unknown quantum state as well as the constrained model complexity of the considered PQC ansatz. Characterizing the excess risk in this setting is an open problem.

One of the important challenges in generalization analysis of generative models is the lack of a clear definition of what it means for a generative model to generalize. In Section~\ref{sec:unsupervisedlearning}, we defined the optimality gap in terms of the closeness of the approximated distribution to the true distribution relative to the closeness of best approximate distribution to true distribution. Intuitively, this measures the ability of generative model to efficiently {\it learn} the unknown distribution. Recent work \cite{gili2022evaluating} advocates an alternate, practical, interpretation of generalization, termed {\it sample-based} generalization, that measures the generalization power of a generative model via its ability to efficiently {\it generate} data samples. Precisely, a QGLM is said to have good sample-based generalization if it can generate novel, high quality,  out-of-training samples. Considering discrete probability distributions, Ref.~\cite{gili2022evaluating} gives a formal definition of sample-based generalization and introduces various performance metrics to characterize it.

\section{Conclusions and Outlook}\label{sec:outlook} 

We have studied the statistical complexity of quantum learning, namely the number of samples and the number of 
copies of each sample to reach a desired accuracy. In the most general setting with quantum data, 
what makes quantum learning different from classical learning is the learner's ignorance about the training states. 
While classically the learner has access to all the information about the training states (e.g., all the bits in an image),
in the quantum case the number of bits that the learner can extract from a single measurement 
is limited.
Moreover, quantum data cannot be copied, so the learner must have access to multiple copies, obtained by 
repeating the state-preparation mechanism.

We have studied different errors depending on the learner's knowledge: (i) the
error when the learner knows 
both the data distribution and the quantum states, which can be addressed via standard statistical techniques, 
without learning methods; (ii) the extra generalization error when the
learner has full classical knowledge of the training data, 
but ignores the data distribution --
the typical setting in classical machine learning -- which can be 
reduced by increasing the amount $N$ of training data; and (iii) the extra
knowledge gap when the learner can only extract partial 
information from the training states, which can be reduced by increasing the average 
number of copies $S$ of such states. 
The total error at the test stage,  when the learnt model is used to make new
predictions, may be decomposed as a sum of these three terms. 

We have reviewed different learning methods proposed in the literature and studied the three sources of errors using a combination 
of quantum information theory and statistical learning theory. All these considered, it is an open question to understand which 
method is ``best'' at exploiting the information available at the training stage. 
For example, classical methods focus on minimizing (i)+(ii), and deep neural
networks trained with gradient descent are known 
to be able to reach remarkably low errors. However, in the quantum setting, e.g., with quantum neural networks, each gradient 
evaluation requires the consumption of copies of the training data, so it is unlikely that quantum neural networks with many parameters 
make an efficient use of their copies and are optimal for error (iii).

In quantum state classification problems, e.g., in quantum state discrimination and hypothesis testing, optimal 
or asymptotically optimal strategies have been found to minimize (i)+(iii), making the most efficient use 
of the available copies. However, identifying the most efficient strategy that minimizes all sources of error, (i)+(ii)+(iii), using the least amount of training 
resources is still an open question. The answer to this question is likely to depend on
the learning scenario, and on whether it is 
simpler to get new quantum data, that is, to make $N$ large, or to repeat the state
preparation procedure to make $S$ large. 

Moreover, the structure of the quantum data is expected to play a key role in determining 
the relative importance of the three types of error as a function of resources  $N$ and $S$. For instance, more distinguishable states are expected to require a 
lower number of copies $S$, while clustered states in the Hilbert space, which display less variation, 
are expected to require a lower number of data points $N$. Therefore, it is important to consider not just the 
scaling with the numbers $N$ and $S$, but also to express the pre-factors using quantities that can be 
simple to interpret.

With the ultimate goal of understanding the accuracy of quantum learning as a
function of the data and copy resources, we have studied
different supervised and unsupervised learning methods and, 
when technically possible, we obtained a bound on the different error terms. 
We have done an extensive analysis of the Helstrom classifier, which is optimal for state discrimination,
according to the 01 loss,
and whose generalization error was studied in \cite{banchi2021generalization}. The above generalization analysis has been extended to study the optimality gap of Helstrom classifiers that discriminate quantum states prepared via a pre-trained PQC in \cite{jose2023transfer}. We have also studied 
how the generalization error changes when the learner is limited to use a finite number of copies of each quantum state, 
both at training and testing stage. 
Moreover, we have shown that the same generalization error also describes learning problems involving 
the optimization of observables with bounded spectrum. 
However, since the Helstrom classifier uses unconstrained operations, it is unlikely to be optimal 
for generalization. We have studied how to introduce constraints via information theoretic techniques, 
and applied this setting to toy problems involving the classification of quantum states or 
quantum phases of matter. 

Finally, we have also considered kernel methods, within the framework of support vector machines, 
and some hybrid methods consisting of a quantum data collection part, followed by a purely classical 
learning process. Moreover, we have reviewed different bounds on parametric
quantum circuits, and, in some case, showed how to obtain them using information-theoretic techniques.

In this paper, we have deliberately omitted any discussion about quantum advantage for machine learning 
for two reasons. Firstly, quantum advantage in learning with real-world problems has several caveats 
\cite{aaronson2015read}. Most results about quantum advantage for machine learning involve
ad-hoc datasets, constructed with the specific aim of beating the best classical approach  
on carefully chosen error metrics. 
Secondly, it is often complicated to define what the best classical approach is so as to make a fair comparison 
between quantum and classical learning.
Nonetheless, we mention here some future directions to explore the possibility of quantum advantage within 
the framework presented in this paper. For instance, 
the generalization error with unconstrained operations 
depends on the amount of shared information between the space of classical inputs and 
the space of quantum states. Since quantum states can be used to compress classical data without losing 
any predictive power \cite{elliott2020extreme,raz1999exponential}, it is tempting to conjecture a link between quantum advantage 
in compressing classical data and quantum advantage in learning, namely in achieving the same accuracy with 
less data.

\appendix 

\section{Useful mathematical results}\label{a:math}

We first introduce some general results from probability theory. 

\begin{lemma}[Hoeffding's inequality]
	Let $X_j$ with $j=1,\dots,n$ be independent identically distributed random variables with 
	mean $\mu = \mathbb E [X_j]$, 
	defined in the interval $a\leq X_j \leq b$. For  $c=b-a$ and arbitrary $t$, we get
	\begin{equation}
		P\left(\frac1n \sum_{j=1}^n X_j - \mu\geq t\right) \leq e^{ -2n t^2/c^2},
	\end{equation}
\end{lemma}

The following theorem is useful for mapping inequalities involving a success probability 
to inequalities involving expectation values.
\begin{thm} \label{thm: prob to expectation}
Suppose that with probability at least $1-\delta$, random variable $X$ follows the inequality
\begin{equation}
    X \leq A + \sqrt{B+C\log(1/\delta)},
\end{equation}
for arbitrary constants (with regard to $\delta$) $A$ and $B$ and positive constant $C$. The expectation value of $X$ is bounded by
\begin{equation}
    \ave[X] \leq A + \sqrt{B} + \frac{1}{2}e^{B/C}\mathrm{erfc}\left[\sqrt{\frac{B}{C}}\right]\sqrt{\pi C},
\end{equation}
where $\mathrm{erfc}$ is the complementary error function.
\end{thm}
\noindent {\it Proof.}  
The cumulative density function (CDF) of $X$ is an increasing function $f(x)$ that takes values from $0$ to $1$ such that $P(X\leq x)=f(x)$. Let us define
\begin{equation}
    x = A + \sqrt{B+C\log(1/\delta)}.\label{eq:x def}
\end{equation}
Then, we can lower bound the CDF of $X$ as $P(X\leq x)\geq 1-\delta$, which is defined for $x\geq A+\sqrt{B}$ (or, equivalently, $\delta\geq 0$). Rearranging Eq.~(\ref{eq:x def}) to give an expression for $1-\delta$ in terms of $x$, we get
\begin{equation}
    1-\delta = 1-\exp\left[\frac{B-(A-x)^2}{C}\right],
\end{equation}
again defined for $x\geq A+\sqrt{B}$. Therefore, the CDF of $x$ is lower bounded by
\begin{equation}
    P(X\leq x) = f(x) \geq g(x) = 1-\exp\left[\frac{B-(A-x)^2}{C}\right].
\end{equation}
To obtain the probability density function (PDF) from the CDF, we can differentiate the CDF, so the (exact) expectation value of $X$ is given by
\begin{equation}
    \ave[X] = \int_{-\infty}^{\infty} x f'(x) dx,\label{eq:CDF to expectation}
\end{equation}
where we note that the lower limit for the true CDF (not the lower bound) may be lower than $A+\sqrt{B}$. Evaluating Eq.~(\ref{eq:CDF to expectation}) for our lower bound on the CDF, we get
\begin{equation}
    \int_{A+\sqrt{B}}^{\infty} x g'(x) dx = A + \sqrt{B} + \frac{1}{2}e^{B/C}\mathrm{erfc}\left[\sqrt{\frac{B}{C}}\right]\sqrt{\pi C}.
\end{equation}
Since we have a lower bound on the CDF, we have an upper bound on the expectation value.

Finally we provide this simple result from Ref.~\cite{banchi2021generalization}. 
\begin{lemma}
	Let $A_i$ be a set of operators and $i$  a random variable with probability distribution $p_i$. Then 
	\begin{equation}
		\ave_{i\sim p} \left(\|A_i\|_1\right) \leq \Tr\sqrt{\ave_{i\sim p} \left(A_iA_i^\dagger\right)}~,
			\label{e:ave1norm}
	\end{equation}
where	$\ave_{i\sim p} f(i) := \sum_i p_i f(i)$. 
\end{lemma}

\subsection{Bounds on Rademacher Complexities}

In this subsection we introduce different bounds on the Rademacher complexity, for
both variables and operators. 

\begin{thm}[Uniform deviation bound\cite{bartlett2021deep}]\label{thm:rada} 
	For any loss in $[0,1]$ and dataset $\mathcal S$ with $N$ pairs
\begin{equation}
	\frac12 R_P(\mathcal F)-\sqrt{\frac{\log 2}{2N}} \leq \ave_{\mathcal
	S}[\mathcal D(\mathcal F,\mathcal S)] \leq 2 R_P(\mathcal F).
\label{rada bound}
\end{equation}
Moreover, for large $N$ the uniform deviation $D(\mathcal F,\mathcal S)$ 
is concentrated around its average, namely
with arbitrarily high probability $1-\delta$
\begin{equation}
	|\ave_{\mathcal S}[\mathcal D(\mathcal F,\mathcal S)] - \mathcal D(\mathcal F,\mathcal S)| \leq 
 \sqrt{\frac{\log(2/\delta)}{2N}}.\label{eq:uniform deviation dist}
\end{equation}
\end{thm}
Equivalently, using Theorem~\ref{thm: prob to expectation}, we can express Eq.~(\ref{eq:uniform deviation dist}) in terms of the expectation value of the distance from the average value as
\begin{equation}
	\ave\left[|\ave_{\mathcal S}[\mathcal D(\mathcal F,\mathcal S)] - \mathcal D(\mathcal F,\mathcal S)|\right] \leq 
 \sqrt{\frac{\log 2}{2N}}+\mathrm{erfc}\left[\sqrt{\log 2}\right]\sqrt{\frac{\pi}{2N}}.
\end{equation}

\begin{thm}[Khintchine inequalities \cite{haagerup1981best}]
Let $x_k$ be real numbers and $\sigma_k=\pm1$ be Rademacher variables. Then
for all integers $p\geq 1$ 
	\begin{equation}
		 A_p \|\bs x\|_2 \leq 
		\left(\ave_{\bs\sigma}\left|\sum_{k=1}^N \sigma_k x_k\right|^p\right)^{1/p} 
		\leq B_p \|\bs x\|_2
	\end{equation}
	where $\|\bs x\|_2 = \sqrt{\sum_k x_k^2}$. Moreover $A_1=1/\sqrt{2}$ and $A_p=1$ for 
	$p\geq 2$, while $B_p=1$ for $p\leq 2$ and $B_{p}  = 2^{-1/4}\sqrt{\pi p/{\rm e}} $
	for $p>2$.
\end{thm}

\begin{thm}[Operator Khintchine inequalities~\cite{lust1986inegalites,candes2012exact}]\label{thm:opekhin}
Let $X_k$ be Hermitian operators and $\sigma_k=\pm1$ be Rademacher variables.
Then for each $\alpha\geq 2$ it holds
\begin{equation}
		\|X_{\mathcal R}\|_\alpha \leq
		\sqrt[\alpha]{\ave_{\bs\sigma} \left\| \sum_k \sigma_k X_k \right\|_\alpha^\alpha}
		\leq B_\alpha \|X_{\mathcal R}\|_\alpha,
		\label{e:khintchinealpha}
\end{equation}
where $X_{\mathcal R} = \sqrt{\sum_k X_k^2}$ and $B_{\alpha}  = 2^{-1/4}\sqrt{\pi\alpha/{\rm e}} $
is a  constant depending on $\alpha$ only. 
On the other hand, for $\alpha=1$
\begin{equation}
	\frac{1}{2\sqrt{\rm e}}	\sqrt{\sum_k \|X_k\|_1^2} \leq 
		\ave_{\bs\sigma} \left\| \sum_k \sigma_k X_k \right\|_1 \leq 
		\|X_{\mathcal R}\|_1,
		\label{e:khintchine1}
\end{equation}
\end{thm}
Note that, since $\|X\|_1\geq \|X\|_2$ and $\|X_{\mathcal R}\|_1 = 
\Tr[X_{\mathcal R}]$ we may simplify the lower bound in the 
last expression and write 
\begin{equation}
	\frac{1}{2\sqrt{\rm e}}	\|X_{\mathcal R}\|_2 \leq 
		\ave_{\bs\sigma} \left\| \sum_k \sigma_k X_k \right\|_1 \leq 
		\|X_{\mathcal R}\|_1,
		\label{e:khintchine}
\end{equation}

\begin{thm}[Tropp inequality~\cite{tropp2015introduction}]\label{thm:tropp}
Let $X_k$ be Hermitian operators of dimension $d$ and $\sigma_k=\pm1$ be Rademacher variables.
Then
\begin{equation}
	\ave_{\bs \sigma}	\left\|\sum_k \sigma_k X_{k}\right\|_\infty \leq \sqrt{\left\|\sum_k X_k^2\right\|_\infty 2 \log d}.
		\label{e:tropp}
\end{equation}
\end{thm}

\begin{lemma}[Contraction lemma\cite{shalev2014understanding}]\label{lemma:contraction}
	Let $\mathcal F$ be a function class and $g$ a $\lambda$-Lipschitz function, then for any set 
	$\mathcal S$
	\begin{equation}
		R(g\circ \mathcal F,\mathcal S) \leq 
		\lambda R(\mathcal F,\mathcal S)
	\end{equation}
	where $g\circ \mathcal F$ denotes the set of functions $x\mapsto g(f(x))$, where $f\in \mathcal{F}$.
\end{lemma}

\begin{thm}[credited to Blanchard\cite{scott2005learning,cohen2020learning}] \label{thm: rada finite} 
	Consider an input space $X$  and 
	a partitioning of $X$ into $K$ disjoint subsets $K_i$, such that $X = \bigcup_{i=1}^K K_i$.
	Let $Y$ be an output space, with two possible outputs $y=\pm1$, and let
	$\mathcal F_\pm$ be the space of functions $f:X\mapsto Y$ that are constant in each subset, i.e. 
	$f(x)=f(x')$ for all $x,x'\in K_i$. 
	Given a training set $\mathcal S =\{(x_n,y_n)\}$, with $n=1,\dots,N$, $x_n\in X$, and $y_n\in Y$, we define the integers $N_i$ (for each $i$ from $1$ to $K$) as the numbers of pairs $(x_n,y_n)$ in $\mathcal{S}$ such that $x_n\in K_i$.
	The empirical Rademacher complexity of the 0-1 loss for function class $\mathcal F_\pm$ and training set $\mathcal{S}$ satisfies 
	\begin{equation}
		\frac1{2\sqrt 2 }\sum_{i=1}^K\sqrt{\frac{N_i}N}
	\leq R(\mathcal F_\pm,\mathcal S) \leq 
	\frac1{2}\sum_{i=1}^K\sqrt{\frac{N_i}N}.
	\end{equation}
\end{thm}

\noindent {\it Proof.}  
The 0-1 loss is such that $\ell_{01}(f,x,y) = \frac{1-yf(x)}2$, hence 
\begin{align}
	R(\mathcal F_\pm,\mathcal S) &= 
	\ave_{\bs \sigma}\left[ \sup_{f\in \mathcal F_\pm} \frac1N\sum_{n=1}^N
	\sigma_n \frac{1-y_n f(x_n)}2\right] = 
	\frac1{2N}\ave_{\bs \sigma}\left[ \sup_{f\in \mathcal F_\pm} \sum_{n=1}^N
\sigma_n f(x_n)\right]\\&=
\frac1{2N}\ave_{\bs \sigma}\left[ \sum_{i=1}^K \sup_{f\in \mathcal F_\pm} f(K_i) \sum_{n: x_n \in K_i }
	\sigma_n \right] = 
\frac1{2N}\ave_{\bs \sigma}\left[ \sum_{i=1}^K \left|\sum_{n: x_n \in K_i }
	\sigma_n\right| \right]
																\\&\leq
	\frac1{2N}\sum_{i=1}^K\sqrt{\ave_{\bs \sigma}\left(\sum_{n: x_n \in K_i } \sigma_n\right)^2} = 
	\frac1{2N}\sum_{i=1}^K\sqrt{N_i},\label{eq: empirical rademacher dictionary}
\end{align}
which completes the upper bound. The lower bound follows by Khintchine inequalities. 
\hfill$\square$

\begin{coro}[adapted from Section~3.3 of Ref.~\cite{cohen2020learning}]\label{coro: entropy}
	In the setting of Theorem~\ref{thm: rada finite}, the Rademacher complexity satisfies 
	\begin{equation}
		R_P(\mathcal F_\pm) =\ave_{\mathcal S}  R(\mathcal F_\pm,\mathcal S) \leq 
		\sqrt{\frac{2^{H_{1/2}(\bs p)}}{4N}}.\label{eq: rada finite}
	\end{equation}
	where $p_j$ is the probability that the input belongs to the set $K_j$ and  
	$H_\alpha(\bs p) = \log_2(\sum_{j=1}^K p_j^\alpha)/(1-\alpha)$ is the $\alpha$-R\'enyi entropy. 
\end{coro}
\noindent{\it Proof. }
   To go from the empirical Rademacher complexity to the true Rademacher complexity, we must evaluate the expectation value of the right hand side of Eq.~(\ref{eq: empirical rademacher dictionary}) over all distributions of the integers $N_j$ such that $\sum N_j = N$. This is the multinomial distribution $P(\{N_j\})$. Then,
    \begin{align}
		\frac{1}{2N}\ave_{P(\{N_j\})} \sum_{j=1}^K\sqrt{N_j} &= \frac{1}{2N}\sum_{\{N_i\}:\sum_i N_i = N} p_1^{N_1} p_2^{N_2} \dots p_K^{N_K} \binom{N}{N_1,N_2,\dots,N_K} \sum_{j=1}^K\sqrt{N_j}\\
		&= \frac{1}{2N}\sum_{j=1}^K\sum_{N_j = 0}^{N} p_j^{N_j}(1-p_j)^{N-N_j} \binom{N}{N_j} \sqrt{N_j}\\
        &= \frac{1}{2N}\sum_{j=1}^K \ave_{N_j\sim B(N,p_j)}\sqrt{N_j},
    \end{align}
    where $B(n,p)$ denotes the binomial distribution and in the second line we focus on a single $N_j$ and 
		sum over all $N_i$ with $i\neq j$. 
		Finally, we can apply Jensen's inequality to write
    \begin{equation}
        \frac{1}{2N}\sum_{j=1}^K \ave_{N_j\sim B(N,p_j)}\sqrt{N_j} \leq \frac{1}{2N}\sum_{j=1}^K \sqrt{\ave_{N_j\sim B(N,p_j)} N_j} = \frac{1}{2N}\sum_{j=1}^K \sqrt{p_j N} = \frac{\sum_{j=1}^K \sqrt{p_j}}{2\sqrt{N}},
    \end{equation}
 and then, by the definition of $H_\alpha(\bs p)$, we recover Eq.~(\ref{eq: rada finite}).
 
\hfill$\square$

We now present a new corollary.
\begin{coro} \label{coro: loss finite}
    Consider an input space $X$ and 
	a partitioning of $X$ into $K$ disjoint subsets $K_i$, such that $X = \bigcup_{i=1}^K K_i$.
	Let $Y$ be an output space, with two possible outputs $y=\pm1$, and suppose we have a training set $\mathcal S =\{(x_n,y_n)\}$, with $n=1,\dots,N$, $x_n\in X$, and $y_n\in Y$. Define the integers $N_i$ as the numbers of pairs $(x_n,y_n)$ in $\mathcal{S}$ such that $x_n\in K_i$.
    Define another output space, $Z$, which consists of the real numbers (not just $\pm1$).
    Let $\mathcal F_{B,p}$ be the space of functions $f:X\mapsto Z$ that are constant in each subset ($f(x)=f(x')$ for all $x,x'\in K_i$) and that obey $\|\bs f\|_p \leq B$, where $\bs f = (f(k_1),\dots,f(k_K))$.
    Finally, consider a loss function of the form $\ell(f,x,y) = \Lambda[yf(x)]$, where $\Lambda$ is a $\lambda$-Lipschitz function. The empirical Rademacher complexity of such a loss satisfies
	\begin{align}
		R(\mathcal F_{B,2},\mathcal S) &\leq \frac{\lambda B}{\sqrt N} &
		R(\mathcal F_{B,\infty},\mathcal S) &\leq \frac{\lambda B}{N}\sum_{i=1}^K \sqrt{N_i}.
	\end{align}
\end{coro}
\noindent {\it Proof.}  
Thanks to the contraction lemma we may write 
\begin{align}
	R(\mathcal F_{B,p},\mathcal S) &\leq \frac{\lambda  }N
	\ave_{\bs \sigma}\left[ \sup_{f\in \mathcal F_{B,p}} \sum_{n=1}^N
	\sigma_n y_n f(x_n)\right] = 
	\frac\lambda{N}\ave_{\bs \sigma}\left[ \sum_{i=1}^K \sup_{f\in \mathcal F_{B,p}} f(K_i) \sum_{n: x_n \in K_i }
\sigma_n \right].
\end{align}
Then, using the Cauchy-Schwartz and Jensen inequalities,
\begin{align}
	R(\mathcal F_{B,2},\mathcal S) &\leq 
\frac{\lambda B}{N}\ave_{\bs \sigma}\sqrt{ \sum_{i=1}^K \left(\sum_{n: x_n \in K_i }
\sigma_n\right)^2 }  
	\leq
	\frac{\lambda B}{N}
	\sqrt{\sum_{i=1}^K\ave_{\bs \sigma}\left(\sum_{n: x_n \in K_i } \sigma_n\right)^2} = 
	\frac{\lambda B}{N}
	\sqrt{\sum_{i=1}^K N_i}
	= \frac{\lambda B}{\sqrt N}.
\end{align}
Otherwise, using the H\"older inequality and proceeding as in Theorem~\ref{thm: rada finite} we get
\begin{align}
	R(\mathcal F_{B,\infty},\mathcal S) &\leq 
	\frac{B\lambda}{N}\ave_{\bs \sigma}\left[ \sum_{i=1}^K \left|\sum_{n: x_n \in K_i }
	\sigma_n\right| \right]  \leq
	\frac{B\lambda}{N}\sum_{i=1}^K\sqrt{N_i}.
\end{align}
\hfill$\square$

\section{Derivation of the upper bounds}

\subsection{Unsupervised Learning under IPM} \label{app:genboundunsup}
In this section, we outline the key techniques to derive the upper bound on the optimality gap $\mathcal{E}_{\rm IPM}(f_{\mathcal{S}, \mathcal{S}_f})$ in \eqref{eq:optgapunsup}. The analysis is inspired by the proof of Theorem 2.1 in  \cite{bai2018approximability} which considers the specific Wasserstein IPM.

An IPM satisfies the \emph{triangle inequality}, and hence we can upper bound the absolute value of the generalization error for any model $f$ as
\begin{align}
    |\mathcal{G}_{{\rm IPM}}(f,\mathcal{S},\mathcal{S}_{f})| \leq \sup_{h \in \mathcal{H}}\Bigl|\ave_{x\sim P(x)}[h(x)]-\frac{1}{N} \sum_{n=1}^N h(x_n) \Bigr| + \sup_{h \in \mathcal{H}}\Bigl|\ave_{x \sim f(x)}[h(x)]-\frac{1}{M} \sum_{m=1}^M h(x_m^f) \Bigr| \label{IPM triangle}.
\end{align} The first term in \eqref{IPM triangle} quantifies the maximum deviation between the true average, under distribution $P(x)$, and the empirical average over functions $h \in \mathcal{H}$ using the training data set, while the second term similarly quantifies the maximum deviation when data is generated according to $f(x)$.  

Using \eqref{IPM triangle}, the uniform deviation in \eqref{eq:unidevunsup} can be upper bounded as
\begin{align}
    \mathcal{D}_{\rm IPM}(\mathcal{F}, \mathcal{S},\mathcal{S}_{\mathcal{F}}) &\leq \sup_{h \in \mathcal{H}}\Bigl|\ave_{x \sim P(x)}[h(x)]-\frac{1}{N} \sum_{n=1}^N h(x_n) \Bigr| + \sup_{h \in \mathcal{H}, f \in \mathcal{F}}\Bigl|\ave_{f(x)}[h(x)]-\frac{1}{M} \sum_{m=1}^M h(x_m^f) \Bigr| \nonumber \\&= {\mathcal{D}}(\mathcal{H}, \mathcal{S})+ {\mathcal{D}}(\mathcal{F} \times \mathcal{H}, \mathcal{S}_{\mathcal{F}}), \label{eq:uniformdeviations_sum}
\end{align} where the first term ${\mathcal{D}}(\mathcal{H}, \mathcal{S})$ is the uniform deviation \eqref{uniform deviation} of the discriminator class under data set $\mathcal{S}$  (set $f=h$, $\mathcal{F}=\mathcal{H}$ and $\ell(f,x,y)= h(x)$ in \eqref{uniform deviation}); while the second term ${\mathcal{D}}(\mathcal{F} \times \mathcal{H}, \mathcal{S}_{\mathcal{F}})$ corresponds to the uniform deviation of the combined function space $\mathcal{F} \times \mathcal{H}=\{(f,h):h \in \mathcal{H}, f \in \mathcal{F}\}$ of discriminator and model under the data set $\mathcal{S}_{\mathcal{F}}$. 

Applying the uniform deviation bound \eqref{rada simeq} separately on each of the uniform deviations in \eqref{eq:uniformdeviations_sum} yields the scaling of \eqref{eq:optgapunsup}.

\subsection{Supervised Learning with Unconstrained POVM}\label{app:unconstrained povm}
In this section, we prove the bounds \eqref{B 01 loss}. 
We first note that 
the optimization over POVM $\mathcal M= \{\Pi_0,\Pi_1\}$ can be cast as an optimization 
over an observable $A$ such that $\Pi_y = (\openone + yA)/2$, $y=\pm$ and $\|A\|_\infty \leq 1$. 
Since the constants are averaged out in the calculation of the Rademacher complexity, 
using the definition \eqref{rademacher} we get 
an exact expression for the empirical Rademacher complexity 
\begin{align}
	R(\mathcal M,\mathcal S) &= \ave_{\bs \sigma} \sup_{\{\Pi_y\} \in \mathcal M} 
	\left(	\frac1{N} \sum_{n=1}^N \sigma_n \Tr[(\openone-\Pi_{y_n}) \rho(x_n)]\right) = 
\ave_{\bs \sigma} \sup_{A : \|A\|_\infty \leq 1 } 
\left(	\frac1{2N} \sum_{n=1}^N \sigma_n {y_n} \Tr[A\rho(x_n)]\right) 
	\nonumber \\ &= \frac1{2N}
	\ave_{\bs \sigma} \Bigg\| \sum_{n=1}^N \sigma_n \rho(x_n)\Bigg\|_1,
	\label{rada M S}
\end{align}
where we used the fact that $\sigma_n$ and $\sigma_n {y_n}$ have the same distribution,
as $y_n=\pm1$, 
and that, by the H\"older inequality,
$\sup_{A : \|A\|_\infty \leq 1} \Tr[AB] = \|B\|_1$.
Thanks to the operator Khintchine inequalities (Theorem~\ref{thm:opekhin} in Appendix~\ref{a:math}) 
we are now ready to find upper and lower bounds 
on the empirical Rademacher complexity as 
\begin{align}
	\frac1{4\sqrt{N {\rm e}}}\leq 
	R(\mathcal M,\mathcal S)  &\leq 
	\frac1{2N} \Tr\sqrt{\sum_{n=1}^N \rho(x_n)^2},
\end{align}
which proves the second inequality in \eqref{B 01 loss}. 
The upper bound was derived in \cite{banchi2021generalization}, while from the lower bound
we note that the empirical Rademacher complexity decreases at least 
as $\mathcal O(1/\sqrt N)$. Finally, the behaviour of the abstract Rademacher complexity in \eqref{B 01 loss} 
is recovered using properties of the operator square root\cite{banchi2021generalization} as
\begin{equation}
	R_N(\mathcal M)  = \ave_{\mathcal S} R(\mathcal M,\mathcal S) 
	\leq \frac1{2N} \Tr\sqrt{\ave_{\mathcal S}\sum_{n=1}^N \rho(x_n)^2}
	=
	\frac1{2\sqrt N} 
\Tr\sqrt{\int dx\,p(x) \rho(x)^2}
\end{equation}
where $p(x) = \sum_y P(x,y)$ is the marginal distribution.

\subsection{Rademacher Complexity of the Majority Rule with Unconstrained Local Measurements}
\label{app:majority rule}

In this section, we prove 
\eqref{B 01 loss V} by proceeding as in Eq.~\eqref{rada M S} 
by writing the Rademacher complexity for the loss \eqref{loss majority vote} 
\begin{align}
	R^V(\mathcal M,\mathcal S) &= \ave_{\bs \sigma} \sup_{\{\Pi_y\} \in \mathcal M} 
	\left(	\frac1{N} \sum_{n=1}^N \sigma_n c_v(\Tr[\Pi_{y_n} \rho(x_n)])\right) 
	\leq L 	R^V(\mathcal M,\mathcal S),
	\label{rada M S V }
\end{align}
with $L\geq c'_v(p)$. The inequality is a simple application of the contraction lemma \eqref{lemma:contraction}, 
where $L$ is the Lipschitz constant of function $c_v(p)$. 
In order to bound such Lipschitz constant, we first note that we may express 
the cumulative distribution with the regularized incomplete beta function $I$ as
$c_v(p) = I_{1-p}(v+1,v+1)$.
Then using the properties of that function 
\begin{equation}
	|c_{v-1}'(p)| = \frac{[p(1-p)]^{v-1}}{B[v,v]} \leq \frac1{4^{v-1}} \frac{\binom{2v}{v}}{2v/v^2} 
	\leq 2\sqrt{\frac{v}\pi},
~~~~~~~
\Longrightarrow
~~~~~~~
L = 2\sqrt{\frac{V+1}{2\pi}},
\end{equation}
where $B[a,b]$ is the beta function and we used the fact that $p(1-p)$ has a maximum for $p=1/2$.

\subsection{Supervised Learning with Constrained Observables} \label{app:rade observables}

We consider the average loss \eqref{eq:A loss leaning},
where $\Lambda$ is a $\lambda$-Lipschitz convex function, and focus 
on families of observables satisfying some norm constraints
\begin{equation}
			\mathcal A_\sharp = \{ A : \| A\|_\sharp \leq c_\sharp \},
\end{equation}
where $\|\cdot\|_\sharp$ is a suitable norm and $c_\sharp$ a fixed constraint. 
Let $\|\cdot\|_\flat$ be the dual norm of $\|\cdot\|_\sharp$, namely the one satisfying
$\|A\|_\flat = \sup_{B:\|B\|_\sharp\leq 1} \Tr[AB]$. Then using the contraction lemma 
\ref{lemma:contraction}
we can bound the Rademacher complexity of class $\mathcal A_\sharp$ as 
\begin{align}
 R(\mathcal A_\sharp,\mathcal S) &\leq  \frac\lambda{2N} 
\ave_{\bs \sigma} \sup_{A \in\mathcal A_\sharp } 
\left(	\sum_{n=1}^N \sigma_n {y_n} \Tr[A\rho(x_n)]\right) \leq
\frac{\lambda c_\sharp}{2N}
	\ave_{\bs \sigma} \Bigg\| \sum_{n=1}^N \sigma_n \rho(x_n)\Bigg\|_\flat. 
\end{align}
For different norms, in particular for all $\|\cdot\|_p$ norms with dual $\|\cdot\|_q$, where 
$1/p+1/q=1$, we can now 
employ use the operator Khintchine inequalities (cf.~Theorem \ref{thm:opekhin}) 
to find a bound on the Rademacher complexity. 
In particular, in $\mathcal A_\infty$ we get the 1-norm, so is equivalent to that obtained 
for quantum states. 
As another example, consider observables with bounded 2-norm. In that case, 
using Jensen's inequality we find 
\begin{equation}
	\mathcal G_2 \leq  O\left( \frac{\lambda c_2}N
\sqrt{\sum_{i=1}^N \Tr[\rho(x_n)^2]}\right) \leq 
O\left( \frac{\lambda c_2}{\sqrt N}\right),
	\label{eq:generalization bound 2}
\end{equation}
since the purity is at most 1. 
Something similar is obtained for $\mathcal A_1$ using the Tropp inequality (cf.~Theorem \ref{thm:tropp}), 
but with an extra factor due to the logarithm of the dimension $d$ of the quantum states. 
\begin{equation}
	\mathcal G_1 \leq  O\left( \frac{\lambda c_1}N
\sqrt{\left\|\sum_{i=1}^N \rho(x_n)^2\right\|_\infty\log d}\right) \leq 
O\left(\lambda c_1 \sqrt{\frac{\log d}{N}}\right).
	\label{eq:generalization bound 1}
\end{equation}

\section{Generalization error with Fourier-like embeddings}\label{app:fourier}

Consider the setting of Section~\ref{sec:fourier}, where 
\begin{equation}
	 \ket{\psi(x)} = \frac{1}{\sqrt{|\Omega|}}\sum_{\omega\in\Omega} e^{i\omega x} \ket{\phi_\omega},
\end{equation}
$\Omega$ defines a set of allowed ``frequencies'', and $|\Omega|$ is the number of frequencies. 
In the expression \eqref{B 01 loss}
for the bound on the empirical Rademacher complexity,  we need to study 
the average states 
\begin{equation}
	\frac 1N \sum_{n=1}^N \ket{\psi(x_n)}\!\!\bra{\psi(x_n)} = \frac1{N|\Omega|}\sum_n \sum_{\omega,\omega'\in\Omega}
	e^{i x_n (\omega-\omega')} \ket{\phi_{\omega}}\!\!\bra{\phi_{\omega'}}.
\end{equation}
Let $F$ be the $|\Omega|\times|\Omega|$ Hermitian matrix
\begin{align}
	F_{\omega,\omega'} &= \frac{1}{N|\Omega|}\sum_n  e^{i x_n (\omega-\omega')}.
										 &
										 F = \frac{1}{N|\Omega|}\sum_n\sum_{\omega,\omega'\in\Omega} e^{i x_n \omega}\ket{\phi_{\omega}}\!\!\bra{\phi_{\omega'}} e^{-i x_n \omega'}.
\end{align}
Diagonalizing it we get $F=S^\dagger \varphi S$. Calling $\ket{\varphi_{\lambda}} = \sum_{\omega \in \Omega} 
S_{\lambda,\omega} \ket{\phi_{\omega}} $ for $\lambda=1,\dots,|\Omega|$ then 
\begin{equation}
	\bar\rho = \frac 1N \sum_{n=1}^N \ket{\psi(x_n)}\!\!\bra{\psi(x_n)} =
\sum_{\lambda=1}^{|\Omega|} 
	\varphi_{\lambda} \ket{\varphi_{\lambda}}\!\!\bra{\varphi_{\lambda}}.
\end{equation}
Since $F$ is a non-negative and $\Tr[F]=1$, $\varphi$ defines a probability distribution. 
From Eq.~\eqref{B 01 loss} we get 
\begin{equation}
	B(\mathcal M,\mathcal S) \leq \Tr[\sqrt{\bar \rho}]^2 = 2^{H_{1/2}(\bar \rho)} \leq 2^{H_{1/2}(\bs\varphi) }
			\leq |\Omega|.
\end{equation}	
The first inequality comes from \eqref{B 01 loss}. The second one comes since 
a mixture of non-orthogonal quantum states has always less entropy than 
the mixture itself. This can be proven easily by adapting a similar proof for the von 
Neumann entropy \cite{Nielsen_Chuang}, by 
defining the pure state 
$\ket{\varphi_{AB}} = \sum_\lambda \sqrt{\varphi_\lambda}\ket{\varphi_\lambda}\ket{\lambda}$, 
and noting that $H_{1/2}(A)=H_{1/2}(B)\leq H_{1/2}({B'})$ where $B'$ is state of  $B$
after performing a projective measurement on the $\ket\lambda$ basis, and
the inequality follows from the data-processing inequality under unital channels.

\section{On Learning to Discriminate: Average-Case Error vs. Worst-Case Error}\label{sec:averagevswc}

As discussed in section~\ref{sec:learning discriminate}, the problem of state discrimination can be formulated as the determination of whether the training and test systems are in either of the states (\ref{eq:overallstate}), where the two component states $\rho_+$ and $\rho_-$ are unknown. As in classical statistics, one can formulate the problem in a \emph{Bayesian}, or \emph{average-case} setting, attaching a prior distribution to the unknown states; or in a \emph{min-max}, or  \emph{worst-case},  setting, in which performance is maximized by assuming the worst-case pair of unknown states  $\rho_+$ and $\rho_-$ (see, e.g., \cite{poor1998introduction}).

In a Bayesian formulation, we represent the learner's prior uncertainty about what the states $\rho_+$ and $\rho_-$ via a classical probability distribution $p(\rho_+,\rho_-)$, which is defined over the space of density matrices $\rho_+$ and $\rho_-$. For instance, if the learner only knows that both states are pure qubit states, the prior $p(\rho_+,\rho_-)$ may be selected as a product  uniform distribution over all pairs of states on the surface of the Bloch sphere. 

Having selected a prior distribution $p(\rho_+,\rho_-)$, the task at hand can be expressed as  the single-shot discrimination between the average states $\sigma_+$ and $\sigma_-$ defined as
\begin{equation}
    \sigma_{\pm} = \int \int p(\rho_+,\rho_-) 
	\underbrace{\rho_{+}^{\otimes S} \otimes \rho_{-}^{\otimes S}}_{\rm training}
		 \otimes\rho_{\pm}^{\otimes V} d\rho_{+} d\rho_{-},
\end{equation}
where the integral can be replaced with a sum if the prior is defined over a discrete set of possible state pairs.

Therefore, the optimal solution that minimizes the average detection probability is given by the Helstrom measurement between states $\sigma_+$ and $\sigma_-$, which is defined by the projection matrices
\begin{equation}\label{eq:Bayesianoptmeas}
    \Pi^{\mathrm{opt}}_{\pm} = \frac{\mathbf{1}\pm \mathrm{sign}(\sigma_+ - \sigma_-)}{2} = \frac{1}{2}\left(\mathbf{1} \pm \int \int p(\rho_+,\rho_-) 
	\rho_{+}^{\otimes S} \otimes \rho_{-}^{\otimes S}
		 \otimes\mathrm{sign}(\rho_{+}^{\otimes V}-\rho_{-}^{\otimes V}) d\rho_{+} d\rho_{-}\right).
\end{equation}
Clearly, the optimal measurement is highly dependent on the a priori distribution, $p(\rho_+,\rho_-)$.

If we were to take an inductive approach, our measurement would need to be
expressed in the inductive-learning form \cite{sentis_quantum_2015,fanizza_universal_2022}, 
via a binary  POVM $\{M,\mathbf{I}-M\}$ operating on system (\ref{eq:overallstate}) with measurement operators of the form \begin{equation}
    M = \sum_{i=1}^{k} A_i\otimes B_i,\label{eq:adaptive measurement}
\end{equation}
where $\{A_1,A_2,...A_k\}$ is a set of measurement operators defining a valid measurement on the training set, whilst each set $\{B_i,\mathbf{I}-B_i\}$, for all $i=1,...,k$, defines a valid measurement on the test states. In words, an inductive learning strategy involves carrying out a measurement (or adaptive set of measurements) on the training set in order to extract information about which measurement to carry out on the test states in order to best discriminate between the possible test states $\rho_+^{\otimes V}$ and $\rho_-^{\otimes V}$. 
There is no guarantee that the optimal measurement from Eq.~(\ref{eq:Bayesianoptmeas}) can be expressed in this form.

For specific a priori distributions, the exact form of the optimal measurement (\ref{eq:Bayesianoptmeas}), along with the minimum average error, are known. For example, when $V=1$ and the prior is  uniform distribution over the surface of the Bloch sphere,  it has been shown that the optimum measurement, whose performance was studied in \cite{hayashi_quantum_2005}, can be achieved with a specific inductive scheme\cite{sentis_quantum_2012}. Similar cases with $V>1$\cite{he_programmable_2007}, mixed states, and qudits\cite{fanizza_optimal_2019} have also been studied in this context.

In a worst-case formulation, optimality is formulated as the maximization  of the minimum  detection probability, where the inner minimum is taken over all possible pairs of states  $\{\rho_+,\rho_-\}$ within some set of interest.
This is the formulation we use in this article.

\medskip\noindent
\textbf{Supporting Information} \par 
Supporting Information is available from the Wiley Online Library or from the author.

\medskip\noindent
\textbf{Acknowledgements} \par
L.B.~acknowledges financial support from PNRR MUR project PE0000023-NQSTI. 
J.P.~acknowledges funding from the U.S. Department of Energy, Office of
Science, Superconducting Quantum Materials and Systems Center (SQMS) under the
Contract No. DE-AC02-07CH11359.  The work of O.S. was supported by the European
Union’s Horizon Europe project CENTRIC (101096379) and by an Open Fellowship
of the EPSRC (EP/W024101/1).

\end{document}